\documentclass[a4paper,titlepage,11pt]{article}
\pdfoutput=1
\usepackage[utf8]{inputenc}
\usepackage{geometry}
\usepackage[english]{babel}
\usepackage{amsmath, amsfonts, graphicx}
\usepackage{mathtools}
\usepackage{graphicx}
\usepackage{float}
\usepackage{mciteplus}
\usepackage{bbold}
\usepackage{pstricks}
\usepackage[labelfont=bf,font={small}]{caption}
\usepackage{pgfplots}
\usepackage{empheq}
\usepackage{tikz}
\usetikzlibrary{positioning, arrows}
\usetikzlibrary{external}
\usepackage{footmisc}
\usepackage{multirow}
\usepackage{url}
\usepackage[breakable, theorems, skins]{tcolorbox}
\usepackage{enumerate}
\usepackage{physics}
\usepackage{bm}
\usepackage{braket}
\usepackage{epstopdf}
\usepackage{hyperref}

\definecolor{blue(pigment)}{rgb}{0.2, 0.2, 0.6}
\definecolor{darkblue}{rgb}{0.0, 0.0, 0.55}

\hypersetup{
    colorlinks,
    citecolor=blue,
    filecolor=blue,
    linkcolor=black,
    urlcolor=blue
}
\usepackage{xfrac}
\newcommand*\widefbox[1]{\fbox{\hspace{0.5em}#1\hspace{0.5em}}}

\def\be{\begin{equation}}
\def\ee{\end{equation}}

\newtagform{normalsize}[\normalsize]{\normalsize(}{\normalsize)}

\def\bea{\begin{eqnarray}}
\def\eea{\end{eqnarray}}

\newcommand{\beq}{\begin{equation}}
\newcommand{\eeq}{\end{equation}}
 
\newcommand{\barr}{\!\begin{array}}
\newcommand{\earr}{\end{array}\!}

\def\be{\bea}
\def\ee{\eea}

	\numberwithin{equation}{section}

\begin{document}
\usetagform{normalsize}

\begin{titlepage}

\setcounter{page}{1} \baselineskip=15.5pt \thispagestyle{empty}

\vfil

${}$
\vspace{1cm}

\begin{center}

\def\thefootnote{\fnsymbol{footnote}}
\begin{changemargin}{0.05cm}{0.05cm} 
\begin{center}
{\Large \bf Branes in JT (super)gravity from group theory}
\end{center} 
\end{changemargin}

~\\[1cm]
{Andreas Belaey,\footnote{{\protect\path{andreas.belaey@ugent.be}}} Francesca Mariani,\footnote{{\protect\path{francesca.mariani@ugent.be}}} Thomas G. Mertens\footnote{{\protect\path{thomas.mertens@ugent.be}}}}
\\[0.3cm]
{\normalsize { \sl Department of Physics and Astronomy
\\[1.0mm]
Ghent University, Krijgslaan, 281-S9, 9000 Gent, Belgium}}\\[3mm]

\end{center}


 \vspace{0.2cm}
\begin{changemargin}{01cm}{1cm} 
{\small  \noindent 
\begin{center} 
\textbf{Abstract}
\end{center} 
In this work, we revisit the end-of-the-world (EOW) brane amplitudes in JT gravity from a BF gauge theoretic perspective. Observing and identifying the correct group theoretic ingredient for a closed EOW brane as a discrete series character, we use the group theory framework as a guide towards formulating the analogous supersymmetric problem. We compute these amplitudes explicitly in the supersymmetric generalizations of JT gravity ($\mathcal{N}=1,2,4$), motivated by the prospective of possibly finite amplitudes. In the process, we develop some of the representation theory of OSp$(2\vert 2,\mathbb{R})$ and PSU($1,1\vert 2)$, relevant for the $\mathcal{N}=2$ and $\mathcal{N}=4$ cases.

}
\end{changemargin}
 \vspace{0.3cm}
\vfil
\begin{flushleft}
\today
\end{flushleft}

\end{titlepage}
\newpage
 \tableofcontents

\newpage
\hypersetup{linkcolor=blue}

\setcounter{footnote}{0}

\section{Introduction}

A detailed understanding of quantum black holes remains one of the biggest goals in the field. 
A particularly attractive model is 1+1d Jackiw-Teitelboim (JT) gravity \cite{Jackiw:1984je, Teitelboim:1983ux}, which captures the near-horizon region of a large class of higher-dimensional nearly extremal black holes, see e.g. \cite{Nayak:2018qej,Iliesiu:2020qvm,Castro:2021csm,Iliesiu:2022kny,Castro:2022cuo}. This model has been heavily investigated at the quantum gravitational level in many recent works, see e.g. \cite{Almheiri:2014cka, Jensen:2016pah, Maldacena:2016upp, Engelsoy:2016xyb, Cotler:2016fpe, Stanford:2017thb, Kitaev:2018wpr, Mertens:2017mtv, Mertens:2018fds, Lam:2018pvp, Harlow:2018tqv, Yang:2018gdb, Blommaert:2018oro, Blommaert:2018iqz, Iliesiu:2019xuh, Saad:2019lba,Saad:2019pqd,Blommaert:2019wfy,Okuyama:2019xbv,Blommaert:2020seb,Saad:2021rcu,Post:2022dfi,Altland:2022xqx,Jafferis:2022wez,Blommaert:2021fob,Griguolo:2023aem} for a selection, and was recently reviewed in \cite{Mertens:2022irh}.
The amount of solvability in this model is unprecedented. This allows us to make real quantitative predictions to long-standing problems. Most notably, coupling the theory to matter, it is able to shed a new light on the Hawking information paradox. By including non-perturbative corrections to the gravitational path integral, the result is a unitary Page curve of the entropy of the Hawking radiation \cite{Penington:2019kki, Almheiri:2019qdq}. In \cite{Penington:2019kki}, the black hole microstates during the evaporation process are modeled by brane-like objects that end spacetime, the so-called end-of-the-world (EOW) branes.
These EOW branes, that we wish to consider in this work, were first introduced in this model in \cite{Kourkoulou:2017zaj} with the aim of geometrically describing pure states in the gravitational quantum Hilbert space. Finally, dynamical EOW branes were considered in \cite{Gao:2021uro}, with the attempt of dealing with a gas of them.

EOW branes can be defined from the (Lorentzian signature) action \cite{Gao:2021uro}: \begin{equation}\label{C5:kolchmeyeraction}
S=\frac{1}{2} \int d^2x\; \phi \sqrt{-g}(R+2)+\int_{\partial\text{AdS}}d\tau\; \phi\sqrt{-g_{\tau\tau}}(K-1)+\int_{\text{EOW}}ds \sqrt{-g_{ss}}(\phi K-\mu).
\end{equation}  
The first two terms are the usual JT gravity actions with $\phi$ the dilaton field and $R$ the Ricci scalar; the second term is the boundary term at a holographic boundary including a counterterm. We omit a factor of $\frac{1}{8\pi G}$ by convention. The coordinates $\tau$ and $s$ represent two timelike parameters with one-dimensional induced metrics $g_{\tau\tau}$ and $g_{ss}$ along the AdS$_2$ boundary and the EOW brane boundary respectively. The parameter $\mu$ denotes the mass of the EOW particle, while $K$ denotes the extrinsic curvature along the two respective boundaries of the spacetime.

Various quantum amplitudes have been obtained using the boundary-particle formalism. The results depend greatly on the topology. In \cite{Penington:2019kki}, several quantum amplitudes of EOW branes attached to the disk partition function have been obtained. The result with one EOW brane boundary can be written as:  
\begin{equation}
\label{C5:fininalEOWdiskpartition}
 \raisebox{-0.45\height}{\includegraphics[height=2.5cm]{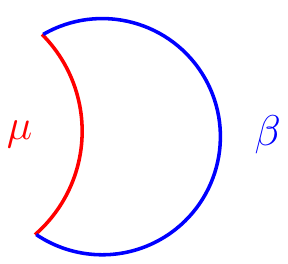}}=\int_{-\infty}^{+\infty}db\;Z_{\text{HH}}(\beta,b)e^{-\mu b},
\end{equation} 
where $Z_{\text{HH}}(\beta,b)$ denotes the Hartle-Hawking state preparing the vacuum, with a geodesic boundary of length $b$ and an asymptotic boundary of length $\beta$: \begin{equation}
    Z_{\text{HH}}(b,\beta)=
    \int_0^\infty dk\; k\sinh{2\pi k}\;e^{-b/2} K_{2ik} (e^{-b/2}) e^{-\beta k^2}.
\end{equation}
We may already guess the appearance of the EOW brane wavefunction $e^{-\mu b}$ from the classical on-shell approximation of the action \eqref{C5:kolchmeyeraction}.

Next to this, \cite{Gao:2021uro} has obtained the quantum amplitude of an EOW brane loop attached to the neck of a single trumpet: \begin{equation}\label{C5:fininalEOWtrumpetpartition}
     \mathcal{A}(\beta;\mu)=\hspace{0.5cm}\raisebox{-0.5\height}{\includegraphics[height=1.65cm]{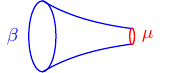}}=\int_0^\infty db\; Z_{\text{trumpet}}(\beta, b)\frac{e^{-\mu b}}{2\sinh(b/2)}.
\end{equation}
\\
As opposed to \eqref{C5:fininalEOWdiskpartition}, this result exhibits an unsettling correction to the classical saddle in the denominator of the EOW brane wavefunction. In particular, there is a UV divergence of this amplitude as the length of the brane circle approaches zero, $b\rightarrow 0$. Since this particular amplitude is the main building block in the gas of EOW branes picture of \cite{Gao:2021uro}, one has to be careful with the interpretation of the $b\to 0$ region. Divergences of this kind, where the neck of a wormhole shrinks to zero size, also appear in matter-coupled JT gravity, as pointed out in \cite{Saad:2019lba}. In \cite{Jafferis:2022wez}, a work-around was presented by $q$-deforming JT gravity in a suitable way.

We are motivated by another possibility of dealing with the $b\to 0$ divergence. As elaborated on in detail in \cite{Moitra:2021uiv}, the $b\to 0$ divergence is actually analogous to the closed string tachyon divergence in bosonic string theory. In fact, as studied in \cite{Saad:2019lba,Mertens:2020hbs,Fan:2021bwt,Goel:2020yxl,Suzuki:2021zbe,Collier:2023cyw,Blommaert:2023wad}, JT gravity can be found as a limiting model of an actual string theory (the minimal string or more general Liouville gravities), so this statement is more than an analogy.
As usual in string theory, the tachyon divergence is dealt with by instead considering superstring theory where the tachyon mode can be consistently projected out of the string spectrum \cite{Gliozzi:1976qd}. We are hence led to investigate the supersymmetric versions of JT gravity to find out whether they allow for finite amplitudes of this kind.
Supersymmetric versions of JT gravity have been defined and studied alongside the bosonic model ever since its conception, see e.g. \cite{Chamseddine:1991fg, Astorino:2002bj} and \cite{ Forste:2017kwy,Forste:2017apw,CamposDelgado:2022cwu} for relevant recent work on the boundary super-Schwarzian descriptions of these models. We distinguish the $\mathcal{N}=1,2,4$ supersymmetric models. Just like the bosonic model is based on the $\mathfrak{sl}(2,\mathbb{R})$ algebra, the supersymmetric versions are based on the $\mathfrak{osp}(1\vert 2,\mathbb{R})$, $\mathfrak{osp}(2\vert 2,\mathbb{R})$, and $\mathfrak{psu}(1,1\vert 2)$ superalgebras for $\mathcal{N}=1,2,4$ respectively. Progressively less is known on the quantum gravitational amplitudes as one increases the amount of supersymmetry. For $\mathcal{N}=1$, we refer the reader to \cite{Mertens:2017mtv,Fan:2021wsb} for the boundary correlators, and to \cite{Stanford:2019vob} for the sum over topologies. For $\mathcal{N}=2$, the boundary correlators were found in \cite{Lin:2022zxd}, and the sum over topologies in \cite{Turiaci:2023jfa}, see \cite{Boruch:2023trc} for interesting applications. Matrix model techniques were applied to these cases in e.g. \cite{Johnson:2019eik,Johnson:2020heh,Johnson:2023ofr}. For $\mathcal{N}=4$, partial results are known on the sum over topologies in the same work \cite{Turiaci:2023jfa}.

In this work, we will compute the particular gravitational EOW brane amplitude \eqref{C5:fininalEOWtrumpetpartition} for all supersymmetric versions of JT gravity. Our calculational method is based on the BF gauge theoretical description of JT (super)gravity \cite{Fukuyama:1985gg, Isler:1989hq,  Chamseddine:1989yz}. In particular, in \cite{Blommaert:2021etf}, it was observed that the form $\frac{e^{-\mu b}}{2\sinh(b/2)}$ of this EOW brane wavefunction coincides with a discrete series character of $\text{SL}(2,\mathbb{R})$. Here we will build on this observation. Our main new results, generalizing \eqref{C5:fininalEOWtrumpetpartition}, can be found in equations \eqref{kolchmeyersucker1}, \eqref{kolchmeyersucker2} for $\mathcal{N}=1$, \eqref{eq:cl1}, \eqref{eq:cl2} for $\mathcal{N}=2$, and \eqref{eq:n4res} for $\mathcal{N}=4$. \\

This work is structured as follows.
In \textbf{section \ref{Section2}}, we aim to elaborate on the above observation, and develop a generic method to arrive at the amplitude \eqref{C5:fininalEOWtrumpetpartition} within the framework of the BF formulation of JT gravity. The main motivation will be to generalize this framework to EOW branes in theories of JT supergravity in the next sections.

Subsequent \textbf{sections \ref{Section3}}, \textbf{\ref{Section4}} and  \textbf{\ref{Section5}} then develop the machinery for the $\mathcal{N}=1,2,4$ JT supergravity models respectively. In each case, we provide appropriate definitions of the EOW branes. As we go up in the amount of supersymmetry, more and more complications will arise that we will have to deal with. Finally, in \textbf{section \ref{sec:conc}} we present some open questions left for future and ongoing work.

The technical framework requires quite a bit of representation theory of these higher supersymmetric models. We develop the required representation theory of the $\mathcal{N}=2$ OSp$(2\vert 2, \mathbb{R})$ supergroup in \textbf{appendix \ref{sec:N2rep}}, and of the $\mathcal{N}=4$ PSU$(1,1\vert 2)$ supergroup in \textbf{appendix \ref{AppendixE}}. These results could be of interest to the reader beyond the current applications. Further technical details are contained in the other appendices.

\section{EOW brane amplitudes in bosonic JT gravity}
\label{Section2}

\subsection{Geodesic description of EOW branes}
\label{section:geodesicdescriptionEOWbranes}
The term corresponding to the action of the EOW brane particle in \eqref{C5:kolchmeyeraction}, written in Euclidean signature is \cite{Gao:2021uro}:
\begin{equation}
\label{eq:ewobran}
    I=\int_{\mathcal{C}} ds\sqrt{g_{ss}}(\mu-\phi K).
\end{equation} 
The first term in this action corresponds to the worldline action of the massive EOW particle. In this context, the mass $\mu$ is often denoted as the tension along the brane. The second term in (\ref{eq:ewobran}) involves the extrinsic curvature $K$. When inserted in the path integral, the value $\phi$ at the EOW brane trajectory acts as a Lagrange multiplier, enforcing the off-shell constraint on the particle's trajectory:\footnote{Similarly to the bulk dilaton field, we need to path integrate along an imaginary contour. This procedure (both in the bulk and here) can be viewed as defining the Euclidean gravitational path integral and in the process resolving the negative conformal mode problem in Euclidean quantum gravity in this set-up.}
\begin{equation}
    K=0.
\end{equation}
The vanishing of the extrinsic curvature trace along the particle's trajectory severely restricts its shape. In particular, it is well-known that $K=0$ trajectories are geodesics. 

In the interest of generalizing this to superspace in the next sections, let us demonstrate this by starting from the geodesic equation of $x^\mu(s)$ labeled by an affine parameter $s$ along the curve $\mathcal{C}$: \begin{align}
    \label{C5:geodesicequation1} 
    U^\alpha \nabla_\alpha U^\mu=0.
\end{align}
Here, $U^\mu(s)\equiv \frac{ dx^\mu}{ds}(s)=\Dot{x}^\mu(s)$ denotes the tangent vector along the curve. The normal vector $n^\mu(s)$ is defined to be orthogonal to the tangent vector along the entire curve: 
\begin{equation}
     U^\alpha(s)n_\alpha(s)\equiv 0.
\end{equation} 
Applying the product rule for covariant derivatives on this definition readily yields a relation between the variation of the tangent vector and the variation of the normal vector: \begin{equation}
    n_\alpha\nabla_\mu U^\alpha =-U^\alpha \nabla_\mu n_\alpha.
\end{equation} 
The geodesic equations follow from the variational solutions of the worldline action along the curve. Since the variation in any direction can be decomposed into its tangential and normal direction, we can restrict to the normal direction $\delta x^\mu=n^\mu$:\footnote{The variation tangential to a given trajectory trivially yields $\delta I=0$, as one can show by taking the covariant derivative of $U_\mu U^\mu =1$. This is intuitive since this is merely a reparametrized version of the same curve and the worldline action is reparametrization-invariant.} 
\begin{align}
    \delta I =-\int ds\; \delta x_\mu (U^\alpha \nabla_\alpha U^\mu)=-\int ds\; n_\mu (U^\alpha \nabla_\alpha U^\mu)=\int ds\; U^\mu U^\alpha \nabla_\alpha n_\mu.
\end{align}
Here, we recognize the definition of the extrinsic curvature trace along the curve $x^\mu(s)$: \begin{equation}\label{C5:extrinsiccurvaturedefinitionbosonic}
    K=U^\mu U^\alpha \nabla_\alpha n_\mu.
\end{equation} The variation of the action is therefore completely specified by the value of the extrinsic curvature: \begin{equation}\label{C5:actionvariation}
    \delta I \sim \int ds\; K.
\end{equation} Hence, on every curve for which $K\equiv 0$, the variation of the worldline action vanishes, constraining it to solutions of the geodesic equation. \\

If we insert (\ref{eq:ewobran}) in the path integral, we can rewrite: \begin{equation}\label{C6:pathintegralrestrictiongeodesic}
    \int_{}\mathcal{D}x\; e^{-\int_{\mathcal{C}} ds\sqrt{g_{ss}}(\mu-\phi K)} \quad \xrightarrow{\text{Integrate over } \phi} \quad \int_{\text{geodesics}}\mathcal{D}x\; e^{-\mu\int_{\mathcal{C}}ds \sqrt{g_{\alpha\beta}\dot{x}^\alpha \dot{x}^\beta}}.
\end{equation} 
Evaluating the worldline path integral on the RHS over geodesics only, is the same as evaluating the worldline path integral in the saddle approximation. The path integral effectively localizes along those classical solutions in the limit of large mass, $\mu \gg1$. 

\subsection{Wilson loops as probe particles}
\label{section:wilsonlinesasprobeparticles}
A crucial identity to interpret EOW branes in a gauge-theoretic formulation is the equivalence between Wilson operator (lines/loops) insertions and probe particles in the second order metric formulation, proposed in the context of AdS$_3$ in e.g. \cite{Ammon:2013hba} and \cite{Castro:2018srf}, and formulated in the context of JT in \cite{Iliesiu:2019xuh}.

Functionally integrating the worldline of a massive particle over a closed path $\mathcal{C}$ results in a trace of the holonomy over the $\text{SL}(2,\mathbb{R})$ spin-$j$ discrete series highest-weight module, which in a BF path integral is just a Wilson loop  insertion: \begin{equation}\label{C5:operatoridentity}
\mathcal{W}_j(\mathbf{A})\;=\;\text{Tr}_j\left(\mathcal{P}\exp-\oint_{\mathcal{C}}\mathbf{A}\right)  \simeq \oint_{\text{paths }\sim \, \mathcal{C}}\mathcal{D}x\;e^{-\mu \int ds\;\sqrt{g_{\alpha\beta}\dot{x}^\alpha\dot{x}^\beta}},
\end{equation}
where the RHS contains all paths diffeomorphic to the curve $\mathcal{C}$ on the LHS. This identity should therefore be understood as an operator equivalence \emph{inside} the BF path integral over flat gauge connections, where infinitesimal gauge transformations on flat gauge fields are equivalent to infinitesimal diffeomorphisms in the gravity theory \cite{Iliesiu:2019xuh}. The precise argument in favor of this equality will be redone later on several occasions when we generalize to the supersymmetric cases.

From generic AdS/CFT considerations, the conformal weight of a primary operator $h$ is related to the mass of the dual scalar field $\mu$ by: \begin{equation}\label{C4:conformalscalingmass}
    h = \frac{1}{2}+\sqrt{\frac{1}{4}+\mu^2}, \qquad\rightarrow\qquad \mu^2=h(h-1).
\end{equation} 
In terms of the representation label $j=-h$ \cite{Blommaert:2018iqz}, this is related to the eigenvalue of the quadratic Casimir
\begin{equation}
\label{C5:masstuning}
\mu^2=h(h-1)=j(j+1)\equiv\mathcal{C}_2.
\end{equation} 

In the limit where we localize along geodesics, we may neglect the linear term in the relation between mass and conformal weight \eqref{C4:conformalscalingmass}, and identify the mass of the probe with the conformal weight of the Wilson operator: \begin{equation}\label{C5:geodesiclimitlarge}
     \mu\approx h, \qquad \mu\gg 1.
\end{equation}

\subsection{Gravitational amplitudes involving EOW branes}

We start by adding the Euclidean EOW brane action \eqref{eq:ewobran} in the gravitational path integral, and path integrate the dilaton along the trajectories of the EOW brane. We may formally write the gravitational amplitude as:
\begin{equation}
\label{C2:gravintegral}
\mathcal{A}(\beta;\mu) \equiv \int\mathcal{D}g\mathcal{D}\phi\oint_{\text{geodesics}}\mathcal{D}x\;e^{-\mu \int ds\;\sqrt{g_{\alpha\beta}\dot{x}^\alpha\dot{x}^\beta}}\;e^{-I_{\text{JT}}[g,\phi]},
\end{equation} 
where we path integrate over all closed geodesic worldlines. Splitting open the integral into geodesics with fixed length $b$ as:
\begin{equation}\label{Eq:integralsplit}
\oint_{\text{geodesics}}\mathcal{D}x \, = \, 
\int_0^{+\infty} db \, \oint_{\text{geodesics with length $b$}}\mathcal{D}x,
\end{equation}
the fixed length integrand evaluates to a Wilson operator insertion, which in the BF language reads:
\begin{equation}\label{C5:bfidentification}
\mathcal{A}(\beta;\mu) = \int_0^{+\infty} db\int_{e^{-\oint A} \, \simeq \, e^{bH}} \mathcal{D}\mathbf{B}\mathcal{D}\mathbf{A}\;\mathcal{W}_{j}(\mathbf{A})e^{-I_{\text{BF}}[\mathbf{B}, \mathbf{A}]}.
\end{equation}
Note that we path integrate over the BF-model with a twisted holonomy constraint as indicated, obtained by exponentiating the hyperbolic Cartan generator $H \in \mathfrak{sl}(2,\mathbb{R})$, $e^{b H}$, which in the fundamental representation reads $\left(\begin{array}{cc}
     e^{b/2}&0\\
      0&e^{-b/2}
    \end{array}\right)$.
This non-local boundary condition implements a standard hyperbolic defect in the BF path integral, which in turn ensures a topological deformation of pure JT on the plane to a single trumpet with geodesic neck length $b$ \cite{Mertens:2019tcm}. 
We can graphically depict this decomposition as:
\begin{equation}
    \raisebox{-0.5\height}{\includegraphics[height=1.7cm]{trumpet_new2.pdf}} =\quad \int_0^{+\infty} db\quad \raisebox{-0.5\height}{\begin{tikzpicture}
 \draw [blue, line width=0.3mm](0,0) circle (1.6);
  \draw [red, line width=0.3mm](0,0) circle (0.8);
  \draw [line width=0.2mm](0,0) circle (0.2);
  \draw [fill, red](-0.1,0.8) -- (0.1,0.9) -- (0.1,0.7) -- cycle;
   \draw (-0.13,0.16) -- (0.13,-0.16) ;
    \draw (0.13,0.16) -- (-0.13,-0.16) ;
  \node at (1.4, 1.4) {{\large\textcolor{blue}{$\beta$}}};
  \node at (0.8, 0.8) {\large\textcolor{red}{$\mu$}};
  \node at (0.3, 0.3) { \large$b$};
\end{tikzpicture}}\;.
\end{equation} \\
The defect can also be regarded as a vertical Wilson line that pierces the two-dimensional disk, and descends from the dimensionally reduced 3d Chern-Simons theory \cite{Blommaert:2018oro,Mertens:2019tcm}. From this perspective, we have two linked Wilson lines, where the encircling one can be viewed as measuring the label $b$ of the inner one. 

Integrating out the dilaton along the EOW brane yields the identification \eqref{C5:geodesiclimitlarge}.
The path integral over the closed EOW brane contour generates a Wilson loop $\mathcal{W}_j(\mathbf{A})$, which evaluates to a trace over a highest-weight discrete series irrep of the holonomy of $\mathbf{A}$ around this contour. 
Due to the twisted boundary condition, the trace of the holonomy $\mathcal{W}_j(\mathbf{A})$ will be evaluated as a hyperbolic character in the highest-weight discrete series representation
\cite{Vilenkin}:  
\begin{equation}
\label{C5:discretehyperboliccharacter}
    \chi_j(\phi)
    =\text{Tr}_j(e^{2\phi H})=\frac{e^{(2j+1) \phi}}{2\sinh(\phi)} \approx \frac{e^{-\mu b}}{2\sinh\frac{b}{2}}, \qquad j=-\frac{1}{2},-1,\hdots
\end{equation}
where we have used the dictionary between the hyperbolic parameter $\phi$ and geodesic length $b=2\phi$, and the limit of large $-j=h\approx\mu \gg1$.
This limit can also be interpreted as the saddle approximation of the worldline path integral, including the one-loop determinant over closed loop trajectories. \\

The amplitude of an EOW brane attached to the neck of a single trumpet can now be readily deduced. As a first step, one should introduce a hyperbolic defect in the bulk, creating a non-trivial monodromy along the thermal boundary circle. The procedure was explained in detail in \cite{Mertens:2019tcm}. The essential takeaway is to introduce a hyperbolic character evaluated in the continuous series representation labeled by a ``momentum'' label $k$ \cite{Vilenkin}:
 \begin{equation}
 \chi_k(\phi)=\frac{\cos(2k\phi)}{\sinh(\phi)}.
 \end{equation}
Removing the Weyl denominator immediately leads to the hyperbolic defect insertion in the trumpet partition function \cite{Mertens:2019tcm}: 
\begin{equation}\label{eq:bosonictrumpet}
    Z_{\text{trumpet}}(\beta, \phi)=\int_0^\infty dk\;\cos(2k \phi) e^{-\beta k^2}.
\end{equation} 
Combined with the EOW brane character of \eqref{C5:discretehyperboliccharacter}, and gluing along positive $b=2\phi$ \eqref{C5:bfidentification}, finally recovers exactly the partition function \eqref{C5:fininalEOWtrumpetpartition} derived by \cite{Gao:2021uro} from the boundary particle formalism:
\begin{equation}\label{C5:finalEOWtrumpetamplitude}
     \mathcal{A}(\beta;\mu)=\raisebox{-0.5\height}{\includegraphics[height=1.5cm]{trumpet_new2.pdf}}\hspace{-0.4cm}=\int_0^{\infty}dk\;e^{-\beta k^2} \int_0^\infty db\;\cos(kb)\frac{e^{-\mu b}}{2\sinh(b/2)}.
\end{equation} 
This opens up a way to extrapolate the notion of EOW branes to more exotic theories of JT supergravity, entirely from their group theoretic formulations.
 
\section{EOW brane amplitudes in $\mathcal{N}=1$ JT supergravity}
\label{Section3}
We start by running the story for $\mathcal{N}=1$ JT supergravity.

\subsection{EOW branes in superspace}
\label{s:eowsuper}
Our goal is to formulate an equivalent boundary action along the lines of \eqref{C5:kolchmeyeraction}, that captures the dynamics of end-of-the-world branes in superspace, and thereby to extend the discussion of the previous section to $\mathcal{N}=1$ JT supergravity. First of all, we recapitulate the JT supergravity action in $2|2$-dimensional superspace formulated in \cite{Chamseddine:1991fg}, with the appropriate $1|1$-dimensional UV boundary term formulated in \cite{Forste:2017kwy}: \begin{equation}\label{C6:totalactionwithoutEOW}
    I_{\text{JT}}^{\mathcal{N}=1}=-\frac{1}{2}\left[\int d^2zd^2\theta\;E\Phi(R_{+-}+2)+2\int_{\partial \text{AdS}}d\tau d\vartheta\; \Phi \mathbf{K}\right].
\end{equation}
The bulk superspace is spanned by two real holomorphic and antiholomorphic coordinates $z$ and $\Bar{z}$, and two fermionic (Grassmann) holomorphic and antiholomorphic coordinates $\theta$ and $\Bar{\theta}$, collectively denoted by $Z^M=\left(z, \overline{z}\mid \theta, \overline{\theta}\right)$. $E$ is the superdeterminant of the frame fields in superspace $E=\text{sdet}(E^A_{\;\;M})$, $R_{+-}$ the scalar supercurvature containing the usual scalar curvature in the $\theta\Bar{\theta}$-term in the superspace expansion \cite{PSHowe_1979}, and $\Phi$ the superdilaton field containing the scalar dilaton field $\phi$ in the bottom component of the superspace expansion. The extrinsic curvature along the UV boundary curve is defined from the first order tangent vectors along the curve $T^A=(\partial_\tau Z^M)E_M^{\;\;A}$ and the variation of the normal vectors defined by $T^An_A=0$ \cite{Forste:2017kwy}: \begin{equation}\label{C6:extrinsiccurvatureads}
    \mathbf{K} \equiv \frac{T^A D_T n_A}{T^AT_A},
\end{equation} in terms of a covariant derivative that acts as a superderivative $D=\partial_{\vartheta}+\vartheta\partial_\tau$ along the boundary equipped with the first order spin connections $\Omega_M$: \begin{equation}\label{C4:covariantderivativespinconnections}
    D_T n_A=D n_A +n_A\frac{\partial Z^M}{\partial\vartheta}\Omega_M+n_A\vartheta \frac{\partial Z^M}{\partial \tau} \Omega_M.
\end{equation} 
An important realization is that the boundary curves are in fact $1|1$-dimensional sheets that are infinitesimally thickened in the fermionic $\vartheta$-direction. I.e., the boundary curve is parameterized in terms of a bosonic $\tau$- and fermionic $\vartheta$-affine coordinate. In Poincaré super upper half-plane (SUHP) coordinates discussed in \cite{Fan:2021wsb}, the boundary curve covers the $1|1$-dimensional boundary sheet in the parametrization
\begin{equation}
    \tau'(\tau,\vartheta), \qquad y'(\tau, \vartheta), \qquad \theta'(\tau, \vartheta), \qquad \Bar{\theta}'(\tau, \vartheta),
\end{equation}
with $z'=\tau'+iy'$, $\Bar{z}'=\tau'-iy'$ and $\theta',\; \overline{\theta}'$ superconformal transformations of the Poincaré SUHP. However, we aim to describe EOW branes as geodesic curves in superspace. These are genuine $1|0$-dimensional curves in the $2|2$-dimensional superspace, describing the trajectory 
\begin{equation}
\label{C6:superspacecurves}
    z'(s), \qquad \Bar{z}'(s), \qquad \theta'(s), \qquad \Bar{\theta}'(s),
\end{equation} 
in terms of a single bosonic worldline parameter $s$, which we may take to be the proper length along the curve.

A natural first step is to add to the JT supergravity action \eqref{C6:totalactionwithoutEOW} a term containing the worldline action in superspace, labeled in terms of this bosonic worldline coordinate  $s$: \begin{equation}\label{C6:freeparticleaction}
    I=\mu \int_{\text{EOW}} ds\;\left(\dot{Z}^M g_{MN}\dot{Z}^N\right)^{1/2}.
\end{equation}
The target space coordinate $Z^M(s)=\left(z'(s), \overline{z}'(s) \mid \theta'(s), \overline{\theta}'(s)\right)$ labels the trajectory in $(2\vert 2)$-dimensional superspace, and the dot indicates differentiation with respect to the worldline parameter $s$. The quantity $g_{MN}$ denotes the metric in superspace, following the conventions of appendix \ref{app:diffgeo}.\footnote{We work in the NW-SE (north-west - south-east) convention, where covectors are constructed by acting with the metric on the left (c.f. (\ref{C6:covectordefinition}))\begin{equation}
    \Dot{Z}_M\equiv g_{MN}\Dot{Z}^N,
\end{equation} and coordinate-invariant contractions appear NW-SE 
\begin{equation}
\Dot{Z}^Ng_{NM}\Dot{Z}^M=\Dot{Z}^M\Dot{Z}_M.
\end{equation}
Similarly, Lorentz-contractions acting on local Lorentz indices, are defined NW-SE with respect to the constrained Cartan-Killing (CK) metric $\kappa_{AB}$, see (\ref{C6:locallorentzmetric}).}

Since our EOW brane trajectories form $1\vert 0$-dimensional worldlines, we need a different definition of the extrinsic curvature than \eqref{C6:extrinsiccurvatureads}. A natural choice would be to simply generalize the definition of the bosonic extrinsic curvature to superspace: \begin{equation}\label{C6:anticipatedresult}
    K=U^\mu U^\alpha \nabla_\alpha n_\mu \quad \rightarrow\quad K=U^N U^M \nabla_M n_N,\qquad U^M\equiv \Dot{Z}^M=\frac{dZ^M}{ds},
\end{equation} 
with the covariant derivative defined from the variation of the worldline action (see appendix \ref{app:diffgeo}). In particular, our convention for the superspace covariant derivative on both vectors and covectors is given in \eqref{C6:contravariantcovariantderivative} and \eqref{C6:covariantderivativeoncovector} in terms of an appropriate definition of the generalized Christoffel symbols \eqref{C6:defchristoffel}.

We define the normal vector in superspace $n_M(s)$ through the condition:
\begin{equation}
    U^M(s)\;n_M(s)\equiv 0.
\end{equation}
 The variation of the worldline action leads to the classical geodesic equations of a superparticle \eqref{eq:supergeodesiceqs}
\begin{equation}
    U^M\nabla_M U^N=0.
\end{equation}
We next show that this generalization directly characterizes geodesic curves as those which have $K=0$.

Taking a covariant derivative of the identity $U^Nn_N\equiv 0$, we can write
 \begin{align}
\nabla_M U^N\;n_N&=-(-)^{MN}U^N\;\nabla_M n_N,
\end{align}
where $M,N$ in the exponent denote the usual fermionic sign factors defined around \eqref{C6:conventies}.
Inserting this into the variation of the worldline action \eqref{eq:supervariation} yields: \begin{align}
     \delta I=-\mu \int ds\;\left(U^M\nabla_M U^N\right) n_N\;=\; \mu \int ds\; U^NU^M \nabla_M n_N \equiv \mu \int ds\; K,
\end{align}
where we have defined the extrinsic curvature along the $1|0$-dimensional curve as: 
\begin{equation}
\label{C6:defextrinsiccurvature}
K=U^NU^M\nabla_M n_N.
\end{equation}
This characterizes completely the variation of the worldline action in superspace. Any superparticle for which $
    K\equiv 0$
along its worldline has a vanishing variation of the worldline action $\delta I=0$, and hence follows its classical geodesic trajectory in superspace.\footnote{
Using the antisymmetry properties of the metric (\ref{C6:anticommutativityconvention}), contractions are commutative in the NW-SE direction: $V^N W_N = V^N g_{NK} W^K = W^K g_{KN} V^N = W^K V_K$. One can thus argue that \begin{equation}
    \nabla_M(V^NW_N) = (\nabla_M V^N) W_N + (\nabla_M W^N) V_N.
\end{equation} This property is compatible with the superspace analogues of the product rule for covariant derivatives (\ref{C6:covariantproductrule}) and the invariance of the metric tensor  (\ref{eq:metricpostulate}). Taking the tangent vectors to be normalized $U^NU_N=1$, it readily follows using the above property that the variation of the worldline action (\ref{eq:supervariation}) along the tangent direction $\delta Z^A=U^A$ again trivially vanishes: \begin{equation}
  \delta I=-\mu \int ds\;\left(U^M\nabla_M U^N\right) U_N = 0.
\end{equation}} \\

We may now extend the total Euclidean action of $\mathcal{N}=1$ JT supergravity \eqref{C6:totalactionwithoutEOW} in the presence of an EOW brane: 
    \begin{align}
    I_{\text{JT}}^{\mathcal{N}=1}=&-\frac{1}{2} \left[\int d^2zd^2\theta\;E\Phi(R_{+-}+2)+2\int_{\partial\text{AdS}}d\tau d\vartheta\; \Phi \mathbf{K}\right]\\
            \label{C6:totalactionwithEOW}
        &+ \int_{\text{EOW}}ds\;\sqrt{\Dot{Z}^Mg_{MN}\Dot{Z}^N}\; \left(\mu -\phi K\right),
    \end{align}
where $\phi $ coincides with the bottom component of the dilaton superfield $\Phi$. 

We emphasize again that the extrinsic curvature along the AdS-boundary $\mathbf{K}$ is different from the extrinsic curvature along the EOW brane $K$. The former is defined along the $1|1$-dimensional boundary curve \eqref{C6:extrinsiccurvatureads}, while the latter is defined along the $1|0$-dimensional brane in \eqref{C6:defextrinsiccurvature}.

Evaluating the quantum mechanical amplitude, including the boundary action \eqref{C6:totalactionwithEOW}, proceeds in the same way as the bosonic case. Path integrating over the dilaton superfield at the EOW brane boundary imposes the extrinsic supercurvature to vanish:
\begin{equation}
    K\equiv 0.
\end{equation} 
By construction of the extrinsic supercurvature above, the worldline path integral localizes along  geodesics in superspace as:
\begin{equation}
\label{eq:pathintegral}
    \int\mathcal{D}Z\; e^{- \int_\mathcal{C} ds\sqrt{\Dot{Z}^Mg_{MN}\Dot{Z}^N}(\mu-\phi K)} \; \xrightarrow{\text{Integrate over } \Phi} \; \int_{\text{geodesics}}\mathcal{D}Z\; e^{-\mu\int_\mathcal{C} ds\sqrt{\Dot{Z}^Mg_{MN}\Dot{Z}^N}}.
\end{equation} 
This localization along geodesics is equivalently achieved by taking $\mu\gg1$ to be large in the worldline path integral. 

To explicitly evaluate \eqref{eq:pathintegral}, we need an analogous identification between a worldline path integral and a Wilson operator insertion in the BF path integral, relevant for $\mathcal{N}=1$ JT supergravity. 

\subsection{Wilson loops as probe particles in superspace}
\label{app:wilsonlinesasprobes}
We first extend the proofs in appendix E of \cite{Iliesiu:2019xuh} and appendix C.3 of \cite{Fan:2021wsb}, to generalize the identification between Wilson loops and worldline path integrals in superspace for arbitrary amount of supersymmetry. The gauge groups of interest here are 2d superconformal groups $G$ for any amount of supersymmetry. They are characterized as having an SL$(2,\mathbb{R})$ subgroup: SL$(2,\mathbb{R}) \subset G$, identified as the gravity subsector, and possibly (for higher supersymmetry) a bosonic R-symmetry group $G_R$. The maximal bosonic subgroup is hence $\text{SL}(2,\mathbb{R}) \otimes G_R$. We work in Euclidean signature.\\

We start by introducing a gauge field for the group $G$, and expand it into the generators of the supergroup in terms of what we will later on identify as the first order superframe fields $E_M^{\;\;\;A}$ and superspin connection $\Omega_M$:
\begin{equation}
\label{C6:gaugefieldexpansion}
    \mathbf{A}_M= E_M^{\;\;\;A}J_A+\Omega_M J_2,\qquad A=0,1,a,\alpha,
\end{equation}
where letters at the beginning of the alphabet $A, B, \hdots $ denote Lorentz frame indices, while letters in the middle of the alphabet $M,N,\hdots$ denote Einstein superspace indices. Latin indices $a,b,\hdots$ denote additional bosonic generators in the (compact) R-symmetry group for higher supersymmetry. Greek indices $\alpha, \beta, \hdots$ denote spinor indices.

The first three bosonic generators $J_0,J_1,J_2$ are taken as the generators of the SL$(2,\mathbb{R})$ subgroup, and are related to the usual Cartan-Weyl basis of $\mathfrak{sl}(2,\mathbb{R})$ generators by \cite{Fan:2021wsb}: 
\begin{equation}
J_0 = -H, \qquad J_1 = \frac{1}{2}(E^- + E^+), \qquad J_2 = \frac{1}{2}(E^- - E^+),
\end{equation}
which in the fundamental representation of the subalgebra $\mathfrak{sl}(2,\mathbb{R})$ look like:
\begin{equation}
    H=\frac{1}{2}\begin{bmatrix}
        1&0\\0&-1
    \end{bmatrix},\qquad E^-=\begin{bmatrix}
        0&0\\1&0
    \end{bmatrix},\qquad E^+=\begin{bmatrix}
        0&1\\0&0
    \end{bmatrix}.
\end{equation}
In particular, $J_2$ is a compact generator, exponentiating into SO$(2) \subset \text{SL}(2,\mathbb{R})$. In addition, we have fermionic generators $J_\alpha$, and R-symmetry generators $J_a$.

The Cartan-Killing metric $\kappa_{AB}$ is defined by the normalization of the generators:
\begin{equation}
\label{C6:repeatedCKmetric}
    \text{STr}(J_AJ_B) =\frac{\kappa_{AB}}{2}.
\end{equation} 
These generators can be chosen to be normalized as:
\begin{align}
    &\kappa_{AB}=\text{diag}(1,1,-1), \quad (A,B=0,1,2), \qquad \kappa_{ab} = \delta_{ab}, \\ 
    &\kappa_{\alpha\beta} = \begin{cases} 2\epsilon_{\alpha\beta}: \text{ if $\alpha$ and $\beta$ are conjugate  pairs,} \\
    0: \text{ otherwise,} \end{cases}
\end{align} 
and all unwritten components zero. The restriction of the Cartan-Killing metric to the directions $A=0,1,\alpha$ coincides with the local Lorentz metric (in Euclidean signature) \cite{Fan:2021wsb}: 
\begin{equation}
\label{C6:locallorentzmetric}
\kappa_{ab}=\delta_{ab}, \qquad \kappa_{\alpha\beta}=2\epsilon_{\alpha\beta}, \qquad\kappa_{a\alpha}=\kappa_{\alpha a}=0.
\end{equation} 
The symmetry properties of the Cartan-Killing metric are summarized as: 
\begin{equation}
\label{C6:kappametricsymmetry}
    \kappa_{AB}=(-)^{AB}\kappa_{BA}.
\end{equation}

We can write a Wilson loop along $\mathcal{C}$ in the discrete series representation labeled by $j$ as a path integral of the first-order action $S_{\boldsymbol{\Lambda}}[g,\mathbf{A}]$ over the closed path $\mathcal{C}$ with dynamical variable $g(s)$:
\begin{equation}\label{C6:prelimidentification}
 \mathcal{W}_j(\mathbf{A})=\int_{\mathcal{C}} \mathcal{D}_\Lambda g\;e^{-S_{\boldsymbol{\Lambda}}[g,\mathbf{A}]}. 
\end{equation}  
The first order action is minimally coupled to a gauge field in superspace $\mathbf{A}_M=A_M{}^AJ_A$:
\begin{equation}\label{C6:specificfirstorderwilsonaction}
    S_{\boldsymbol{\Lambda}}[g,\mathbf{A}]=\int_\mathcal{C} ds\;\text{STr}\left(\boldsymbol{\Lambda} g^{-1}D_Ag\right),
\end{equation} 
with the covariant derivative defined symbolically as:  
\begin{equation}\label{eq:covariantdef}
D_A=\partial_s+\mathbf{A}_s.
\end{equation}
The gauge field along the curve is defined as $\mathbf{A}_s=\Dot{Z}^M \mathbf{A}_M(Z(s))$.
The argument is well-known and goes under the name of the Borel-Weil-Bott theorem. It appeared in the 3d Chern-Simons context in \cite{Moore:1989yh}, as nicely explained more recently in \cite{Beasley:2009mb,Fan:2018wya}; and has been generalized in the supergroup context in 
\cite{Penkov:1986ht,Mikhaylov:2014aoa}. 

The action \eqref{C6:specificfirstorderwilsonaction} has a gauge redundancy, being invariant under (local) left multiplication by elements of $G$:
\begin{equation}
\label{eq:gred}
g \to U g, \qquad \mathbf{A}_s \to U \mathbf{A}_s U^{-1} - \partial_s U U^{-1}.
\end{equation}
The vector $\boldsymbol{\Lambda}$ is the highest-weight vector of the spin-$j$ representation in the $\mathfrak{g}$-algebra. The precise choice of $\boldsymbol{\Lambda}$ can be changed by conjugation since the action is invariant under the (global) right-multiplication:
\begin{equation}
\label{eq:tf}
g \to g V, \qquad \boldsymbol{\Lambda} \to V^{-1}\boldsymbol{\Lambda} V.
\end{equation}

The adjoint action on $\boldsymbol{\Lambda}$ as in \eqref{eq:tf} does not change any invariant tensors or Casimir operators of the algebra. The representation is fixed by choosing the values of all Casimir operators. 

\subsection{Wilson loops as probe particles: rank-1 groups}
\label{s:rank1}

To proceed, we need to construct all Casimirs of the Lie superalgebra, which depend on the specific choice of $\mathfrak{g}$. For the $\mathcal{N}=0$ $\mathfrak{sl}(2,\mathbb{R})$ and $\mathcal{N}=1$  $\mathfrak{osp}(1|2,\mathbb{R})$ (super)algebra, there is only the quadratic Casimir (and a trivial R-symmetry) and we review how the argument works \cite{Iliesiu:2019xuh,Fan:2021wsb}, adding some clarifications. In a later section, we will have to revisit this argument to deal with higher amounts of supersymmetry (non-trivial R-symmetry group) and multiple Casimir operators.

Since, up to the value of the quadratic Casimir, the precise choice of the weight vector  $\boldsymbol{\Lambda}$ is irrelevant, we can average over this choice by introducing a functional integral over the latter:
\begin{equation}
\int \mathcal{D}\boldsymbol{\Lambda} \mathcal{D}\Theta \mathcal{D}_\Lambda g\;e^{-S_{\boldsymbol{\Lambda}}[g,\mathbf{A},\Theta]},
\end{equation}
with
\begin{equation}
S_{\boldsymbol{\Lambda}}[g,\mathbf{A},\Theta]=\int_\mathcal{C} ds\;\left[\text{STr}(\boldsymbol{\Lambda} g^{-1}D_Ag)+\frac{i}{2}\Theta(\Lambda^A\Lambda_A -4\mathcal{C}_2)\right],
\end{equation}
and with fixed value of the quadratic Casimir as:
\begin{align}\label{eq:casimirfixing}
    \frac{1}{2}\text{STr}\left(\boldsymbol{\Lambda}^2\right)&=\frac{1}{4}\Lambda^I\kappa_{IJ}\Lambda^J
    = \mathcal{C}_2\equiv j(j+1/2),
\end{align}
where the real number $\mathcal{C}_2$ is specifying the representation.

Performing the Gaussian integral over all components of $\boldsymbol{\Lambda}$ leads to the action:
\begin{equation}
S[g,\mathbf{A},\Theta] = \frac{i}{2}\int_\mathcal{C} ds\;\left[\frac{1}{2\Theta}\text{STr}(g^{-1}D_A g g^{-1}D_Ag)-4\Theta\mathcal{C}_2\right].
\end{equation}
If the connection in the BF path integral is flat, it can be absorbed into the group element $g$ by \eqref{eq:gred}, and the action finally becomes:
\begin{equation}
\label{eq:pog}
S[g,\mathbf{A},\Theta] = \frac{i}{2} \int_\mathcal{C} ds\;\left[\frac{1}{2\Theta}\text{STr}(g^{-1}\partial_s g g^{-1}\partial_sg)-4\Theta\mathcal{C}_2\right],
\end{equation}
which describes a particle moving on the group $G$ manifold, with its canonical Cartan-Killing metric $ds^2_{\text{CK}} \equiv 2\text{STr}[(g^{-1}dg)^2]$. However, this is not gravity. We have not yet used the dictionary between how the vielbein and spin connection are encoded into the components of the gauge field \eqref{C6:gaugefieldexpansion}. At the group theoretical level, we have described a particle moving on the parent superconformal group $G$ manifold, whereas we want a particle moving on the right coset manifolds $G/H$ describing hyperbolic space. In our particular case for $\mathcal{N}=0$ or $\mathcal{N}=1$, we have $G=\text{SL}(2,\mathbb{R})$ or $\text{OSp}(1\vert 2, \mathbb{R})$ respectively, and $H = \text{U}(1)\simeq \text{SO}(2)$ (generated by the compact generator $J_2$ for Euclidean signature as here) or $\text{SO}(1,1)$ (generated by the non-compact generator $H$ for Lorentzian signature):
\begin{alignat}{2}
\text{H}_{2} \, &\simeq \, \frac{\text{PSL}(2,\mathbb{R})}{\text{U}(1)}, \qquad
&&\text{H}_{2 \vert 2} \, \simeq \,\frac{\text{OSp}(1\vert 2,\mathbb{R})}{\text{U}(1)}, \\
\text{AdS}_{2} \, &\simeq \, \frac{\text{PSL}(2,\mathbb{R})}{\text{SO}(1,1)}, \qquad
&&\text{AdS}_{2 \vert 2} \, \simeq \,\frac{\text{OSp}(1\vert 2,\mathbb{R})}{\text{SO}(1,1)}.
\end{alignat}

To implement this in the procedure, we note that the transformation \eqref{eq:tf} allows us to choose $\boldsymbol{\Lambda}$ to be of the restricted form:
\begin{equation}
\label{C6:expansionwithout2}
    \boldsymbol{\Lambda}=\Lambda^0 J_0 +\Lambda^1 J_1 + \Xi^\alpha J_\alpha.
\end{equation}
We have chosen here to set to zero the component of $\boldsymbol{\Lambda}$ along the $J_2$-direction (and all components along the R-symmetry group generators for higher supersymmetry). 

At the coset level, we hence reintroduce a functional integral over the restricted set of weights $\boldsymbol{\Lambda}$ as: 
\begin{equation}
\int \mathcal{D}\boldsymbol{\Lambda} \mathcal{D}\Theta \mathcal{D}_\Lambda g\;e^{-S_{\boldsymbol{\Lambda}}[g,\mathbf{A}]},
\end{equation}
where $\Theta$ is a scalar bosonic Lagrange multiplier enforcing the constraint $(\Lambda^0\Lambda^0 + \Lambda^1\Lambda^1+\Xi^\alpha\epsilon_{\alpha\beta}\Xi^\beta)\equiv 4\mathcal{C}_2$: 
\begin{equation}
S_{\boldsymbol{\Lambda}}[g,\mathbf{A},\Theta]=\int_\mathcal{C} ds\;\left[\text{STr}(\boldsymbol{\Lambda} g^{-1}D_Ag)+\frac{i}{2}\Theta(\Lambda^0\Lambda^0 + \Lambda^1\Lambda^1 + \Xi^\alpha\epsilon_{\alpha\beta}\Xi^\beta-4\mathcal{C}_2)\right].
\end{equation}
Note again that we integrate over adjoint elements $\Lambda^A$ that live in a subvectorspace of the algebra (by excluding the $J_2$ (and R-symmetry for higher rank) components in the expansion \eqref{C6:expansionwithout2}). This means that the components of $g^{-1}D_Ag$ along these directions are absent. At the level of the Cartan-Killing metric of the particle on the group manifold \eqref{eq:pog}, this implements a coset condition along the $J_2$-direction as:
\begin{equation}
\label{eq:metcosnew}
ds^2_{\text{coset}} = \left.2\text{STr}[(g^{-1}dg)^2]\right\vert_{\neq J_2}.
\end{equation}
Also note that the classical variables $\Xi^\alpha$ are treated as Grassmann-variables in this path integral.

We fix the gauge redundancy in $g$, which induces a transformation of $\mathbf{A}$, by setting $g\equiv 1$ along the entire curve $\mathcal{C}$ and smoothly extending this gauge into the bulk \cite{Iliesiu:2019xuh}. This gauge transfers the information of the metric from the group variables into the gauge fields according to \eqref{C6:gaugefieldexpansion}. The frame fields are then captured entirely by the covariant derivative in the Lagrangian as: 
\begin{equation}
g^{-1}D_Ag=\mathbf{A}_s=\Dot{Z}^M E_{M}^{\;\;\;A}J_A.
\end{equation}
The total action thus becomes:
\begin{equation}
\label{C6:intermediatepathintegral}
S_{\boldsymbol{\Lambda}}[g,\mathbf{A}]=\frac{1}{2}\int_\mathcal{C} ds\left[\Lambda^A\kappa_{AB}\Dot{Z}^ME_M^{\;\;\;B}+i\Theta(\Lambda^0\Lambda^0 + \Lambda^1\Lambda^1 + \Xi^\alpha\epsilon_{\alpha\beta}\Xi^\beta-4\mathcal{C}_2)\right].
\end{equation}
Path integrating over all (non-zero) components of the weight vector $\boldsymbol{\Lambda}$ ($\Lambda^0, \Lambda^1, \Xi^\alpha$) yields a reduced action:
\begin{equation}
S[Z,E] = \frac{i}{2}\int_\mathcal{C} ds\; \left[\frac{1}{4\Theta}\Dot{Z}^M E_M^{\;\;\;A}\kappa_{AB}\Dot{Z}^NE_N^{\;\;\;B}-4\Theta \mathcal{C}_2\right].
\end{equation}
The metric tensor is, by definition, related to the frame fields as: 
\begin{equation}
\label{C6:metricintermsofframe}
g_{MN} = E_M^{\;\;\;A}\kappa_{AB}E_{\;\;\;N}^B.
\end{equation}
The frame field should satisfy the symmetry property $ E_M^{\;\;\;A}=(-)^{M+MA}E^A_{\;\;\;M}$ in order to obey the required symmetries \eqref{C6:anticommutativityconvention}. 
Then we can rewrite the gravitational coset metric $ds^2_{\text{coset}}$ as a spacetime metric
\begin{align}
ds^2_{\text{coset}}=\Dot{Z}^M E_M^{\;\;\;A}\kappa_{AB}\Dot{Z}^N E_N^{\;\;\;B} 
=\Dot{Z}^Mg_{MN}\Dot{Z}^N,
\end{align}
to obtain:
\begin{equation}
\label{eq:act3}
S[Z,g_{MN}] = \frac{i}{2}\int_\mathcal{C} ds\; \left[\frac{1}{4\Theta}\Dot{Z}^M g_{MN}\Dot{Z}^N-4\Theta \mathcal{C}_2\right].
\end{equation}

The Gaussian integrals over the non-zero components of $\Lambda$ also pick up the path integral measure factor
\begin{equation}
\label{eq:PImeasure}
\sim \prod_{s} \frac{1}{\Theta(s)^{D/2}},
\end{equation}
where $D = 2-2$ is the super-dimension (bosonic minus fermionic dimension) of the bulk superspace H$_{2 \vert 2}$. 
This is precisely the required measure in order to guarantee 1d reparametrization invariance along the worldline of a particle in superspace \eqref{eq:act3}. One pedestrian way to see this, is to consider the free non-relativistic particle path integral in the discretized language:
\begin{align}
\label{eq:discr}
\frac{1}{(2\pi \epsilon e_{N+1})^{D/2}}\prod_{n=1}^{N} \int \frac{d^Dx_n}{(2\pi \epsilon e_n)^{D/2}} e^{-\sum_{n=1}^{N+1}(x_n-x_{n-1})^2/2 \epsilon e_n} = \frac{1}{(2\pi T)^{D/2}} e^{-\frac{(x_f-x_i)^2}{2 T}},
\end{align}
where $e_n$ is the discretized worldline einbein, and $T=\sum_{n=1}^{N+1}\epsilon e_n$ is the total physical proper time as measured along the worldline. 1d reparametrization invariance is manifest on the RHS, and is implemented on the LHS in that the einbein only appears in the combination $\epsilon e_n$. In particular, we note the $1/e_n^{D/2}$ in the path integral measure. This argument is purely on the worldline, and hence generalizing to curved target spacetime is immediate.\footnote{Although of course the integral is no longer Gaussian.} Generalizing to target superspace is also immediate by incorporating Gaussian integrals of Grassmann variables, leading indeed to the advertised measure \eqref{eq:PImeasure}.

Optionally, we can now integrate out the field $\Theta$. Choosing the upper branch solution of $\Theta$,
\begin{equation}
\Theta=\frac{i}{4\mathcal{C}_2^{1/2}}\sqrt{\Dot{Z}^M g_{MN}\Dot{Z}^N},
\end{equation} 
gives: 
\begin{equation}\label{C6:terminssquareroot}
S[Z,g_{MN}] = \mathcal{C}_2^{1/2}\int ds\sqrt{\Dot{Z}^M g_{MN}\Dot{Z}^N}.
\end{equation}
Here one recognizes the more familiar form of the worldline action. This is exact also off-shell, but requires introduction of the proper path integration measure in the square-root action, as usual. 

Recognizing that flat field gauge transformations in the BF path integral are equivalent to superdiffeomorphisms in the metric formulation finally proves the equivalence between a Wilson loop operator insertion in the BF path integral, and the worldline path integral in the metric formulation:
\begin{equation}
\label{C6:wilsonlineresult}
\mathcal{W}_j(\mathbf{A})=\text{STr}_j\left(\mathcal{P}\exp-\oint_{\mathcal{C}}\mathbf{A}\right)\simeq\oint_{\mathcal{C}} \mathcal{D}Z\;e^{-\mathcal{C}_2^{1/2}\int ds\sqrt{\Dot{Z}^Mg_{MN}\Dot{Z}^N}}.
\end{equation}
 
\subsection{Supergravitational amplitudes involving EOW branes}
\label{s:amn1}
To compute supergravitational amplitudes, we will have need for the analogous characters of the representations. The relevant group to describe $\mathcal{N}=1$ JT supergravity in its BF formulation is $\text{OSp}(1|2,\mathbb{R})$, some of the representation theory was developed in \cite{Fan:2021wsb}.

Defects are implemented in the supergravitational amplitudes by inserting a suitably normalized continuous series character in the disk partition function. Since we will be interested in EOW branes ending at the neck of a supersymmetric trumpet, we consider the insertion of a hyperbolic character, obtained by exponentiating the Cartan generator $H\in \mathfrak{osp}(1|2,\mathbb{R})$  \cite{Fan:2021wsb}:
    \begin{equation}\label{C6:hyperbolicgausselement}
    g(\phi)=e^{2\phi H}\simeq\left[\begin{array}{cc|c}
     e^{\phi}&0&0\\
      0&e^{-\phi}&0\\
      \hline
      0&0&\pm 1
    \end{array}\right],
\end{equation} 
for the respective \textbf{R} $(+)$ and \textbf{NS} $(-)$ sectors. These two sectors of the supergroup are not continuously connected to each other. 
The technical details to proceed can be found in \cite{Fan:2021wsb} and lead to two distinct characters depending on the relevant spin-structure sector 
\begin{equation}
    \chi_k^{\textbf{NS}}(\phi)=i \frac{\cos(2k\phi)}{\sinh(\phi/2)}, \qquad \chi_k^{\textbf{R}}(\phi)=i\frac{\sin(2k\phi)}{\cosh(\phi/2)},
\end{equation}
for the principal series representation label $j=-1/4 + ik, \, k \in \mathbb{R}^+$.
Removing the Weyl denominator immediately yields the appropriate defect insertions in the super-gravitational disk partition function:
\begin{align}
    Z^{\textbf{NS}}_{\text{trumpet}}(\beta, \phi)&=\int_0^\infty dk\; \cos(2\phi k)e^{-\beta k^2}=\frac{1}{2}\sqrt{\frac{\pi}{\beta}}e^{-\phi^2/\beta}\label{C6:trumpet1},\\Z^{\textbf{R}}_{\text{trumpet}}(\beta, \phi)&=\int_0^\infty dk\;\sin(2\phi k)e^{-\beta k^2}= \frac{1}{2}\sqrt{\frac{\pi}{\beta}}e^{-\phi^2/\beta} \text{erfi}\left(\frac{\phi}{\sqrt{\beta}}\right).
    \label{C6:trumpet2}
\end{align}
The $\textbf{NS}$ trumpet is one-loop exact and recovers the bosonic trumpet partition function \eqref{eq:bosonictrumpet}. The $\mathbf{R}$ trumpet amplitude is \emph{not} one-loop exact, and has the following perturbative expansion:
\begin{equation} \label{C6:ramondtrumpetexpansion}
    Z^{\textbf{R}}_{\text{trumpet}}(\beta,\phi)=\left(\frac{\phi}{\beta}+\frac{\phi^3}{3\beta^2}+\frac{\phi^5}{10\beta^3}+\frac{\phi^7}{42\beta^4}+\dots\right)e^{-\phi^2/\beta}.
\end{equation}
In particular, we observe that the one-loop component vanishes, but higher-loop contributions do not. The precise relation with a super-Schwarzian evaluation of this amplitude is not clear to us, as these are usually one-loop exact \cite{Stanford:2017thb}, with the Ramond sector zero at one loop due to a fermion zero-mode \cite{Stanford:2019vob}. 

The NS sector is in this case the one-loop exact component, whose monodromy is disconnected from the identity element. A similar situation will turn up in the connected gravitational sector versus disconnected sector of $\mathcal{N}=2$ OSp$(2|2,\mathbb{R})$ representation theory. In that case, we observe that the component connected to the identity is one-loop exact, whereas the component disconnected from the identity yields a similar all-order perturbative expansion. The latter is treated in appendix \ref{sec:othercomponent}.\\

From the previous discussion, the mass of the EOW particle is related to the quadratic Casimir by: 
\begin{equation}
    \mu^2\equiv\mathcal{C}_2=j\left(j+1/2\right).
\end{equation}
For a highest-weight discrete series module with $h=-j$ \cite{Fan:2021wsb}, the relation between mass and conformal scaling dimension $h$ leads to: \begin{equation}\label{C6:lmrelation}
    h(h-1/2)=\mu^2.
\end{equation}
Integrating over the dilaton field along the EOW brane, we have the identification in superspace:
\begin{equation}
    \begin{aligned}
        & \int \mathcal{D}g\mathcal{D}\Phi\oint_{\text{geodesics}}\mathcal{D}Z\;e^{-\mu \int ds\;\sqrt{\dot{Z}^Mg_{MN}\dot{Z}^N}}\;e^{-I_{\text{JT}}^{\mathcal{N}=1}[g,\Phi]} \\
        =&\int_0^{+\infty} db\int_{e^{-\oint A} \, \simeq \, e^{bH}} \mathcal{D}\mathbf{B}\mathcal{D}\mathbf{A}\;\mathcal{W}_{j}(\mathbf{A})e^{-I_{\text{BF}}^{\mathfrak{osp}(1|2,\mathbb{R})}[\mathbf{B}, \mathbf{A}]},
    \end{aligned}
\end{equation}
where we have again split the integral over geodesic paths according to \eqref{Eq:integralsplit}. The (bottom component of the) geodesic length $b$ is related to the holonomy by the twisted boundary condition, implementing a hyperbolic deformation of the disk.

Due to the relation between the mass parameter $\mu$ and the conformal weight $h$, the geodesic approximation instructs us to identify 
\begin{equation}
\label{C6:relationlmu}
    h\approx \mu, \qquad \mu \gg1.
\end{equation} Note that we consider positive mass $\mu$, and highest weight modules are defined for $h>0$ \cite{Fan:2021wsb}.\\

Equation \eqref{C6:wilsonlineresult} demonstrates that the worldline path integral is the hyperbolic character evaluated in a highest-weight discrete series representation module labeled in terms of the tension parameter $\mu$. 
The character depends only on the conjugacy class of the group element, which for OSp$(1\vert 2,\mathbb{R})$ consists of a real hyperbolic variable $\phi$ and a $\mathbb{Z}_2$ choice, distinguishing between the Ramond (\textbf{R}) and Neveu-Schwarz (\textbf{NS}) sectors. 
The resulting characters have been calculated explicitly in \cite{Fan:2021wsb}, and take the form:
\begin{align}
    \chi_j^{\textbf{NS}}(\phi)=
    \frac{e^{(2j+1/2) \phi}}{2\sinh\left(\phi/2\right)}\label{C6:character1}, \qquad
     \chi_j^{\mathbf{R}}(\phi)
     =\frac{e^{(2j+1/2) \phi}}{2\cosh(\phi/2)}.
\end{align}

The transition from the group theoretical language to gravity is again achieved by replacing $b=2\phi$ \cite{Fan:2021wsb}. We further identify the mass tension with the conformal scaling dimension $-j=h\approx \mu\gg 1$  within the geodesic approximation, yielding:
\begin{align}
\oint_{\mathcal{C} _\textbf{NS}}\mathcal{D}Z\;e^{-\mu \oint_\mathcal{C}ds\;\sqrt{\Dot{Z}^Mg_{MN}\Dot{Z}^N}}&\simeq\frac{e^{-\mu b}}{2\sinh(b/4)}, \label{C6:oneloopcorrection1}\\ \oint_{\mathcal{C}_\textbf{R}}\mathcal{D}Z\;e^{-\mu \oint_\mathcal{C}ds\;\sqrt{\Dot{Z}^Mg_{MN}\Dot{Z}^N}}&\simeq\frac{e^{-\mu b}}{2\cosh(b/4)}.\label{C6:oneloopcorrection2}
\end{align} 
It is again interesting to note that the denominator can be interpreted as a one-loop correction to the classical (geodesic) saddle approximation. Gluing the worldline particle amplitudes along the geodesic length at the neck of the relevant spin-structured trumpets (equations \eqref{C6:trumpet1}, \eqref{C6:trumpet2}) finally yields: 
\begin{align}
 \mathcal{A}^\textbf{NS}(\beta;\mu)&=\int_0^\infty dk \;e^{-\beta k^2}\int_0^\infty db \cos(bk)\frac{e^{-\mu b}}{2\sinh(b/4)},\label{kolchmeyersucker1}\\\mathcal{A}^\textbf{R}(\beta;\mu)&=\int_0^\infty dk \;e^{-\beta k^2}\int_0^\infty db \sin(bk)\frac{e^{-\mu b}}{2\cosh(b/4)}.\label{kolchmeyersucker2}
\end{align}
An immediate realization is that the spurious UV divergence for small $b\rightarrow0$ is only present in the \textbf{NS} sector. On the other hand, the \textbf{R} sector is perfectly regular in the UV. The Weyl denominator of the discrete series character is explicitly visible since the worldline path integral evaluates to a genuine character in group theory, and can be viewed as the culprit for this possible UV-divergence.

\section{EOW brane amplitudes in $\mathcal{N}=2$ JT supergravity}
\label{Section4}
In this section, we attempt to further generalize our group-theoretic construction to incorporate $\mathcal{N}=2$ JT supergravity. Starting at this level of supersymmetry, the relevant superalgebra is higher rank which adds new subtleties as we elaborate on. 

\subsection{EOW branes in superspace}
The Euclidean worldline action for the EOW brane is readily generalized. Indeed, our analysis for $\mathcal{N}=1$ in subsection \ref{s:eowsuper} was written such that the discussion there immediately applies. We hence retake
\begin{align}
\label{eq:acteow}
I_{\text{EOW}}=\int_{\text{EOW}}ds\;\sqrt{\Dot{Z}^Mg_{MN}\Dot{Z}^N}\; \left(\mu -\phi K\right)
\end{align}
in a $(2 \vert 4)$-dimensional target space as our proposal for the EOW brane action. Our main question is whether we can reproduce the corresponding worldline path integral from representation theory.

\subsection{Wilson loops as probe particles: higher rank groups}
\label{s:higherank}
Starting with $\mathcal{N}=2$ supersymmetry, the superalgebra has higher rank which complicates the derivation of subsection \ref{s:rank1}. We describe how to deal with this here. We are focused on $\mathcal{N}=2$ but some of our notation directly generalizes to $\mathcal{N}=4$.

For general topological supergravity models, we write the gauge connection in superspace as:
\begin{align}
\label{eq:n2exp}
\mathbf{A}_M &\equiv e_M{}^a J_a + f_M{}^\alpha J_\alpha + \Omega_M J_2 + \sigma_M{}^b J^{\text{R}}_b, \quad a=0,1, \quad b=\text{dim R-symmetry group} \nonumber \\
&= E_M{}^A J_A + \Omega_M J_2 + \sigma_M{}^b J^{\text{R}}_b,
\end{align}
where in the second line we have combined the two bosonic zweibein components with their superpartners into the super-zweibein $E_M{}^A \equiv (e_M{}^a \vert f_M{}^{\alpha})$, and have introduced the notation $\sigma_M$ for the components of $\mathbf{A}_M$ along the R-symmetry generators. 

In the $\mathcal{N}=2$ case, we work with the $(4 \vert 4)$-dimensional superalgebra $\mathfrak{osp}(2 \vert 2,\mathbb{R})$. For $\mathcal{N}=2$ supergravity, there are hence four fermionic coordinates, one R-symmetry generator, and the $A$-index takes on $2\vert 4$ values. The associated BF description was written down in component form in \cite{Gomis:1991cc}. In Appendix \ref{app:tsg}, we provide some details of the direct superspace description of this model, in particular elaborating on how gauge transformations decompose gravitationally into diffeomorphisms, local Lorentz transformations and U(1) gauge transformations.

Exponentiating this Lie superalgebra, we obtain the supergroups OSp$(2\vert 2,\mathbb{R})$ and SU$(1,1 \vert 1)$, where the first contains two components out of which only the one connected to the identity group element is appropriate for supergravity. The SU$(1,1 \vert 1)$ description on the other hand directly produces supergravity. Nonetheless, in this work we will choose to work explicitly with the real supergroup OSp$(2\vert 2,\mathbb{R})$.\footnote{We have two reasons. Firstly, the required representation theory of SL$(2,\mathbb{R}$) and OSp$(1 \vert 2, \mathbb{R})$ will generalize most straightforwardly if we work with a real supergroup. Secondly, as we will discuss in the concluding section \ref{sec:conc}, (super)gravity is not precisely equal to a gauge theory. In the BF formalism, one can improve gauge theory by adding a suitable positivity condition, which can only be done if we work with a real fundamental representation to begin with.}${}^{,}$\footnote{This group plays a similar role in the 3d Chern-Simons formulation of 3d pure $\mathcal{N}=2$ supergravity, see e.g. \cite{Merbis:2023uax} for a recent explicit description.}

Picking a weight vector $\boldsymbol{\Lambda}$ that satisfies $\frac{1}{2}\text{STr}\boldsymbol{\Lambda}^2 = \mathcal{C}_2 = j^2-q^2$ and $\text{STr}\boldsymbol{\Lambda}^3 = \mathcal{C}_3 = q(j^2-q^2)$, the character in the irrep $(j,q)$ is obtained as a path integral over $g$ with action:
\begin{equation}
S_{\boldsymbol{\Lambda}}[g, \mathbf{A}] = \int ds \operatorname{STr}(\boldsymbol{\Lambda}g^{-1}D_A g), \quad D_A\equiv \partial_s + \mathbf{A}_s, \quad \mathbf{A}_s(s)\equiv \dot{Z}^M(s)\mathbf{A}_M(Z(s)).
\end{equation}
We now fix a gauge on the disk in which $g = 1$ along the curve $\mathcal{C}$, such that $g^{-1}D_A g$ reduces to $\mathbf{A}_s = \dot{Z}^M(E_M{}^A J_A + \Omega_M J_2 + \sigma{}_M J^{\text{R}})$ and the action becomes

\begin{equation}
\frac{1}{2}\int ds\left[\Lambda^A \kappa_{AB} \dot{Z}^ME_M{}^B + \Lambda_2 \dot{Z}^M\Omega_M  + \Lambda \dot{Z}^M \sigma_M  \right].
\end{equation}
The last term describes the minimal coupling to a U(1) gauge field $\sigma_M$.\\ 

We now restrict the weight vector $\boldsymbol{\Lambda}$ to $\Lambda=0=\Lambda_2$ and with quadratic \emph{and} cubic Casimir fixed as $\mathcal{C}_2$ and $\mathcal{C}_3$ respectively. Since the above holds for any weight vector $\boldsymbol{\Lambda}$, this step is merely a constructive step that will lead to the final result we want. The first restriction is done to effectively reduce the target space of the particle to the right supercoset where we mod out by the right action of the one-parameter subgroups generated by $J_2$ and $Z$. This corresponds to the fact that the bulk superspace is not a supergroup itself, but is instead a homogeneous space that can be written as a supercoset of the original gauge group. Indeed, similarly as in the bosonic and $\mathcal{N}=1$ case, the hyperbolic superspaces are then constructed as a supercoset:
\begin{alignat}{2}
\label{eq:supads}
\text{H}_{2 \vert 4} \, &\simeq \,\frac{\text{OSp}(2\vert 2,\mathbb{R})}{\text{U}(1)_{\text{Lorentz}} \otimes \text{U}(1) }, \qquad 
\text{AdS}_{2 \vert 4} \, &\simeq \,\frac{\text{OSp}(2\vert 2,\mathbb{R})}{\text{SO}(1,1) \otimes \text{U}(1) },
\end{alignat}
where we divide by a subgroup that is larger now compared to $\mathcal{N}=0,1$. 

This particular subgroup can also be characterized geometrically as the relevant tangent space group in $\mathcal{N}=(2,2)$ supergravity as follows. The tangent space group (or local Lorentz group) can usually be defined by the ambiguity in the definition of the zweibein $g_{MN} = E_M^{\;\;\;A}\;\kappa_{AB}\;E_{\;\;\;N}^B$, given the metric. For the $(2\vert 4)$-dimensional supermetric, the above relation is preserved by local transformations $E^A \to L^A{}_B E^B$ belonging to OSp$(2\vert 4,\mathbb{R})$. However, as well-known (see e.g. \cite{West:1990tg}), in superspace supergravity it turns out one should restrict this ``maximal'' tangent space group to a specific subgroup in order to make contact with the component formulation. In our case, we need to focus on the tangent space subgroup
\begin{equation}
\text{U}(1)_{\text{Lorentz}} \otimes \text{U}(1) \,\, \subset \,\, \text{OSp}(2 \vert 4,\mathbb{R}),
\end{equation}
where U(1)$_{\text{Lorentz}}$ and U(1) act on the basis $E^A \equiv (e^\pm \vert f^\pm,\bar{f}^\pm)$ precisely as described in Appendix \ref{app:tsg} (as parametrized there by $l$ and $s$ respectively). Hence the (abelian) productgroup $\text{U}(1)_{\text{Lorentz}} \otimes \text{U}(1)$ serves as the superspace tangent group of relevance here for the construction of aAdS $\mathcal{N}=2$ JT supergravity.

More generally, the relevant superspace tangent space group for any 2d superconformal algebra is $\text{U}(1)_{\text{Lorentz}} \otimes G_R$, where one includes the complete R-symmetry group. E.g. for $\mathcal{N}=4$ JT supergravity, the tangent space group is $\text{U}(1)_{\text{Lorentz}} \otimes \text{SU}(2)$, and we will see this group appear in the coset construction of the $\mathcal{N}=4$ supersymmetric hyperbolic 2-plane in section \ref{Section5}.

The restriction $\Lambda=0=\Lambda_2$ effectively corresponds to considering those components of the gauge connection \eqref{eq:n2exp} which are \emph{not} $\Omega$ and $\sigma$. This can be viewed as a partial gauge-fixing of the symmetry \eqref{eq:tf} of the system.

We implement the constraints on the Casimirs through two Lagrange multipliers as 
\begin{equation}
\int ds\left[\frac{1}{2}\Lambda^A \kappa_{AB}\dot{Z}^ME_M{}^B  
+ i\Theta_1 (\text{STr}\boldsymbol{\Lambda}^2 - 2\mathcal{C}_2) + i\Theta_2 (\text{STr}\boldsymbol{\Lambda}^3 - \mathcal{C}_3) \right].
\label{eq:eowN2}
\end{equation}
For bosonic higher rank algebras, such actions were written down in \cite{Ammon:2013hba}.\footnote{More generally, for a rank $r$ group $G$ we would add $r$ constraints to fix all higher Casimirs, and fix the irrep. 
} \\ 

To proceed and explicitly compute quantum gravitational amplitudes, we require explicit expressions for the discrete and principal series characters of OSp$(2 \vert 2,\mathbb{R})$. Unfortunately, the representation theory has not been worked out sufficiently for our purposes, so we first develop the required mathematical background explicitly, presenting the specifics in a detailed Appendix \ref{sec:N2rep}. We can summarize our results as:

\begin{itemize}
\item We construct the principal series irreducible representations of OSp$(2 \vert 2,\mathbb{R})$ by parabolic induction, generalizing the explicit construction from PSL$(2,\mathbb{R})$ \newline ($\mathcal{N}=0$) and OSp$(1\vert 2, \mathbb{R})$ ($\mathcal{N}=1)$. As before, these irreps are physically important since they describe the states that propagate in the bulk Hilbert space of the supergravity model.
\item We explicitly compute the characters in these representations to be \eqref{eq:n2char}:
\begin{equation}
\label{eq:charNN2}
   \chi_{k,q}^{\mathcal{N}=2}(\phi,\theta)=2\cos (2k\phi)e^{2iQ\theta} \frac{\cosh\phi-\cos\theta}{\sinh\phi},
\end{equation}
where $\phi \in \mathbb{R}^+$ and $\theta \in [0,2\pi)$ label conjugacy classes of OSp$(2 \vert 2,\mathbb{R})$ (they parametrize the 2d Cartan subalgebra), and $k\in \mathbb{R}^+$ and $Q \in \mathbb{R}$ are labeling the irrep itself. One can think of these as a momentum label $k$ and charge label $Q$. The last factor in \eqref{eq:charNN2} descends from the usual Weyl denominator of the superalgebra \eqref{eq:WeylDen}. This character is physically important as it implements a hyperbolic defect in the super-JT bulk, which is geometrically interpretable as a trumpet with a geodesic boundary. For those purposes, as earlier, the Weyl denominator is stripped off \cite{Mertens:2019tcm}, leading to the trumpet partition function: \begin{equation}
    Z_{\text{trumpet}}(\beta; \phi,\theta)=\int_0^{+\infty} dk \sum_{Q \in \mathbb{N}/2} e^{-\beta (k^2+Q^2)} e^{2iQ\theta} \cos (2k\phi),
\end{equation}
where the $Q$ charge quantum number is in principle discretized.
The energy variable $k^2+Q^2$ is (up to a sign) the quadratic Casimir of the principal series representations \eqref{eq:quadcas}.
\item 
We compute the character in the highest weight discrete series irreps to be \eqref{eq:N=2highestweightchar}:
\begin{equation}
\label{eq:dchar}
   \chi_{j,q}^{\mathcal{N}=2}(\phi,\theta)=e^{2iq\theta}e^{2j \phi }\frac{\cosh\phi-\cos\theta}{\sinh\phi}.
\end{equation}
We will once again define the EOW branes as inserting this character into the gravitational amplitude.
\end{itemize}
The reader who is willing to believe these statements can safely skip the results reported in Appendix \ref{sec:N2rep}.

Armed with these expressions, we can write down some explicit amplitudes. However, in order to make contact with the superspace action \eqref{eq:acteow}, we will need to distinguish two possible definitions of branes, which we denote as class-1 and class-2 EOW branes, and discuss consequently.

\subsubsection{Class-1 EOW branes}
We cannot integrate over $\boldsymbol{\Lambda}$ exactly in the action \eqref{eq:eowN2}, but it is still interesting to analyze the resulting Lagrangian perturbatively in $\Theta_2$. One could hope to rewrite the terms in perturbation theory fully in terms of the second order metric $g_{MN}$ only. We attempt to do so in Appendix \ref{app:Perturbationtheory}, where we note that this is impossible: the brane action is a purely first-order construction.

Nonetheless, we can proceed and \emph{define} an EOW brane with ``mass'' label $j$ and ``charge'' label $q$ through the first-order action \eqref{eq:eowN2}. Using \eqref{eq:dchar} and the stripped version of \eqref{eq:charNN2} (removing the Weyl denominator), the final expression for the amplitude with one asymptotic boundary and one EOW-brane boundary with mass $\mu$ and charge $q$ can then be compiled as:
\begin{align}
\label{eq:cl1}
\mathcal{A}_{\text{I}}&(\beta;\mu,q) =\hspace{0.5cm}\raisebox{-0.5\height}{\includegraphics[height=1.6cm]{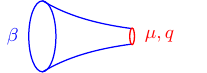}} \\
&=\int_0^{+\infty} dk \sum_{Q \in \mathbb{N}/2} e^{-\beta (k^2+Q^2)} \int_0^{\infty} db \int_0^{2\pi} d\theta e^{2iQ\theta} \cos (kb)  e^{2iq\theta} e^{-\mu b} \frac{\cosh \frac{b}{2} - \cos \theta}{\sinh \frac{b}{2}},\nonumber
\end{align}
 where the hyperbolic parameter is replaced by the geodesic length parameter $b=2\phi$ in the same way as before. Furthermore, we take the highest weight label to be proportional to the mass parameter $-j=h=\mu \gg 1$ in the geodesic limit.

Next, we investigate convergence of this amplitude. In the IR where $b\to +\infty$ (large EOW brane circle), the integral over $b$ converges due to the $e^{-\mu b}$ suppression. 
Far more interesting is the UV region where $b\to 0$. As a consequence of the Weyl denominator expression, we find the typical UV-divergences due to the pinching of the brane length as:
\begin{equation}
\sim \int_0 db \frac{\cosh \frac{b}{2} - \cos \theta}{\sinh \frac{b}{2}}.
\end{equation}
This is divergent due to the $b \to 0$ region unless simultaneously $\theta=0$. Physically restricting to $\theta=0$ means fixing the U(1) gauge field holonomy along the brane worldline to vanish. The gravitinos pick up an Aharonov-Bohm phase upon circling the tube $\sim e^{i \theta}$. Hence setting $\theta=0$ removes this phase, and makes the gravitinos periodic. This is the Ramond sector. Setting on the other hand $\theta=\pi$, leads to the other extreme where the fermions are anti-periodic. This is the Neveu-Schwarz sector. This range of $\theta$ is continuously connected by the spectral flow operation.
Performing the integral over $\theta$ first, it is readily seen that the amplitude $\mathcal{A}_{\text{I}}(\beta;\mu,q)$ again diverges.

\subsubsection{Class-2 EOW branes}
We next define a second type of EOW brane. This brane type can be directly formulated in the bulk superspace in the second order formalism, and as such is perhaps a more natural analogous definition. If we integrate the path integral with action \eqref{eq:eowN2} over $\mathcal{C}_3$, we enforce $\Theta_2=0$ and the resulting path integral becomes Gaussian again. Path integrating over $\boldsymbol{\Lambda}$, we obtain the effective action (cfr \eqref{eq:act3}):
\begin{equation}
\label{eq:act1}
\frac{i}{2}\int ds\left[\frac{1}{4\Theta} \dot{Z}^M g_{MN} \dot{Z}^N - 4\Theta \mathcal{C}_2 \right],
\end{equation}
where we have again defined the supermetric $g_{MN} = E_M^{\;\;\;A}\kappa_{AB}E_{\;\;\;N}^B$ 
, with indices taking on $2 \vert 4$ values. Additionally, we produce a measure factor
\begin{equation}
\label{eq:PImeasureN}
\sim \prod_{s} \frac{1}{\Theta(s)^{D/2}},
\end{equation}
where $D = 2-4$ is the super-dimension (bosonic minus fermionic dimension) of the bulk superspace H$_{2 \vert 4}$, precisely as required for the superspace worldline particle path integral as discussed around \eqref{eq:discr}. 

At the classical level (or with a suitable path-integral measure at the quantum level as well), the worldline einbein $\Theta$ can be eliminated by the equations of motion, leading once again to the natural coupling of a massive particle to supergravity (cfr \eqref{C6:terminssquareroot}):
\begin{equation}
\label{eq:act2}
S[Z,g_{MN}] = \mathcal{C}_2^{1/2}\int ds\sqrt{\dot{Z}^M g_{MN} \dot{Z}^N}.
\end{equation}

The resulting final action \eqref{eq:act1} (or \eqref{eq:act2}) has the form of a particle moving on the supermanifold with metric $ds^2 = dZ^M\,g_{MN}\, dZ^N$, and ``mass'' $\mathcal{C}_2^{1/2}$.
The result is an EOW brane where we are forgetting about the eigenvalue of the cubic Casimir (morally the U(1) charge).

At a technical level, we can get the final amplitude by integrating the earlier result for class-1 EOW branes \eqref{eq:cl1} over the cubic Casimir eigenvalue $\mathcal{C}_3$:
\begin{align}
\label{eq:cl2}
\mathcal{A}_{\text{II}}(&\beta;\mathcal{C}_2) = 
\int_0^{+\infty} dk \sum_{Q \in \mathbb{N}/2} e^{-\beta (k^2+Q^2)} \nonumber \\
\times &\int_0^{\infty} db \int_0^{2\pi} d\theta\; e^{2iQ\theta} \cos (kb) \frac{\cosh \frac{b}{2} - \cos \theta}{\sinh \frac{b}{2}} \int_{-\infty}^{+\infty} d\mathcal{C}_3 \, e^{-\sqrt{\mathcal{C}_2 + \frac{\mathcal{C}_3^2}{\mathcal{C}_2^2}} b}e^{2i\frac{\mathcal{C}_3}{\mathcal{C}_2}\theta},
\end{align}
where the quadratic and cubic Casimirs are related to the representation labels $j,q$ by \eqref{eq:expressionsquadraticandcubiccasimirs}.

Note that the $\mathcal{C}_3$-integral converges, although it cannot be done analytically. Convergence of the $b$-integral on the other hand is modified compared to the class-1 EOW brane from the previous subsection. Due to the behavior of the integrand at large $\mathcal{C}_3$  as $e^{- \vert \mathcal{C}_3 \vert b}$, there is an additional $\sim 1/(b-2i\theta/\mathcal{C}_2)$ factor in the integral, causing the $b$-integral to diverge even when $\theta=0$.

\section{Towards EOW brane amplitudes in $\mathcal{N}=4$ JT supergravity}
\label{Section5}
The methods described in the previous sections \ref{Section2}, \ref{Section3} and  \ref{Section4} can in principle be generalized to the more complicated case of $\mathcal{N}=4$ JT supergravity, described in terms of the $\text{PSU}(1,1|2)$ supergroup. 

The worldline action is very similar to that of $\mathcal{N}=2$ JT supergravity detailed in formula \eqref{eq:acteow} in subsection \ref{s:higherank}. The main new thing is the presence of a non-abelian R-symmetry group SU$(2)$ that is completely modded out in the worldline description. This corresponds to the description of the hyperbolic superspaces as supercosets:
\begin{equation}
\text{H}_{2 \vert 8} \,\simeq \,\frac{\text{PSU}(1,1\vert 2)}{\text{U}(1)_{\text{Lorentz}} \otimes \text{SU}(2) }, \qquad \text{AdS}_{2 \vert 8} \,\simeq \,\frac{\text{PSU}(1,1\vert 2)}{\text{SO}(1,1) \otimes \text{SU}(2) }.
\end{equation}
This is a right coset of a $6 \vert 8$ dimensional space by a $4 \vert 0$ dimensional subspace, leading indeed to the $2 \vert 8$ dimensional superspace H$_{2 \vert 8}$.
Moreover, there are two Casimir operators: the usual quadratic one and a higher rank Casimir that we sketch in Appendix \ref{app:hrcas}.

Some aspects of the representation theory of the relevant supergroup PSU$(1,1\vert 2)$ and Lie superalgebra $\mathfrak{psu}(1,1|2)$ are discussed in Appendix \ref{AppendixE}. Here we simply report the discrete series character and the principal series character, which can be used to write down the EOW brane amplitudes in $\mathcal{N}=4$ JT supergravity. For this rank 2 supergroup, we again label conjugacy classes by two real variables $\phi \in \mathbb{R}^+$ and $\theta \in [0,\pi)$. \\

The principal series character is given by \eqref{eq:charN4}:
\begin{align}
\chi_{k,j_2}^{\mathcal{N}=4}(\phi,\theta) = 4\cos(2k\phi) \sin (2j_2+1)\theta\frac{\left(\cos\theta-\cosh\phi\right)^2}{\sinh\phi \sin\theta},
\end{align}
where the last factor descends from the Weyl denominator \eqref{eq:N=4weyl}. This character without the Weyl denominator is then used to derive the single trumpet amplitude as:
\begin{equation}
Z_{\text{trumpet}}(\beta;\phi, \theta) \sim \int_0^{+\infty} dk \sum_{J \in \mathbb{N}/2}\cos(2k\phi) \sin (2J+1)\theta\; e^{-\beta (k^2 + J(J+1))},
\end{equation}
where the energy variable $k^2 + J(J+1)$ is the quadratic Casimir in the principal series representation, quickly found by substituting the correct value of $j_1= -1/2 + ik$ in \eqref{eq:N=4highestweightcasimir} (and subtracting a zero-point energy term).

For the highest weight discrete series representation, we have \eqref{eq:N=4discretecharacter}:
\begin{equation}
     \chi^{\mathcal{N}=4}_{j_1,j_2}(\phi,\theta)=2e^{(2j_1+1) \phi}\sin(2j_2+1)\theta\frac{\left(\cos\theta-\cosh\phi\right)^2}{\sinh\phi \sin\theta}.
\end{equation}
 This character is again used to define the relevant (class-1) EOW-brane insertion.\\

Going through the same logic as before, this then leads to the expression for the amplitude with a single circular class-1 EOW-brane with mass $\mu$ and SU(2) spin $j$:

\begin{align}
\label{eq:n4res}
&\mathcal{A}(\beta;\mu,j) =\hspace{0.5cm}\raisebox{-0.5\height}{\includegraphics[height=1.6cm]{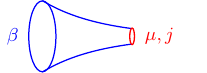}} =\int_0^{+\infty} dk \sum_{J\in\mathbb{N}/2} e^{-\beta (k^2+J(J+1))} \nonumber \\
&\times \int_0^{\infty} db \int_0^\pi d\theta \cos (kb) \sin(2J+1)\theta\; e^{-\mu b}\sin (2j+1)\theta \frac{(\cosh \frac{b}{2} - \cos \theta)^2}{\sinh \frac{b}{2} \sin \theta}. 
\end{align}
As for the $\mathcal{N}=2$ case discussed below \eqref{eq:cl1}, this expression is UV-divergent unless one can somehow restrict to $\theta=0$, the Ramond (periodic) sector. 

A further open (but technical) problem, is to integrate this expression over the second (higher-rank) Casimir to obtain what we denoted as class-2 EOW branes, which are manifestly described by a worldline action \eqref{eq:acteow} on the supermanifold with ``mass'' $\mathcal{C}_2^{1/2}$. Expressions for the quadratic and higher Casimir in terms of $j_1$ and $j_2$ appear in Appendix \ref{AppendixE}.

\section{Concluding remarks}
\label{sec:conc}
In this work we have heavily utilized group-theoretical methods to compute and generalize (super)gravitational trumpet amplitudes that contain an asymptotic boundary and a second end-of-the-world brane boundary. Such amplitudes are the building blocks for more sophisticated amplitudes that we leave for future work.

In this section we highlight several tangential routes, for which a thorough study is postponed to the future.

\subsection{Immediate extensions}
Our treatment of $\mathcal{N}=4$ JT supergravity and its underlying PSU$(1,1\vert 2)$ framework (in Appendix \ref{AppendixE}) have been somewhat less developed than those of the other cases $\mathcal{N}=0,1,2$. We leave this as a problem for the future.

Furthermore, in this work, we have focused solely on circular EOW branes. However, the disk EOW brane segment amplitudes (as shown for the bosonic case in \eqref{C5:fininalEOWdiskpartition}), are also interesting. We are in the process of understanding these amplitudes for the various supersymmetric versions of JT gravity.

As a final possible extension, amplitudes with a single circular EOW brane have been computed in the context of double-scaled SYK in \cite{Okuyama:2023byh}. It would be interesting to understand those calculations directly in the language of $q$-representation theory of the underlying quantum algebra U$_q(\mathfrak{su}(1,1))$, where $q=e^{-\lambda}$ denotes the double-scaling parameter, as developed in this context in e.g. \cite{Berkooz:2018jqr,Berkooz:2022mfk,Blommaert:2023opb,Lin:2022rbf,Goel:2023svz,Lin:2023trc}. Relatedly, given the formal relation between this quantum group and the one relevant for Liouville gravity \cite{Mertens:2022aou,Fan:2021bwt}, also similar amplitudes can be expected to be computable for Liouville gravity and the minimal string. It would be interesting to understand these better.

\subsection{Gravity and the positive semi-group}
It is known that actual (super)gravity cannot be entirely described by a BF gauge theory based on any of the groups we have studied in this work. The discrepancy has to do with the fact that perfectly fine gauge connections can correspond to singular geometries in Euclidean signature, which are not taken into account in the Euclidean gravitational path integral. In terms of the BF gauge theory, this corresponds to the moduli space of flat gauge connections containing several disconnected components, of which one (or two) actually correspond to possible geometries, the Teichm\"uller component. These considerations played no role in the story we have developed in this work at the level of the characters.

In previous works, and heavily inspired by an analogous story in the $q$-deformed setting developed by Teschner and collaborators (see e.g. \cite{Ponsot:1999uf,Teschner:2001rv,Teschner:2013tqy}), we have proposed to instead change the gauge group $G$ to its positive subsemigroup $G^+$. We have worked out this proposal and representation theory for the bosonic case, based on SL$^+(2,\mathbb{R})$ in \cite{Blommaert:2018iqz} and the $\mathcal{N}=1$ case based on OSp$^+(1\vert 2,\mathbb{R})$ in \cite{Fan:2021wsb}. It is not hard to make an analogous proposal for higher supersymmetries: when working with a real representation of the gauge group, one simply demands positivity of the four supernumbers in the bosonic subgroup in the fundamental representation. For $\mathcal{N}=2$, this would be the supernumbers $a,b,c,d$ in the parametrization \eqref{eq:OSp2matrix}, defining the positive semi-supergroup OSp$^+(2\vert 2,\mathbb{R})$. If one works with a complex representation instead (such as the fundamental representation of PSU$(1,1\vert 1)$), one has to transfer this positivity condition through the isomorphism between both formulations. Notice that the positivity condition only applies to the ``gravitational'' SL$(2,\mathbb{R})$-like subsector of the supergroup element: there is no constraint on the U(1) gauge connection for instance in order to be a physical $\mathcal{N}=2$ supergravity configuration. 

\subsection{Gas of branes and better UV behavior?}
This work initiated with the question on whether supersymmetric versions of JT gravity could have ameliorated behavior in the UV, as $b\to 0$ and the wormhole neck shrinks to zero size. There is reason to expect this since these amplitudes are essentially worldsheet amplitudes in string theory. The $b\to 0$ divergence is the usual closed string tachyon divergence. Transferring to supergravity means one transfers to the superstring where these divergences can be absent. Indeed, in all cases the Ramond sector was convergent, familiar from worldsheet string theory again. The main question we are hence led to is whether one can restrict to these kinds of branes in a dynamical way.\footnote{See e.g. \cite{Balasubramanian:2020jhl} for a related analysis.} 

It is useful to characterize the appearance of the UV-divergence in more algebraic terms. We have seen in all examples that the culprit is the numerator of the Weyl denominator, defined generically for a superalgebra in \eqref{eq:wd}. Now, for any superconformal algebra (for which the maximal bosonic subalgebra is $\mathfrak{sl}(2,\mathbb{R}) \oplus \mathfrak{g}_{\text{R}}$ for compact $\mathfrak{g}_{\text{R}}$), there is one Cartan element in the $\mathfrak{sl}(2,\mathbb{R})$ algebra, which automatically leads to a $\sim \sinh b/2$ in the denominator of the integrand of the amplitude. This means that there will generically be a divergence in the amplitude, unless a conspiracy happens. We have seen this to happen when specifying to the Ramond ($\theta=0$) subsector.

To achieve finiteness of the gravitational amplitudes, worldsheet string theory suggests we should implement the analogue of the GSO projection and effectively project out the closed string tachyon. It is our hope that we have provided the technical methods and expressions that would allow us to tackle this question. We leave a more in-depth study of this problem in JT supergravity amplitudes to future work.

\section*{Acknowledgments}
We thank A. Blommaert, Y. Fan, M. Heller, G.J. Turiaci  and Q. Wu for discussions. TM thanks ETH Zurich and Matthias Gaberdiel in particular for hospitality during the early stages of this work. The authors acknowledge financial support from the European Research Council (grant BHHQG-101040024). Funded by the European Union. Views and opinions expressed are however those of the author(s) only and do not necessarily reflect those of the European Union or the European Research Council. Neither the European Union nor the granting authority can be held responsible for them.

\appendix

\section{Superspace differential geometry conventions}\label{app:diffgeo}

\subsection{Conventions}

In general, we locally parameterize the $\mathcal{N}=1$ superspace by a pair of bosonic and fermionic coordinates $z^m,\; \theta^\mu$ ($m=0,1$, $\mu=0,1$) respectively. The latter satisfy the anti-commutative Grassmann algebra $\{\theta^\mu,\theta^\nu\}=0$. Along the lines of \cite{PSHowe_1979}, the $2|2$-dimensional bulk supergeometry is easy to describe by a pair of holomorphic and anti-holomorphic coordinates 
\begin{equation}
    Z^M=(z^m,\theta^\mu)=(z,\overline{z},\theta,\overline{\theta}).
\end{equation}
Due to the anticommutative nature of the fermionic partners, care has to be taken when swapping two superspace coordinates (or their differentials) 
\begin{equation}\label{C6:conventies}
Z^MZ^N=(-)^{MN} Z^NZ^M,\,\, dZ^M \otimes dZ^N = (-)^{MN}dZ^N \otimes dZ^M,\,\, dZ^MZ^N=(-)^{MN}Z^NdZ^M.
\end{equation} 
We imagine the numbers $M,N$ in the exponent $(-)^{MN}$ to be $\mathbb{Z}_2$-valued (0,1), and are either even or odd if the respective coordinate is bosonic or fermionic.

Next, we define a metric in superspace $g_{MN}$. This has a bosonic block for $(M+N)\mod 2=0$, and a fermionic block for $(M+N)\mod 2=1$, with $M,N$ the $\mathbb{Z}_2$-valued indices labeling the bosonic ($0$) or fermionic ($1$) parity of the superfield. We further define an inverse metric tensor $g^{MN}$, satisfying \begin{equation}
\label{C6:supermetricinverse}
g^{NK}g_{KM}=g_{MK}g^{KN}=\delta^{\;N}_M.
\end{equation} 
For consistency between these definitions, we take the conventions
\begin{equation}\label{C6:anticommutativityconvention}
g^{MN}=(-)^{MN}\;g^{NM},\qquad \text{and} \qquad g_{MN}=(-)^{MN+M+N}g_{NM}.
\end{equation}
Concretely, this means that both the fermionic and doubly fermionic block of the metric tensor $g_{MN}$ are antisymmetric with respect to interchanges in the indices. This symmetry property is consistent with the expression for the invariant distance in superspace, written in the symmetric way as $ds^2=dZ^Mg_{MN}dZ^N$. 

Moreover, since this line element $ds^2=dZ^Mg_{MN}dZ^N$ is by definition coordinate invariant, we define a covariant vector according to: \begin{equation}\label{C6:covectordefinition}
\Dot{Z}_M\equiv g_{MN}\Dot{Z}^N.
\end{equation}
Contractions are thereby defined in the NW-SE (north-west - south-east) convention: 
\begin{equation}
\Dot{Z}^M g_{MN}\Dot{Z}^N = \Dot{Z}^M\Dot{Z}_M.
\end{equation} 
We define a coordinate transformation on covariant vectors from the left 
\begin{equation}\label{C6:covarianttransformation}
\Dot{Z}_M{}'\equiv \frac{\partial Z^N}{\partial Z^M{}'}\Dot{Z}_N,
\end{equation}
where coordinate-invariant contractions again appear in the NW-SE ordering. The transformation rule on covectors \eqref{C6:covarianttransformation} is consistent with the coordinate transformation of a gradient in superspace $\partial_M$, where the (fermionic) chain rule acts also from the left: 
\begin{equation}
        \partial_M\;\rightarrow\; \partial_M{}'=\frac{\partial Z^N}{\partial Z^M{}'}\partial_N.
\end{equation}

Due to the coordinate invariant (NW-SE) structure, a coordinate transformation acts on a contravariant vector $\Dot{Z}^M$ from the right: \begin{equation}\label{C6:contravarianttransformation}
\Dot{Z}^M{}'=\Dot{Z}^N\frac{\partial Z^M{}'}{\partial Z^N}.
\end{equation} 
$\frac{\partial Z^M{}'}{\partial Z^N}$ is defined as the inverse of $\frac{\partial Z^N}{\partial Z^M{}'}$ within the NW-SE structure: \begin{equation}
    \frac{\partial Z^M{}'}{\partial Z^N} \frac{\partial Z^K}{\partial Z^M{}'}= \delta^K_N.
\end{equation}

\subsection{Geodesic equations in superspace}
 We start by writing the worldline action in superspace symbolically as: \begin{equation}
    I=\mu \int ds\;\sqrt{\Dot{Z}^M g_{MN}\Dot{Z}^N}=\mu\int \sqrt{dZ^M g_{MN}dZ^N},
\end{equation} 
by formally absorbing the measure of the affine parameter $ds$ inside the square root. Varying the action yields:
\begin{align*}
    \delta I&=\mu \int \;\delta\left(\sqrt{dZ^Mg_{MN}dZ^N}\right)\\&=\mu \int \frac{1}{2\sqrt{ dZ^Mg_{MN}dZ^N}} \;\left(\delta dZ^M\;g_{MN}dZ^N+ dZ^M\;\delta g_{MN} \; dZ^N+dZ^M\;g_{MN}\;\delta dZ^N\right)\\&=\frac{\mu}{2}\int ds\; \left( (\delta \Dot{Z}^M)\:g_{MN}\Dot{Z}^N+\Dot{Z}^M\;\delta g_{MN}\; \Dot{Z}^N+\Dot{Z}^Mg_{MN}\;(\delta \Dot{Z}^N)\right),
\end{align*}where we have used the standard chain rule in the bosonic line element $ds=\sqrt{dZ^Mg_{MN}dZ^N}$. Furthermore, the variation $\delta$ obeys the bosonic product rule in the convention that it acts from the left.
With our convention \eqref{C6:anticommutativityconvention}, the first and last terms are in fact equal. 
Furthermore, using the natural chain rule within the NW-SE convention, we may write $\delta g_{MN}=\delta Z^P \partial_P g_{MN}$ and obtain: 
\begin{align*}
    \delta I&=\mu\int ds\; \left((\delta\Dot{Z}^M)\;g_{MN}\Dot{Z}^N+\frac{1}{2}\Dot{Z}^M\;\delta Z^P\;\partial_P g_{MN}\Dot{Z}^N\right)\\&=\mu\int ds\; \left((\delta\Dot{Z}^M)\;g_{MN}\Dot{Z}^N+(-)^{M}\frac{1}{2}\delta Z^P \;\partial_P g_{MN}\Dot{Z}^N\Dot{Z}^M\right)\\&\simeq -\mu \int ds\; \left(\delta Z^M\;g_{MN}\Ddot{Z}^N+\delta Z^M\; \Dot{Z}^P\partial_P g_{MN}\Dot{Z}^N-(-)^{M}\frac{1}{2}\delta Z^P \;\partial_P g_{MN}\Dot{Z}^N\Dot{Z}^M\right),
\end{align*}
where we have partially integrated in the last line for the bosonic derivative with respect to $s$. Separating out $\delta Z^Pg_{PK}$, and using our definition of the inverse metric $g_{PK}g^{KR}=\delta_P^{\;R}$: 
\begin{align*}
    \delta I&=-\mu \int ds\;\delta Z^Pg_{PK}\left(\Ddot{Z}^K+g^{KR}\Dot{Z}^M \partial_M g_{RN}\Dot{Z}^N-\frac{1}{2}(-)^{M}g^{KR}\partial_Rg_{MN}\Dot{Z}^N\Dot{Z}^M\right)\\ &=-\mu \int ds\; \delta Z^Pg_{PK}\left(\Ddot{Z}^K+\frac{1}{2}g^{KR}\left(2 (-)^{M(1+R)}\partial_Mg_{RN}-(-)^{M}\partial_Rg_{MN}\right)\Dot{Z}^N \Dot{Z}^M\right).
\end{align*}
Symmetrizing the first term within brackets as: \begin{align*}
    2 (-)^{M(1+R)}\partial_Mg_{RN}\Dot{Z}^N\Dot{Z}^M&=(-)^{M(1+R)}\partial_Mg_{RN}\Dot{Z}^N\Dot{Z}^M+(-)^{N(1+R)}\partial_N g_{RM}\Dot{Z}^M\Dot{Z}^N\\ &=\left((-)^{M(1+R)}\partial_Mg_{RN}+(-)^{N(1+R+M)}\partial_N g_{RM}\right)\Dot{Z}^N\Dot{Z}^M,
\end{align*}
eventually yields: 
\begin{align}
    \delta I&= -\mu \int ds\;\delta Z^Pg_{PK}\;\Big(\Ddot{Z}^K\nonumber\\&\qquad+\frac{1}{2}g^{KR}\left((-)^{M(1+R)}\partial_Mg_{RN}+(-)^{N(1+R+M)}\partial_N g_{RM}-(-)^{M}\partial_R g_{MN}\right)\Dot{Z}^N\Dot{Z}^M\Big) \nonumber\\&\equiv -\mu \int ds\;\delta Z^Pg_{PK}\left(\Ddot{Z}^K+\Gamma^K_{MN}\Dot{Z}^N \Dot{Z}^M\right).
\end{align}
In the last line, we have introduced an appropriate definition of the Christoffel symbol in superspace: 
\begin{equation}
\label{C6:defchristoffel}
    \Gamma^K_{MN}\equiv\frac{1}{2}g^{KR}\left((-)^{M(1+R)}\partial_Mg_{RN}+(-)^{N(1+R+M)}\partial_N g_{RM}-(-)^{M}\partial_R g_{MN}\right).
\end{equation}
This definition of the Christoffel symbol matches with the one introduced in \cite{Shuji}.
By our definition of the metric tensor \eqref{C6:anticommutativityconvention} and the covector \eqref{C6:covectordefinition}, we can write the last line in a more suggestive coordinate-invariant NW-SE way as:
\begin{align}
    \delta I&=-\mu \int ds\;\left(\Ddot{Z}^K+\Gamma^K_{MN}\Dot{Z}^N \Dot{Z}^M\right) g_{KP}\:\delta Z^P\nonumber\\&=-\mu \int ds\;\left(\Ddot{Z}^K+\Gamma^K_{MN}\Dot{Z}^N \Dot{Z}^M\right)\:\delta Z_K.\label{C6:finalvariation}
\end{align}

To proceed, we rewrite the geodesic equation more compactly by introducing a covariant derivative in superspace in terms of the superspace Christoffel symbol. Acting on a contravariant vector $U^M\equiv\Dot{Z}^M$, we define \begin{equation}\label{C6:contravariantcovariantderivative}
    \nabla_MU^K\equiv \partial_M U^K+(-)^{M(K+1)}\;\Gamma^K_{MN}U^N.
\end{equation}
One can check that this is indeed consistent: \begin{equation} \label{eq:supergeodesiceqs}
    U^M\nabla_M U^K=\Dot{U}^K+\Gamma^K_{MN} U^NU^M.
\end{equation}
We may therefore write the variation of the action more suggestively as: 
\begin{equation}\label{eq:supervariation}
    \delta I=-\mu \int ds\;\left( U^M\nabla_M U^K\right)\delta Z_K,
\end{equation}
where all contractions appear in a manifestly coordinate-invariant NW-SE ordering. 
This unambiguously fixes the transformation of $\nabla_M U^K$ under general coordinate transformations in order to preserve this structure, \begin{equation} \label{C6:transformationcovariantderivative1}
    \nabla_M U^K\; \rightarrow \; \nabla_M' U^K{}'= \frac{\partial Z^L}{\partial Z'^M}\left(\nabla_L U^R \right)\frac{\partial Z'^K}{\partial Z^R}.
\end{equation} 

We introduce a covariant derivative acting on covariant vectors by demanding that the covariant derivative acting on a scalar structure reduces to the standard (possibly Grassmann) derivative: $\nabla_M X\equiv \partial_M X$. Acting on the scalar product $U^Nn_N$, it should obey the standard (fermionic) product rule: \begin{equation}\label{C6:firstline}
    \nabla_M\left(U^Nn_N\right)\equiv \partial_M\left(U^Nn_N\right)=\left(\partial_M U^N\right)n_N+(-)^{MN} U^N\left( \partial_M n_N\right).
\end{equation} On the other hand, we \emph{define} the covariant derivative on covectors such that it obeys a similar product rule: \begin{align}
    \nabla_M\left(U^Nn_N\right)&\equiv \left(\nabla_M U^N\right)n_N+(-)^{MN} U^N\left(\nabla_M n_N\right)\label{C6:covariantproductrule}\\&=\left(\partial_M U^N+(-)^{M(N+1)}\;\Gamma^N_{MK}U^K\right)n_N+(-)^{MN}U^N\left(\nabla_M n_N\right).\label{C6:lastline}
\end{align}
Compared to the previous line \eqref{C6:firstline}, this fixes the covariant derivative on a general covector $n_A$: \begin{equation}
\label{C6:covariantderivativeoncovector}
\nabla_M n_N\equiv\partial_M n_N-(-)^{M(1+K)+N(1+K)}\;\Gamma^K_{MN}n_K,
\end{equation} 
in terms of the Christoffel symbol \eqref{C6:defchristoffel}. Additionally, one can check that the metric is invariant under the covariant derivative  \begin{equation}\label{eq:metricpostulate}
     \nabla_M U_N=(-)^{M(N+K)}g_{NK}\nabla_MU^K. 
\end{equation}

Since the LHS in \eqref{C6:lastline} and the first term on the RHS are manifestly covariant under the NW-SE convention, the transformation rule of the covariant derivative acting on covectors under general coordinate transformations is a posteriori fixed to: \begin{equation}
    \nabla_M n_N\;\rightarrow\;\nabla_M' n_N'\;=\;\frac{\partial Z^K}{\partial Z'^M}\;\nabla_K\;\left(\frac{\partial Z^P}{\partial Z'^N}n_P\right)\;\equiv \; (-)^{K(P+N)} \frac{\partial Z^K}{\partial Z'^M}\;\frac{\partial Z^P}{\partial Z'^N}\;\nabla_K n_P.
\end{equation} This transformation is fine-tuned to keep the second term in  \eqref{C6:covariantproductrule} covariant:
\begin{align}
    (-)^{MN}U^N\left(\nabla_M n_N\right)\;\rightarrow\; &(-)^{MN+K(P+N)} U^L\;\frac{\partial Z'^N}{\partial Z^L}\; \frac{\partial Z^K}{\partial Z'^M}\;\frac{\partial Z^P}{\partial Z'^N}\;\nabla_K n_P\nonumber\\=&(-)^{MP} U^L\;\frac{\partial Z'^N}{\partial Z^L}\frac{\partial Z^P}{\partial Z'^N}\frac{\partial Z^K}{\partial Z'^M}\;\nabla_K n_P=(-)^{MP} U^P \frac{\partial Z^K}{\partial Z'^M}\nabla_Kn_P \nonumber\\=&(-)^{PK}\frac{\partial Z^K}{\partial Z'^M} U^P\nabla_Kn_P.\label{C6:consistentderiv}
\end{align}

\section{OSp$\left(2\vert 2,\mathbb{R}\right)$ Representation Theory}
\label{sec:N2rep}

In this appendix, we give an overview of OSp$\left(2\vert 2,\mathbb{R}\right)$ representation theory. We give the details of the computation of the principal series character used in section 4. Our methods are largely based on those used in Appendix E of \cite{Fan:2021wsb}, for $\mathcal{N}=1$ JT supergravity. 
\subsection{OSp$\left(2|2,\mathbb{R}\right)$ Supergroup and Lie superalgebra}
The relevant group for $\mathcal{N}=2$ JT supergravity is the supergroup OSp$\left(2|2,\mathbb{R}\right)$,
which is defined as the subgroup of GL$\left(2|2,\mathbb{R}\right)$ matrices 
\begin{equation}
g=\left[\begin{array}{c c | c c} 
	a & b & \alpha_1 & \beta_1\\ 
    c & d & \gamma_1 & \delta_1\\	
 \hline 
	\alpha_2 & \beta_2 & w & y\\
    \gamma_2 & \delta_2 & z & u\\
\end{array}\right],
\label{eq:OSp2matrix}
\end{equation}
with 8 bosonic variables $a,b,c,d,w,y,z,u$ and 8 fermionic variables $\alpha_{1,2},\beta_{1,2},\gamma_{1,2},\delta_{1,2}$,
that preserve the orthosymplectic form $\Omega$: $g\Omega g^{{st}^3}= \Omega$:
\begin{equation}
    \left[\begin{array}{c c | c c} 
	a & b & \alpha_1 & \beta_1\\ 
    c & d & \gamma_1 & \delta_1\\	
 \hline 
	\alpha_2 & \beta_2 & w & y\\
    \gamma_2 & \delta_2 & z & u\\
\end{array}\right]\left[\begin{array}{c c | c c} 
	0 &-1 & 0 & 0\\ 
    1 & 0 & 0 & 0\\	
 \hline 
	0 & 0 & 1 & 0\\
    0 & 0 & 0 & 1\\
    
\end{array}\right] \left[\begin{array}{c c | c c} 
	a & c & \alpha_2 & \gamma_2\\ 
    b & d & \beta_2 & \delta_2\\	
 \hline 
	-\alpha_1 & -\gamma_1 & w & z\\
    -\beta_1 & -\delta_1 & y & u\\
\end{array}\right]=\left[\begin{array}{c c | c c} 
	0 &-1 & 0 & 0\\ 
    1 & 0 & 0 & 0\\	
 \hline 
	0 & 0 & 1 & 0\\
    0 & 0 & 0 & 1\\
    
\end{array}\right]~,
\label{eq:generators1}
\end{equation}
where $g^{{st}^3}$ is the matrix obtained starting form a matrix of the form (\ref{eq:OSp2matrix}) after applying the supertransposition three consecutive times.\footnote{The supertranspose operation consists in flipping the sign of one block of fermionic variables so that the property $(g_1g_2)^{st}=g_2^{st}g_1^{st}$ is preserved,
\begin{equation}
    \left[\begin{array}{c | c} 
	 A & B\\ 	
 \hline 
	 C & D\\
    
\end{array}\right]^{st}=\left[\begin{array}{c | c} 
	 A^T & -C^T\\ 	
 \hline 
	 B^T & D^T\\
    
\end{array}\right]~.
\end{equation} }

The matrix inverse has a particularly simple form:
\begin{equation}
\label{eq:inv}
g^{-1}=\left[\begin{array}{c c | c c} 
d & -b & -\beta_2 & -\delta_2\\ 
-c & a & \alpha_2 & \gamma_2 \\	
 \hline 
\gamma_1 & -\alpha_1 & w & z\\
\delta_1 & -\beta_1 & y & u
\end{array}\right].
\end{equation}

The $\mathfrak{osp}(2|2)$ Lie superalgebra is a $4 \vert 4$ dimensional super-vectorspace, with bosonic generators $H,E^\pm, Z$ and fermionic generators $F^\pm$, $\bar{F}^\pm$. In the Cartan-Weyl/Chevalley basis, these generators satisfy the superalgebra relations (see e.g. \cite{Frappat:1996pb}):
\begin{alignat}{3}
\label{eq:superalgebra}
[H,E^\pm] &= \pm E^\pm, \qquad &[H,F^\pm] &= \pm \frac{1}{2} F^\pm, \quad &[H, \bar{F}^\pm ] &= \pm \frac{1}{2} \bar{F}^\pm, \nonumber \\
[Z,H] &= [Z,E^\pm] =0, \quad &[Z,F^\pm] &= \frac{1}{2} F^\pm, \quad &[Z,\bar{F}^\pm] &= - \frac{1}{2}\bar{F}^\pm, \nonumber \\
[E^\pm,F^\pm] &= [E^\pm,\bar{F}^\pm] =0, \quad &[E^\pm,F^\mp] &=-F^\pm, \quad &[E^\pm,\bar{F}^\mp] &= \bar{F}^\pm, \nonumber \\
\{F^\pm,F^\pm\} &= \{\bar{F}^\pm,\bar{F}^\pm\} = 0, \qquad &\{F^\pm,F^\mp\} &= \{\bar{F}^\pm,\bar{F}^\mp\} =0, \qquad &\{F^\pm,\bar{F}^\pm\} &= E^\pm, \nonumber \\
[E^+,E^-] &= 2H, \quad &\{F^\pm,\bar{F}^\mp\} &= Z\mp H. \quad &
\end{alignat}
The Cartan subalgebra is spanned by the two generators $H $ and $Z$, whose eigenvalues can be raised and lowered by half a unit by acting with the different $F^\pm$, $\bar{F}^\pm$, and the eigenvalue of $H$ by a full unit by acting with $E^\pm$.

This form of the algebra is related to the more familiar global NS sector $\mathcal{N}=2$ superconformal algebra $(L_0, L_{\pm 1}, G_{\pm \frac{1}{2}}, \bar{G}_{\pm \frac{1}{2}})$:
\begin{alignat}{2}
[L_{0},L_{\pm 1}] &= \mp L_{\pm1}, \qquad &[L_+,L_-] &= 2 L_0, \\
[L_\pm, G_{\mp \frac{1}{2}}] &= \pm G_{\pm \frac{1}{2}}, \qquad &[L_\pm, \bar{G}_{\mp \frac{1}{2}}] &= \pm \bar{G}_{\pm \frac{1}{2}}, \\
[L_0, G_{\pm \frac{1}{2}}] &= \mp \frac{1}{2} G_{\pm \frac{1}{2}}, \qquad &[L_0, \bar{G}_{\pm \frac{1}{2}}] &= \mp \frac{1}{2} \bar{G}_{\pm \frac{1}{2}}, \\
[J, G_{\pm \frac{1}{2}}] &= G_{\pm \frac{1}{2}}, \qquad &[J, \bar{G}_{\pm \frac{1}{2}}] &= - \bar{G}_{\pm \frac{1}{2}}, \\
\{G_{\pm \frac{1}{2}}, \bar{G}_{\pm \frac{1}{2}}\} &= 2 L_{\pm 1}, \qquad &\{G_{\pm \frac{1}{2}}, \bar{G}_{\mp \frac{1}{2}}\} &= 2L_0 \pm J,
\end{alignat}
by the relations:
\begin{gather}
L_0 = H, \qquad L_{+1} = -E^-, \qquad L_{-1} = E^+, \qquad J=2 Z, \\
G_{+\frac{1}{2}} = \sqrt{2} F^-, \qquad G_{-\frac{1}{2}} = \sqrt{2} F^+, \qquad \bar{G}_{+\frac{1}{2}} = -\sqrt{2} \bar{F}^-, \qquad \bar{G}_{-\frac{1}{2}} = \sqrt{2}\bar{F}^+. \nonumber
\end{gather}

The fundamental representation of the $\mathfrak{osp}(2|2)$ algebra is given by:
\begin{align}
\label{eq:generators}
H &= \left[\begin{array}{c c | c c} 
\frac{1}{2} & 0 & 0 & 0 \\
0& -\frac{1}{2} & 0 & 0 \\
\hline 
0 & 0 & 0 & 0 \\
0 & 0 & 0 & 0
\end{array}\right], 
E^+ = \left[\begin{array}{c c | c c} 
0 & 1 & 0 & 0 \\
0& 0 & 0 & 0 \\
\hline 
0 & 0 & 0 & 0 \\
0 & 0 & 0 & 0
\end{array}\right], 
E^- = \left[\begin{array}{c c | c c} 
0 & 0 & 0 & 0 \\
1& 0 & 0 & 0 \\
\hline 
0 & 0 & 0 & 0 \\
0 & 0 & 0 & 0
\end{array}\right],\\
Z &= \left[\begin{array}{c c | c c} 
0 & 0 & 0 & 0 \\
0& 0 & 0 & 0 \\
\hline 
0 & 0 & 0 & \frac{i}{2} \\
0 & 0 & -\frac{i}{2} & 0
\end{array}\right], 
F^+ = \left[\begin{array}{c c | c c} 
0 & 0 & \frac{1}{2} & -\frac{i}{2} \\
0& 0 & 0 & 0 \\
\hline 
0 & \frac{1}{2} & 0 & 0 \\
0 & -\frac{i}{2} & 0 & 0
\end{array}\right],
\bar{F}^+ = \left[\begin{array}{c c | c c} 
0 & 0 & \frac{1}{2} & \frac{i}{2} \\
0& 0 & 0 & 0 \\
\hline 
0 & \frac{1}{2} & 0 & 0 \\
0 & \frac{i}{2} & 0 & 0
\end{array}\right],\\
F^- &= \left[\begin{array}{c c | c c} 
0& 0 & 0 & 0 \\
0 & 0 & -\frac{1}{2} & \frac{i}{2} \\
\hline 
\frac{1}{2} & 0  & 0 & 0 \\
-\frac{i}{2}& 0  & 0 & 0
\end{array}\right],
\bar{F}^- = \left[\begin{array}{c c | c c} 
0& 0 & 0 & 0 \\
0 & 0 & \frac{1}{2} & \frac{i}{2} \\
\hline 
-\frac{1}{2}  & 0 & 0 & 0 \\
-\frac{i}{2} & 0 & 0 & 0
\end{array}\right].
\end{align}

As a rank 2 Lie superalgebra, there are two independent Casimir operators spanning the center of the universal enveloping algebra. 

The Cartan-Killing (CK) metric $\text{STr}(J_IJ_J)=\kappa_{IJ}/2$ is determined from the overlap of the generators $J_I$ and $J_J$. The inverse CK metric $\kappa^{IJ}\kappa_{JL}=\delta_{\;\;L}^I$ then defines the quadratic Casimir according to the standard definition $\mathcal{C}_2=J_I\kappa^{IJ}J_J$:
\begin{equation}
\label{eq:Casimir1}
    \mathcal{C}_2=H^2-Z^2+E^-E^++F^-\bar{F}^+-\bar{F}^-F^+.
\end{equation}
We write the cubic Casimir following \cite{Frappat:1996pb}:
\begin{gather}
\label{eq:cubicCasimir}
    \mathcal{C}_3=(H^2-Z^{2})Z+E^-E^+(Z-\tfrac{1}{2})-\tfrac{1}{2}F^-\bar{F}^+(H-3Z+1)\\
    -\tfrac{1}{2}\bar{F}^-F^+(H+3Z+1)+\tfrac{1}{2}E^-\bar{F}^+F^++\tfrac{1}{2}\bar{F}^-F^-E^+.\nonumber
\end{gather}
Both $\mathcal{C}_2$ and $\mathcal{C}_3$ are seen to commute with all generators.\\

The link between the group and the algebra is given by exponentiation. One convenient parametrization of the group element $g$ is the \textit{Gauss-Euler} decomposition:
\begin{align}
\label{eq:GEcomplex}
g = e^{\bar{\theta}^- \bar{F}^-} e^{\theta^- F^-} e^{\beta E^-} e^{2\phi H} e^{2i \theta Z}e^{\gamma E^+} e^{\theta^+ F^+} e^{\bar{\theta}^+ \bar{F}^+},
\end{align}
where the complex Grassmann variables $\theta_\pm$ and $\bar{\theta}_\pm$ are complex conjugates.

It is convenient to transfer to new real fermionic generators defined as:
\begin{equation}\label{eq:generatorstransf}
F^+_x \equiv \frac{F^++\bar{F}^+}{\sqrt{2}}, \quad F^+_y \equiv \frac{F^+-\bar{F}^+}{\sqrt{2}i}, \quad F^-_x \equiv \frac{F^--\bar{F}^-}{\sqrt{2}}, \quad F^-_y \equiv \frac{F^-+\bar{F}^-}{\sqrt{2}i},
\end{equation}
satisfying $\{F^+_x,F^+_x\} = \{F^+_y,F^+_y\} = E^+$ and $\{F^-_x,F^-_x\} = \{F^-_y,F^-_y\} = -E^-$. They form the $x$- and $y$-components of a vector that is rotated by the generator $Z$. We can now use the identities:
\begin{align}
e^{\theta^+ F^+} e^{\bar{\theta}^+\bar{F}^+} &= e^{i \theta_x^+\theta_y^+ E^+}e^{\sqrt{2}\theta^+_{x} F^+_x} e^{\sqrt{2}\theta^+_{y} F^+_y}, \\
e^{\bar{\theta}^-\bar{F}^-}e^{\theta^- F^-}  &= e^{\sqrt{2}\theta^-_{y} F^-_y} e^{\sqrt{2}\theta^-_{x} F^-_x} e^{i \theta_x^-\theta_y^- E^-},
\end{align}
where $\theta^+ \equiv \theta_x^+ - i \theta_y^+$ and $\bar{\theta}^+ \equiv \theta_x^+ + i \theta_y^+$ and $\theta^- \equiv \theta_x^- - i \theta_y^-$ and $\bar{\theta}^- \equiv -\theta_x^- - i \theta_y^-$.
Hence using real variables, we can write the Gauss-Euler decomposition as (with a shifted coordinate $\beta$ and $\gamma$ to absorb the above Grassmann-valued offset):
\begin{align}
\label{eq:GE}
\boxed{
g = e^{\sqrt{2}\theta^-_{y} F^-_y} e^{\sqrt{2}\theta^-_{x} F^-_x} e^{\beta E^-} e^{2\phi H} e^{2i \theta Z}e^{\gamma E^+} e^{\sqrt{2}\theta^+_{x} F^+_x} e^{\sqrt{2}\theta^+_{y} F^+_y}}~.
\end{align}
For later convenience we write it down explicitly in matrix form:
\tiny
\begin{align}
\label{eq:GEexpli}
\left[\begin{array}{c c | c c} 
\hspace{-0.1cm} e^\phi & e^\phi \gamma & e^\phi \theta^+_x & - e^{\phi} \theta^+_{y} \\
\hspace{-0.1cm} e^\phi \beta & \hspace{-0.1cm} e^{-\phi} \hspace{-0.1cm} + \hspace{-0.05cm}\beta \gamma e^{\phi} \hspace{-0.1cm} - \hspace{-0.05cm}(\theta^-_x \hspace{-0.05cm}\cos\theta \hspace{-0.05cm}- \hspace{-0.05cm}\theta^-_y \hspace{-0.05cm} \sin\theta) \theta^+_x \hspace{-0.1cm}- (\theta^-_x \hspace{-0.05cm} \sin\theta + \theta^-_y \hspace{-0.05cm} \cos\theta) \theta^+_y \hspace{-0.1cm} & \hspace{-0.05cm} e^\phi \beta \theta^+_x \hspace{-0.05cm} -\hspace{-0.05cm}\theta^-_x \hspace{-0.05cm}\cos\theta \hspace{-0.05cm}+\hspace{-0.05cm}\theta^-_y \hspace{-0.05cm}\sin\theta & -e^\phi \beta \theta^+_y \hspace{-0.05cm} + \hspace{-0.05cm} \theta^-_x \hspace{-0.05cm}\sin\theta \hspace{-0.05cm}+\hspace{-0.05cm}\theta^-_y \hspace{-0.05cm}\cos\theta \hspace{-0.1cm}\\
\hline
\hspace{-0.1cm} e^\phi \theta^-_x & e^\phi \gamma \theta^-_x + \theta^+_x \cos\theta + \theta^+_y \sin \theta & e^\phi \theta^-_x\theta^+_x + \cos \theta & -e^{\phi} \theta^-_x \theta^+_y - \sin\theta \\
\hspace{-0.1cm} -e^\phi \theta^-_y & -e^\phi \gamma \theta^-_y + \theta^+_x \sin\theta - \theta^+_y \cos \theta & -e^{\phi} \theta^-_y \theta^+_x + \sin\theta & e^\phi \theta^-_y\theta^+_y + \cos \theta 
\end{array} \right]
\end{align}
\normalsize

\subsection{Euler angle decompositions}
Next to the Gauss-Euler decomposition \eqref{eq:GEcomplex}, one can also find two other parametrizations of the group element OSp$(2\vert 2,\mathbb{R})$, forming an analogue of the Euler ``angle'' decompositions of the SL$(2,\mathbb{R})$ group element. These are useful when constructing hyperbolic space and the Lorentzian signature AdS superspaces as a coset of this (super)group as we do further on.

We start with \eqref{eq:GEcomplex}, and consider only the bosonic subgroup: 
\begin{equation}
\left[\begin{array}{c c | c c} 
a & b & 0 & 0 \\
c& d & 0 & 0 \\
\hline 
0 & 0 & \cos\theta & -\sin\theta \\
0 & 0 & \sin \theta & \cos \theta
\end{array}\right],
\end{equation}
which is a direct product SL$(2,\mathbb{R})\otimes \text{U}(1)$. Using the ``hyperbolic'' Euler angle parametrization of SL$(2,\mathbb{R})$:
\begin{equation}
\label{eq:eu1}
e^{\varphi H} e^{\eta(E^+ + E^-)} e^{\psi H} = \left[\begin{array}{c c} e^{\varphi/2} &0 \\ 0 & e^{-\varphi/2} \end{array} \right] \left[\begin{array}{c c} \cosh \eta & \sinh \eta \\ \sinh \eta & \cosh \eta \end{array} \right]  \left[\begin{array}{c c} e^{\psi/2} &0 \\ 0 & e^{-\psi/2} \end{array} \right],
\end{equation}
where $\eta$ is the hyperbolic angle, we can write the full supergroup element:
\begin{align}
g = e^{\bar{\theta}^- \bar{F}^-} e^{\theta^- F^-} e^{\varphi H} e^{\eta(E^+ + E^-)} e^{\psi H} e^{2i\theta Z} e^{\theta^+ F^+} e^{\bar{\theta}^+ \bar{F}^+}~.
\end{align}
Next, we move $e^{\psi H} e^{2i\theta Z}$ all the way to the right, and $e^{\varphi H}$ all the way to the left. From the superalgebra relations, we see that this procedure leads to a simple replacement of the Grassmann-coordinates by:
\begin{alignat}{2}
\theta^+ &\to \theta^+ e^{\psi/2+i\theta}, \qquad &&\bar{\theta}^+ \to \bar{\theta}^+ e^{\psi/2-i\theta}, \\
\theta^- &\to \theta^- e^{\varphi/2}, \qquad &&\bar{\theta}^- \to \bar{\theta}^- e^{\varphi/2},
\end{alignat}
such that we obtain the ``hyperbolic'' Euler-angle decomposition (where we absorb the rescalings again into the Grassmann-variables):
\begin{align}
\label{eq:EA}
\boxed{
g = e^{\varphi H} e^{\bar{\theta}^- \bar{F}^-} e^{\theta^- F^-}  e^{\eta(E^+ + E^-)} e^{\theta^+ F^+} e^{\bar{\theta}^+ \bar{F}^+}e^{\psi H} e^{2i\theta Z}} ~.
\end{align}
Note that we chose to move the factor $e^{2i\theta Z}$ to the right here. One can make other choices, but the current choice is very convenient to describe the anti de Sitter superspace AdS$_{2 \vert 2}$ as a supercoset as we do in the next subsection. \\

Finally, there is the following ``elliptic'' Euler angle decomposition of SL$(2,\mathbb{R})$:
\begin{equation}
\label{eq:ellEulangl}
e^{\varphi (E^+-E^-)/2} e^{\eta(E^+ + E^-)} e^{\psi (E^+-E^-)/2} \hspace{-0.1cm}= \hspace{-0.1cm}\left[\begin{array}{c c} \hspace{-0.1cm}\cos\frac{\varphi}{2} & \hspace{-0.1cm} \sin \frac{\varphi}{2} \\ \hspace{-0.1cm}-\sin \frac{\varphi}{2} & \hspace{-0.1cm} \cos \frac{\varphi}{2}\end{array} \right] \hspace{-0.1cm}\left[\begin{array}{c c} \hspace{-0.1cm}\cosh \eta & \hspace{-0.1cm}\sinh \eta \\ \hspace{-0.1cm}\sinh \eta & \hspace{-0.1cm}\cosh \eta \end{array} \right]  \hspace{-0.1cm}\left[\begin{array}{c c} \hspace{-0.1cm}\cos\frac{\psi}{2} & \hspace{-0.1cm}\sin \frac{\psi}{2} \\ \hspace{-0.1cm}-\sin \frac{\psi}{2} & \hspace{-0.1cm}\cos \frac{\psi}{2} \end{array} \right].
\end{equation}
Conjugating with the SL$(2,\mathbb{C})$ matrix $t=\frac{1}{\sqrt{2}}\left[\begin{array}{c c} 1 & i \\ i & 1 \end{array} \right]$ implements the isomorphism of $\text{SL}(2,\mathbb{R}) \simeq \text{SU}(1,1)$. Then the above decomposition \eqref{eq:ellEulangl} is mapped into the $\text{SU}(1,1)$ decomposition:
\begin{equation}
e^{i\varphi H} e^{\eta(E^+ + E^-)} e^{i\psi H} = \left[\begin{array}{c c} e^{i\varphi/2} &0 \\ 0 & e^{-i\varphi/2} \end{array} \right] \left[\begin{array}{c c} \cosh \eta & \sinh \eta \\ \sinh \eta & \cosh \eta \end{array} \right]  \left[\begin{array}{c c} e^{i\psi/2} &0 \\ 0 & e^{-i\psi/2} \end{array} \right],
\end{equation}
which is formally the analytic continuation $\varphi \to i \varphi,\, \psi \to i \psi$ of \eqref{eq:eu1}. This is important since the analytic continuation of the $\psi$-coordinate will precisely correspond to the Wick-rotation to swap metric signature between AdS and hyperbolic (super)space.

Inserting this into the full matrix element for OSp$(2\vert 2,\mathbb{R})$, one can again move the left and right rotation matrices (over $\phi$ and $\psi$) through the fermionic generators. This description is useful for describing Euclidean hyperbolic superspace as a supercoset.

\subsection{AdS$_{2 \vert 4}$ and H$_{2 \vert 4}$ space as supercosets}
The group element $g$ describes the full OSp$(2\vert 2,\mathbb{R})$ supermanifold. 
If we consider the equivalence classes $g \sim g \cdot (\text{SO}(1,1) \otimes \text{U}(1))$ where the abelian subgroup $\text{SO}(1,1) \otimes \text{U}(1)$ is generated by $H$ and $Z$, we get the superanalogue of AdS space. It is represented as the right coset of the full isometry group by the isotropy subgroup. The above ``hyperbolic'' Euler angle decomposition \eqref{eq:EA} allows a quick implementation of this procedure by e.g. picking a representative of the equivalence class as:
\begin{align}
g_{\text{coset}} = e^{\varphi H} e^{\bar{\theta}^- \bar{F}^-} e^{\theta^- F^-}  e^{\eta(E^+ + E^-)} e^{\theta^+ F^+} e^{\bar{\theta}^+ \bar{F}^+}
\end{align}
by setting $\psi = 0 = \theta$. Notice that the one-parameter group $e^{\psi H}$ is non-compact, corresponding to a Lorentzian SO$(1,1)$ boost local Lorentz group. 
This means that what we are describing here is the coset in Lorentzian signature as will be clear below. This describes a $2 \vert 4$-dimensional supermanifold that is the right supercoset:
\begin{equation}
\text{AdS}_{2 \vert 4} \simeq \frac{\text{OSp}(2\vert 2,\mathbb{R})}{\text{SO}(1,1) \otimes \text{U}(1) }.
\end{equation}
We emphasize that this superspace is found by dropping the coordinates associated to the two generators $H$ and $Z$. This will be found as well in the main text in the physical context of the BF description of $\mathcal{N}=2$  JT supergravity.

From the particle-on-group Lagrangian, the coset condition is implemented by setting two of the components of the conserved currents to zero:
\begin{equation}
(g^{-1}dg)\vert_{H,Z} = 0.
\end{equation}
This implies in particular that the Cartan-Killing metric $ds^2_{\text{CK}} \equiv 2\text{STr}[(g^{-1}dg)^2]$ on the full group manifold has a reduced structure on the coset: 
\begin{equation}
\label{eq:metcos}
ds^2_{\text{coset}} = \left.2\text{STr}[(g^{-1}dg)^2]\right\vert_{\neq H,Z},
\end{equation}
where the notation means we simply do not add the contribution from $H$ and $Z$ in the computation. 
In order to concretely implement this, we note that the Euler angle description \eqref{eq:EA} has the following properties:
\begin{itemize}
\item
The component of the Cartan-Killing form $g^{-1}dg$ along $d\psi$ resp. $d\theta$ is only in the $H$ resp. $Z$ direction in the algebra.\footnote{This is readily seen by plugging in the product decomposition (\ref{eq:EA}) in the one-form $g^{-1}dg$.} 
\item 
Translations $\psi \to \psi + a$, $\theta \to \theta +b$, implemented by right multiplication of \eqref{eq:EA} by a constant matrix in the abelian subgroup $e^{a H}e^{2 i b Z}$, leaves the metric $ds^2_{\text{CK}}$ invariant. This means the metric components have no non-trivial $\psi$ and $\theta$ dependence.
\end{itemize}
These properties directly imply that the reduced CK metric on the coset space \eqref{eq:metcos} has no non-zero $d\psi$ or $d\theta$ components, nor any dependence on these coordinates.

For illustration, if one applies this procedure to the bosonic subgroup, the resulting coset metric is then:
\begin{equation}
ds^2_{\text{coset}} = d \eta^2 - \sinh \eta^2 d\varphi^2,
\end{equation}
which is the Lorentzian AdS$_{2}$ metric with radial coordinate $\eta \in \mathbb{R}^+$, and time-coordinate $\varphi$. In these coordinates, the AdS boundary is at $\eta \to +\infty$.

If one is instead interested in describing the Euclidean signature hyperbolic superspace $H_{2 \vert 4}$, one needs to instead consider the right coset by the \emph{compact} rotation subgroup SO(2) $\subset$ SL$(2,\mathbb{R})$:
\begin{equation}
\text{AdS}_{2 \vert 4} \simeq \frac{\text{OSp}(2\vert 2,\mathbb{R})}{\text{SO}(2) \otimes \text{U}(1) }.
\end{equation}
This is usefully described in the coordinatization \eqref{eq:ellEulangl} by modding out the $\psi$- and $\theta$-subgroups. Pragmatically, this corresponds to changing to the SU$(1,1)$ description, where one just analytically continues $\varphi \to i \varphi$ and $\psi \to i \psi$ from the hyperbolic description, and we can then keep on using the same ``hyperbolic'' Euler angle decomposition.

\subsection{Finite-dimensional representations}
\label{Section:finite_dim_rep}
Finite-dimensional representation were classified in e.g. \cite{Scheunert:1976wj,Gotz:2005jz}. We will not review the specifics. Our focus here instead is on the characters and branching rules in the maximal bosonic subgroup, since this is the part that will be used in the main text to describe end-of-the-world brane amplitudes in supergravity models. 

A typical finite-dimensional irreducible representation is labeled by two parameters $j$ and $q$, corresponding physically to the energy and the charge respectively. These are also the eigenvalues of the Cartan generators $H$ and $Z$ on the highest weight state in the representation. They are related to the Casimir eigenvalues \eqref{eq:Casimir1}, \eqref{eq:cubicCasimir} in the irrep as
\begin{equation}
\mathcal{C}_2 = j^2-q^2, \qquad \mathcal{C}_3 = q(j^2-q^2).
\end{equation} 
We can decompose these irreps into those of the bosonic subalgebra $\mathfrak{sl}(2,\mathbb{R}) \oplus \mathfrak{u}(1)$ as a direct sum:
\begin{equation}
\label{eq:branch2}
(j,q) \,\oplus \, (j-\frac{1}{2},q-\frac{1}{2}) \,\oplus \, (j-\frac{1}{2},q+\frac{1}{2})\,\oplus \, (j-1,q).
\end{equation}
The corresponding character is denoted as $\chi_{j,q}^{\mathcal{N}=2}(\phi,\theta)$. The character depends only on the conjugacy class of a group element. Group elements in the hyperbolic conjugacy class are labeled by two real parameters $\phi \in \mathbb{R}$ and $\theta \in [0,2\pi)$. SL$(2,\mathbb{R})$ and U(1) characters for the finite-dimensional representations are of the form
\begin{equation}
\chi_j^{\mathfrak{sl}(2,\mathbb{R}) }(\phi) = \text{Tr}_j e^{2\phi H} = \frac{\sinh (2j+1)\phi}{\sinh \phi}, \qquad \chi^{\mathfrak{u}(1)}_q(\theta) = e^{2iq \theta}.
\end{equation} 
The above decomposition \eqref{eq:branch2} leads to a mirrored character decomposition formula:\footnote{The minus signs between the terms in (\ref{eq:charactersupertrace}) are related to the fact that the fermionic states get a minus sign in the supertrace.}
\begin{align}
\label{eq:charactersupertrace}
&\chi_{j,q}^{\mathcal{N}=2}(\phi,\theta) \equiv \text{STr} \left[ e^{2\phi H}e^{2i\theta Z}\right] \nonumber \\
&= \chi_j^{\mathfrak{sl}(2,\mathbb{R}) }(\phi) \chi^{\mathfrak{u}(1)}_q(\theta) - \chi_{j-1/2}^{\mathfrak{sl}(2,\mathbb{R}) }(\phi) \chi^{\mathfrak{u}(1)}_{q-1/2}(\theta) - \chi_{j-1/2}^{\mathfrak{sl}(2,\mathbb{R}) }(\phi) \chi^{\mathfrak{u}(1)}_{q+1/2}(\theta) + \chi_{j-1}^{\mathfrak{sl}(2,\mathbb{R}) }(\phi) \chi^{\mathfrak{u}(1)}_{q}(\theta) \nonumber \\
&=  2\sinh \left(2j\phi\right)e^{2iq\theta} \frac{(\cosh \phi - \cos \theta)}{\sinh \phi}. 
\end{align}
A simple check is that as $\phi \to 0$ and $\theta \to \pi$, we find $\chi_{j,q}^{\mathcal{N}=2}(\phi,\theta) \to (-)^{2q}8j$, which is (up to a possible sign) the dimension of the representation.\footnote{We assume $q\in\mathbb{N}/2$ as would be appropriate for the compact group U(1).} Indeed, in this case, the insertion in the supertrace of the charge $Z$ leads to $(-)^{2Z} = \pm (-)^F$ ($F$ is the fermion number), which converts the minus sign for fermionic states back to a plus sign; hence adding all states and producing the dimension of the representation. 

The last factor in \eqref{eq:charactersupertrace} does not depend on the representation labels $(j,q)$ and can be identified as the Weyl denominator$^{-1/2}$ in analogy with an ordinary Lie algebra. Explicitly, the Weyl denominator $\Delta(\phi,\theta)$ can be computed for the component of the Lie (super)group connected to the identity, by using:\footnote{It was proven in \cite{Fan:2021wsb} that this is also precisely the measure appearing in Weyl's integration formula on supergroups, just like for ordinary Lie groups.}
\begin{equation}
\label{eq:wd}
\Delta(t) \equiv \frac{\prod_{\alpha_B} \vert e^{\alpha_B(t)}-1 \vert}{\prod_{\alpha_F} e^{\alpha_F(t)}-1}, \qquad t \in \mathfrak{h},
\end{equation}
in terms of the bosonic resp. fermionic roots $\alpha_B$ and $\alpha_F$, and where $t$ is in a Cartan subalgebra $\mathfrak{h}$. In our specific case, the maximal torus has two bosonic coordinates $t=(\phi,\theta)$, where we have 2 bosonic and 4 fermionic roots:
\begin{equation}
\alpha_B(t) = \pm 2\phi, \quad \alpha_F(t) = \pm \phi \pm i \theta.
\label{eq:roots}
\end{equation}
We hence obtain
\begin{align}
\Delta(\phi,\theta) &= \frac{(e^{2\phi}-1)(1-e^{-2\phi})}{(e^{\phi+i\theta}-1)(e^{\phi-i\theta}-1)(e^{-\phi+i\theta}-1)(e^{-\phi-i\theta}-1)} \label{eq:WeylDen} 
&= \frac{\sinh^2 \phi}{(\cosh \phi - \cos \theta)^2}.
\end{align}

\subsection{Highest and lowest weight representations}
Highest and lowest weight representations can be constructed as well. The towers of states decompose just as for the finite-dimensional representations into representations of the bosonic subgroup SL$(2,\mathbb{R})$ according to the same branching rule \eqref{eq:branch2} (since this is just an algebraic procedure).
We can use the (hyperbolic conjugacy class) highest weight irrep character for SL$(2,\mathbb{R})$:\footnote{We assume $\phi >0$ here. If not, an absolute value of $\phi$ should be used.}
\begin{equation}\label{eq:sl2Rcharderiv}
    \chi^{\mathfrak{sl}(2,\mathbb{R})}_j(\phi)=\text{Tr}_j(e^{2\phi H})
    =\frac{e^{(2j+1)\phi}}{2\sinh \phi}, \qquad j=-\frac{1}{2},-1,\hdots
\end{equation} 
The highest weight $\mathcal{N}=2$ character, evaluated in representation $(j,q)$ is then readily evaluated: 
\begin{align}\label{eq:N=2highestweightchar}
    \chi^{\mathfrak{osp}(2|2)}_{j,q}(\phi,\theta)&=\frac{e^{\phi(2j+1)}}{2\sinh\phi} e^{2iq\theta}+\frac{e^{\phi(2j-1)}}{2\sinh\phi} e^{2iq\theta}-\frac{e^{2j\phi}}{2\sinh\phi} e^{2i(q-1/2)\theta}-\frac{e^{2j\phi}}{2\sinh\phi} e^{2i(q+1/2)\theta}\nonumber\\ &=e^{2j\phi}e^{2iq\theta}\frac{\cosh \phi-\cos\theta}{\sinh\phi}.
\end{align}
The highest weight state in this module has $H=j$ and $Z=q$.
The Casimirs are given by the same expression as before:
\begin{equation}\label{eq:expressionsquadraticandcubiccasimirs}
\mathcal{C}_2 = j^2-q^2, \qquad \mathcal{C}_3 = q(j^2-q^2).
\end{equation}

\subsection{Principal Series Representations}
This section is devoted to the construction of the principal series representations of OSp$(2\vert 2,\mathbb{R})$. We take inspiration from the textbook case of SL$(2,\mathbb{R})$ \cite{Vilenkin}, and its generalization in the $\mathcal{N}=1$ case OSp$(1\vert 2,\mathbb{R})$ as worked out in \cite{Fan:2021wsb}.

The final aim of the construction is to find an action of the group on functions on the real superline $\mathbb{R}^{1\vert 2}$. So the carrier space of the representation will be $L^2(\mathbb{R}^{1\vert 2})$. The group itself acts on the coordinates of $\mathbb{R}^{1\vert 2}$ in terms of a super-M\"obius transformation. Let us first write down these super-M\"obius transformations. In order to find superconformal transformations for a bosonic variable $\tau$ and two fermionic variables $\vartheta_1, ~\vartheta_2$, we act with $g^{st^3}$ on the homogeneous superspace vector $(x,y \vert\vartheta_1,\vartheta_2)$, identified up to rescalings, where $x,y$ are bosonic variables and $\vartheta_1,\vartheta_2$ are fermionic variables.
We can equivalently work with a vector obtained after dividing all the entries of the previous one by $y$. After defining the new coordinates $\tau\equiv \frac{x}{y}$, $\vartheta_{1,2}\equiv \frac{\vartheta_{1,2}}{y}$, we get 
 \begin{equation}
g^{st^3} \left[\begin{array}{c}
\tau\\
1\\ 
\hline
\vartheta_1\\
\vartheta_2\\
 \end{array}\right]=\left[\begin{array}{c}
c+a\tau+\alpha_2\vartheta_1+\gamma_2\vartheta_2\\
d+b\tau+\beta_2\vartheta_1+\delta_2\vartheta_2\\ 
\hline
-\gamma_1+w\vartheta_1+z\vartheta_2-\alpha_1\tau\\
-\delta_1+y\vartheta_1+u\vartheta_2-\beta_1\tau\\
 \end{array}\right].
 \label{eq:gst_vector}
 \end{equation}
Normalizing the resulting vector again to have its second entry as $1$, we obtain the linear fractional transformations:
 \begin{align} 
 (\tau'\vert \vartheta_1',\vartheta_2') \hspace{-0.05cm} = \hspace{-0.1cm} \left(\frac{c+a\tau+\alpha_2\vartheta_1+\gamma_2\vartheta_2}{d+b\tau+\beta_2\vartheta_1+\delta_2\vartheta_2} \right\vert \left. \frac{-\gamma_1+w\vartheta_1+z\vartheta_2-\alpha_1\tau}{d+b\tau+\beta_2\vartheta_1+\delta_2\vartheta_2}, \frac{-\delta_1+y\vartheta_1+u\vartheta_2-\beta_1\tau}{d+b\tau+\beta_2\vartheta_1+\delta_2\vartheta_2}\right),
 \end{align}
 acting on the coordinates $(\tau\vert \vartheta_1,\vartheta_2)$ of $\mathbb{R}^{1\vert 2}$.

\subsubsection{Parabolic Induction}
We first work towards the definition of the principal series representation using the method of parabolic induction \cite{GelfandNaimark} (see e.g. the textbooks \cite{knapptrapa,Jeffrey:2000}). The following discussion will be somewhat technical, but it is necessary in order to construct the representation correctly. Our final result is reported in equation \eqref{eq:gact} below.

To set this up, we need to identify some important subgroups of OSp($2\vert 2,\mathbb{R})$. The upper- and lower ``triangular'' parabolic subgroups $N$ and $\bar{N}$ are: 
\begin{align}
N = \left[\begin{array}{c c | c c} 
1& \gamma & \theta^+_x & -\theta^+_y \\
0 & 1 & 0 & 0 \\
\hline 
0  & \theta^+_x & 1 & 0 \\
0 & -\theta^+_y & 0 & 1
\end{array}\right], \quad \bar{N} = \left[\begin{array}{c c | c c} 
1 & 0 & 0 & 0 \\
\beta & 1 & -\theta^-_x & \theta^-_y \\
\hline 
\theta^-_x  & 0 & 1 & 0 \\
-\theta^-_y & 0 & 0 & 1
\end{array}\right]. 
\end{align}
The subgroup $A$ is abelian and ``maximally noncompact''. It is of the form:
\begin{align}
A = \left[\begin{array}{c c | c c} 
a& 0 & 0 & 0 \\
0 & a^{-1} & 0 & 0 \\
\hline 
0  & 0 & 1 & 0 \\
0 & 0 & 0 & 1
\end{array}\right], \qquad a \in \mathbb{R}^+,
\end{align}
and the maximally compact subgroup $K$ is:
\begin{align}
K = \left[\begin{array}{c c | c c} 
\cos \eta & -\sin\eta & 0 & 0 \\
\sin \eta & \cos \eta & 0 & 0 \\
\hline 
0  & 0 & \cos\theta & -\sin\theta \\
0 & 0 & \sin\theta & \cos\theta
\end{array}\right], \qquad \eta,\theta \in [0,2\pi).
\end{align}
The subgroup $M = Z_K(A)\in K$ is the centralizer of $A$ in $K$, i.e. all matrices in $K$ that commute with all of $A$. This group has two disconnected components:\footnote{We restrict to the component of OSp($2\vert 2,\mathbb{R})$ connected to the identity, of relevance to gravity. Otherwise, we would have $M = \mathbb{Z}_2 \otimes \text{O}(2) \simeq \mathbb{Z}_2 \otimes \mathbb{Z}_2 \otimes \text{SO}(2)$.}
\begin{align}
M = \left[\begin{array}{c c | c c} 
\pm 1 & 0 & 0 & 0 \\
0 & \pm 1 & 0 & 0 \\
\hline 
0  & 0 & \cos\theta & -\sin\theta \\
0 & 0 & \sin\theta & \cos\theta
\end{array}\right] = \mathbb{Z}_2 \otimes \text{SO}(2).
\end{align}
The full group $G$ can be decomposed as $G = \bar{N}MAN$, which is essentially the Gauss-Euler decomposition \eqref{eq:GE}.

The idea of parabolic induction is to take a non-trivial representation of the subgroup $H = MAN \in G $ and consider functions in $L^2(G)$ satisfying the additional constraint 
\begin{equation}
\label{eq:induce}
f(gh) = D(h)^{-1}f(g), \quad h \in H,
\end{equation}
where $D(h)$ is a representation matrix of the subgroup $H$. If one picks the trivial representation of $N$, we have concretely:
\begin{equation}
f(gman) = D_{M}(m)^{-1}D_A(a)^{-1}f(g), \qquad m\in M,\quad a\in A, \quad n\in N.
\end{equation}
This restricted function space defines a representation of $g$ through the usual left action: $f(g_0) \to f(g^{-1}g_0)$, as the property \eqref{eq:induce} is preserved since the left action does not ``interfere'' with \eqref{eq:induce}. Because of \eqref{eq:induce} we can w.l.g. restrict to $g_0=\bar{n} \in \bar{N}$. The concrete problem is then to decompose $g^{-1}\bar{n} = \bar{n}'p = \bar{n}' m'a'n'$ in terms of some elements $\bar{n}'\in \bar{N},\;m'\in M,\;a'\in A$ and $n'\in N$, after which we recover
\begin{equation}\label{eq:generalparabolicrep}
f(\bar{n}) \to f(g^{-1}\bar{n})= f(\bar{n}' m'a'n') = D_M(m')^{-1}D_A(a')^{-1}f(\bar{n}'),
\end{equation}
giving a concrete formula for the representation. So we just need to choose representations of $A$ and $M$, and solve the technical problem 
\begin{equation}
\label{eq:tosolve}
g^{-1}\bar{n} = \bar{n}'p = \bar{n}' m'a'n',
\end{equation}
determining explicit expressions for $m',a',n'$ in terms of the group element $g$ and the initial coordinates $\bar{n}$.

We fix the representation of the maximally noncompact abelian group $A$ as $D_A(a)=e^{-2ik \ln a}=a^{-2ik}$, where $\ln a$ acts as the generator of the 1d irreps of $A$, and $k\in \mathbb{R}$ labels the representation. Due to the $i$ in the exponent, we will be inducing from a unitary representation of $A$. 

For the representation of the central element $m \in \mathbb{Z}_2\otimes \text{SO}(2)$, we take the character $\sigma(m)$ of $\mathbb{Z}_2$ as $\sigma(m)=(1,1)$ or $\sigma(m)=(1,-1)$ for $m=\mathbb{1}_{2\times 2}$ or $-\mathbb{1}_{2\times 2}$ respectively, where we denote the former as the trivial representation with $\epsilon=0$, and the latter as the non-trivial representation with $\epsilon=1$.
In addition, we define a charge $q$ SO(2) rotation matrix as 
\begin{equation}\label{eq:mrotationmatrix}
D_M(m^{-1}) \equiv M^{(q)}(\theta) = \left[\begin{array}{c c} 
\cos 2q \theta & -\sin 2q \theta \\
\sin 2q \theta & \cos 2q \theta
\end{array}\right],
\end{equation}
specifying the SO(2) representation.

Next, we explicitly solve \eqref{eq:tosolve}. Inserting the matrix inverse \eqref{eq:inv}, the first equality is in detail:
\begin{align} 
\label{eq:gminus1n}
&\left[\begin{array}{c c | c c} 
d & -b & -\beta_2 & -\delta_2\\ 
-c & a & \alpha_2 & \gamma_2 \\	
 \hline 
\gamma_1 & -\alpha_1 & w & z\\
\delta_1 & -\beta_1 & y & u
\end{array}\right]\left[\begin{array}{c c | c c} 
1 & 0 & 0 & 0 \\
-\tau & 1 & \vartheta_1 & \vartheta_2 \\
\hline 
-\vartheta_1  & 0 & 1 & 0 \\
-\vartheta_2 & 0 & 0 & 1
\end{array}\right] \\
&= \left[\begin{array}{c c | c c} 
1 & 0 & 0 & 0 \\
-\frac{a\tau+c+\alpha_2 \vartheta_1 + \gamma_2 \vartheta_2}{b\tau+d+\beta_2\vartheta_1 + \delta_2 \vartheta_2} & 1 & -\frac{\alpha_1 \tau + \gamma_1 -w \vartheta_1 - z\vartheta_2}{b\tau+d+\beta_2\vartheta_1 + \delta_2 \theta_2} & -\frac{\beta_1\tau + \delta_1 - y \vartheta_1 - u \vartheta_2}{b\tau+d+\beta_2\vartheta_1 + \delta_2 \vartheta_2} \\
\hline 
\frac{\alpha_1 \tau + \gamma_1 -w \vartheta_1 - z\vartheta_2}{b\tau+d+\beta_2\vartheta_1 + \delta_2 \vartheta_2}  & 0 & 1 & 0 \\
\frac{\beta_1\tau + \delta_1 - y \vartheta_1 - u \theta_2}{b\tau+d+\beta_2\vartheta_1 + \delta_2 \vartheta_2} & 0 & 0 & 1
\end{array}\right] \times \nonumber \\
&\left[\begin{array}{c c | c c} 
\hspace{-0.1cm} b\tau+d+\beta_2\vartheta_1 + \delta_2 \vartheta_2 & -b & \hspace{-0.1cm}-b\vartheta_1-\beta_2 & \hspace{-0.1cm} -b\vartheta_2-\delta_2 \\
0 & (b\tau+d+\beta_2\vartheta_1 + \delta_2 \vartheta_2)^{-1} & 0 & 0 \\
\hline 
0  & \hspace{-0.75cm} \frac{-b\vartheta_1-\beta_2}{b\tau+d+\beta_2\vartheta_1 + \delta_2 \vartheta_2} \cos\psi - \frac{b\vartheta_2+\delta_2}{b\tau+d+\beta_2\vartheta_1 + \delta_2 \vartheta_2} \sin \psi 
& \cos \psi & \sin \psi  \\
0 & \hspace{-0.75cm} -\frac{-b\vartheta_1-\beta_2}{b\tau+d+\beta_2\vartheta_1 + \delta_2 \vartheta_2}  \sin\psi - \frac{b\vartheta_2+\delta_2}{b\tau+d+\beta_2\vartheta_1 + \delta_2 \vartheta_2} \cos \psi & -\sin \psi & \cos \psi
\end{array}\right] \nonumber
\end{align}
where we have defined a supernumber angle $\psi$ by:
\begin{align}
\label{eq:cos_psi}
\cos \psi &= -\beta_1 \vartheta_2 + u + \frac{\beta_1 \tau + \delta_1 - y \vartheta_1 - u\vartheta_2}{b\tau+d+\beta_2\vartheta_1 + \delta_2 \vartheta_2} (b\vartheta_2 + \delta_2), \\
\label{eq:sin_psi}
\sin \psi &= -\alpha_1 \vartheta_2 + z + \frac{\alpha_1 \tau + \gamma_1 - w \vartheta_1 - z\vartheta_2}{b\tau+d+\beta_2\vartheta_1 + \delta_2 \vartheta_2} (b\vartheta_2 + \delta_2).
\end{align}
This last matrix can then be further decomposed as $p = m'a'n'$:
\begin{align}
&\left[\begin{array}{c c | c c} 
\hspace{-0.1cm}b\tau+d+\beta_2\vartheta_1 + \delta_2 \vartheta_2 & -b & \hspace{-0.1cm}-b\vartheta_1-\beta_2 & \hspace{-0.1cm}-b\vartheta_2-\delta_2 \\
0 & (b\tau+d+\beta_2\vartheta_1 + \delta_2 \vartheta_2)^{-1} & 0 & 0 \\
\hline 
0  & \hspace{-0.8cm}\frac{-b\vartheta_1-\beta_2}{b\tau+d+\beta_2\vartheta_1 + \delta_2 \vartheta_2} \cos\psi - \frac{b\vartheta_2+\delta_2}{b\tau+d+\beta_2\vartheta_1 + \delta_2 \vartheta_2} \sin \psi 
& \cos \psi & \sin \psi  \\
0 & \hspace{-0.8cm}-\frac{-b\vartheta_1-\beta_2}{b\tau+d+\beta_2\vartheta_1 + \delta_2 \vartheta_2}  \sin\psi - \frac{b\vartheta_2+\delta_2}{b\tau+d+\beta_2\vartheta_1 + \delta_2 \vartheta_2} \cos \psi & -\sin \psi & \cos \psi
\end{array}\right] \nonumber \\
&= \left[\begin{array}{c c | c c} 
\text{sgn}(b\tau+d+\beta_2\vartheta_1 + \delta_2 \vartheta_2) & 0 & 0 & 0 \\
0 & \text{sgn}(b\tau+d+\beta_2\vartheta_1 + \delta_2 \vartheta_2)^{-1} & 0 & 0 \\
\hline 
0  & 0 
& \cos \psi & \sin \psi  \\
0 & 0 & -\sin \psi & \cos \psi
\end{array}\right] \nonumber \\
&\times \left[\begin{array}{c c | c c} 
\vert b\tau+d+\beta_2\vartheta_1 + \delta_2 \vartheta_2 \vert & 0 & 0 & 0 \\
0 & \vert b\tau+d+\beta_2\vartheta_1 + \delta_2 \vartheta_2 \vert^{-1} & 0 & 0 \\
\hline 
0  & 0 
& 1 & 0  \\
0 & 0 & 0 & 1
\end{array}\right] \nonumber \\
&\times \left[\begin{array}{c c | c c} 
1 & \frac{-b}{b\tau+d+\beta_2\vartheta_1 + \delta_2 \vartheta_2} & \frac{-b\vartheta_1-\beta_2}{b\tau+d+\beta_2\vartheta_1 + \delta_2 \vartheta_2} & - \frac{b\vartheta_2+\delta_2}{b\tau+d+\beta_2\vartheta_1 + \delta_2 \vartheta_2} \\
0 & 1 & 0 & 0 \\
\hline 
0  & \frac{-b\vartheta_1-\beta_2}{b\tau+d+\beta_2\vartheta_1 + \delta_2 \vartheta_2} 
& 1 & 0  \\
0 & - \frac{b\vartheta_2+\delta_2}{b\tau+d+\beta_2\vartheta_1 + \delta_2 \vartheta_2}  & 0 & 1
\end{array}\right].
\label{eq:explicitdecomposition}
\end{align}

To induce a unitary representation, we need to include the correct ``half-density'' as follows. Using that the super-Jacobian in the decomposition of the Haar measure $dg$ into $d\Bar{n}$ and $d(man)$ is a signed Berezinian
(where the prime means we take the absolute value of the determinant of the top left bosonic subblock when computing the Berezinian), the precise super-Jacobian can be worked out as a product:
\begin{equation}
\label{prodjac}
\Delta(man) = \operatorname{sdet}'(\operatorname{Ad}_{\mathfrak{g}/\mathfrak{man}}(man)) = \operatorname{sdet}'(\operatorname{Ad}_{\mathfrak{g}/\mathfrak{man}}(m)) \operatorname{sdet}'(\operatorname{Ad}_{\mathfrak{g}/\mathfrak{man}}(a)) \operatorname{sdet}'(\operatorname{Ad}_{\mathfrak{g}/\mathfrak{man}}(n)),
\end{equation}
where generally
\begin{equation}
\operatorname{sdet}'(\operatorname{Ad}_{\mathfrak{g}/\mathfrak{man}}(a)) = a^{2 (\rho_B-\rho_F)}, \qquad \operatorname{sdet}'(\operatorname{Ad}_{\mathfrak{g}/\mathfrak{man}}(n)) = 1.
\end{equation}
Here, $\rho_B=\frac{1}{2}\sum_{i\in \Delta_B^+}\alpha_i$ and $\rho_R=\frac{1}{2}\sum_{i\in \Delta_F^+}\alpha_i$ are the Weyl vectors of (positive) bosonic and fermionic roots. In our case $\rho_B=\rho_F$, and only the first factor $\operatorname{sdet}'(\operatorname{Ad}_{\mathfrak{g}/\mathfrak{man}}(m))$ is the non-trivial one. The $1\vert 2$ dimensional super-vectorspace $\mathfrak{g}/\mathfrak{man}$ is spanned by the three generators $E^-,F^-,\bar{F}^-$. For the two components of $m_\pm \equiv \text{diag}(\pm \mathbb{1}_{2\times 2},\text{SO}(2))$, we have the adjoint action:
\begin{alignat}{3}
m_{+}^{-1} E^- m_{+} &= E^- ,\quad &&m_-^{-1} E^- m_- &&= E^-, \\
m_+^{-1} F^- m_+ &= e^{-i\theta} F^-, \quad &&m_-^{-1} F^- m_- &&= -e^{-i\theta}F^-, \\
m_+^{-1} \bar{F}^- m_+ &= e^{i\theta} \bar{F}^+, \quad &&m_-^{-1} \bar{F}^- m_- &&= -e^{i\theta} \bar{F}^-,
\end{alignat}
leading to\footnote{\label{fnO2} If we would consider the other component of $\text{O}(2)\simeq \mathbb{Z}_2 \times \text{SO}(2)$ not connected to the identity, we would find a minus sign here instead, leading to the overall factor $\text{sign}\left(\text{det}\left[\begin{array}{cc} w & y \\ z & u \end{array} \right]\right)$, leading to a factor $\text{sign}\left(\text{det}\left[\begin{array}{cc} w & y \\ z & u \end{array} \right]\right)^{\epsilon'-1/2}$ in the principal series definition where $\epsilon'=0,1$ is the additional irrep label of $\mathbb{Z}_2$. Restricting to SO(2), this factor is absent altogether.}
\begin{equation}
\operatorname{sdet}'(\operatorname{Ad}_{\mathfrak{g}/\mathfrak{man}}(m_\pm)) = 1.
\end{equation}
Thus, the modular function $\Delta(man)$ for the parabolic subgroup $P$ leads to a trivial Jacobian in the decomposition of the Haar measure into $d\Bar{n}$ and $d(man)$. The square root of the modular function $\Delta(man)^{1/2}$ is absorbed in the inner product in the transformation of $f$, which then automatically induced a unitary representation.

Identifying the coordinates of $f(\tau,\vartheta_1,\vartheta_2)$ with the entries of the lower triangular parabolic element $\Bar{n}$ according to \eqref{eq:gminus1n}, and the group elements $m',a',n'$ with the general decomposition \eqref{eq:explicitdecomposition}, the group action \eqref{eq:generalparabolicrep} finally leads to the definition of the principal series irreps of OSp$(2|2,\mathbb{R})$ ($i=1,2$):
\begin{align}
f_i(\tau,&\vartheta_1,\vartheta_2) \to\sum_{k=1}^{2} M_{il}^{(q)}(\psi) \, \text{sgn}(b\tau+d+\beta_2\vartheta_1 + \delta_2 \vartheta_2)^{\epsilon} \vert b\tau+d+\beta_2\vartheta_1 + \delta_2 \vartheta_2 \vert^{2j} \\
&\times f_l\left(\frac{a\tau+c+\alpha_2 \vartheta_1 + \gamma_2 \vartheta_2}{b\tau+d+\beta_2\vartheta_1 + \delta_2 \vartheta_2}, -\frac{\alpha_1 \tau + \gamma_1 -w \vartheta_1 - z\vartheta_2}{b\tau+d+\beta_2\vartheta_1 + \delta_2 \vartheta_2}, -\frac{\beta_1\tau + \delta_1 - y \vartheta_1 - u \vartheta_2}{b\tau+d+\beta_2\vartheta_1 + \delta_2 \vartheta_2}\right) ,\nonumber
\end{align}
in terms of the label $j=ik$ (coming from the chosen representation of $A$) and the labels ($\epsilon,q$) (coming from the chosen representation of $M$).
For a hyperbolic monodromy matrix $e^{2\phi H}e^{2i\theta Z}$, we have that $\cos \psi = u = \cos \theta$, $\sin \psi = z= \sin \theta$, and hence $\psi$ equals the $\theta$-variable of $g$ itself.

One can decouple the charge $+q$ and $-q$ sectors by complexifying the function space by transferring to $ f \equiv f_1+ i f_2, \, \bar{f} = f_1-if_2$, leading to: 
\begin{empheq}[box=\widefbox]{align}
\label{eq:gact}
f(\tau,&\vartheta_1,\vartheta_2) \to e^{2iq \psi} \, \text{sgn}(b\tau+d+\beta_2\vartheta_1 + \delta_2 \vartheta_2)^{\epsilon} \vert b\tau+d+\beta_2\vartheta_1 + \delta_2 \vartheta_2 \vert^{2j} \\
&\times f\left(\frac{a\tau+c+\alpha_2 \vartheta_1 + \gamma_2 \vartheta_2}{b\tau+d+\beta_2\vartheta_1 + \delta_2 \vartheta_2}, -\frac{\alpha_1 \tau + \gamma_1 -w \vartheta_1 - z\vartheta_2}{b\tau+d+\beta_2\vartheta_1 + \delta_2 \vartheta_2}, -\frac{\beta_1\tau + \delta_1 - y \vartheta_1 - u \vartheta_2}{b\tau+d+\beta_2\vartheta_1 + \delta_2 \vartheta_2}\right) \nonumber
\end{empheq}

\vspace{-1cm}
\noindent 
If one picks the trivial representation in $M$, the representation is called the \emph{spherical principal series representation}. In this case, it corresponds to zero charge $q=0$ and $\epsilon=0$.
In most of what follows, we will focus on the representations with $\epsilon=0$.

\subsubsection{Unitarity}
\label{sec:unitarity}
Finally, we want to check explicitly that we require $j=ik, \, k\in\mathbb{R}^+$ and $q\in\mathbb{R}$ in order to have induced a unitary representation.\footnote{The representation label $q$ will be discretized $q\in \mathbb{N}/2$ to correspond to the compact R-symmetry group U(1), but this is not forced on us by unitarity.} I.e. we want:
\begin{align}
\label{eq:uni}
\int d\tau d\vartheta_1 d\vartheta_2 F(\tau,\vartheta_1,\vartheta_2)^* (g \cdot G) (\tau,\vartheta_1,\vartheta_2) = \int d\tau d\vartheta_1 d\vartheta_2 (g^{-1} \cdot F)(\tau,\vartheta_1,\vartheta_2)^* G(\tau,\vartheta_1,\vartheta_2)
\end{align}
for any functions $F$ and $G$ in $L^2(\mathbb{R}^{1\vert 2})$, with the group action \eqref{eq:gact}. If true, this property shows that the inverse action is equal to the adjoint action and hence the representation would be unitary. The proof in the $\mathcal{N}=0$ case is elementary, and the brute force proof in the $\mathcal{N}=1$ case was given in appendix E of \cite{Fan:2021wsb}. The main ingredient of this calculation is the Jacobian/Berezinian of the coordinate transformation between $(\tau,\vartheta_1,\vartheta_2)$ and $(\tau',\vartheta_1',\vartheta_2')$, where we the inverse transformations are:  
\begin{align} 
 (\tau\vert \vartheta_1,\vartheta_2) \hspace{-0.05cm} = \hspace{-0.12cm} \left(\hspace{-0.12cm}\frac{d\tau'\hspace{-0.05cm}-c+\gamma_1\vartheta_1'+\delta \vartheta_2'}{-b\tau'\hspace{-0.05cm}+a-\alpha_1\vartheta_1'-\beta_1\vartheta_2'} \right\vert\hspace{-0.12cm} \left. \frac{\beta \tau'-\alpha_2+w\vartheta_1'+y\vartheta_2'}{-b\tau'+a-\alpha_1\vartheta_1'-\beta_1\vartheta_2'},\hspace{-0.05cm} \frac{\delta_2\tau'\hspace{-0.05cm}-\gamma_2+z\vartheta_1'+u\vartheta_2'}{-b\tau'+a-\alpha_1\vartheta_1'-\beta_1\vartheta_2'}\hspace{-0.05cm}\right)\hspace{-0.12cm}.
 \end{align}For computations, we utilize a trick by exploiting projective coordinates, which is easily generalized to the supersymmetric cases. We start with $\mathcal{N}=0$, and consider the following manipulations:
\begin{align}
\int d\tau f(\tau,1) &= \int dx dy f(x,y) \delta(y-1) \\
&= \int dx' dy' f(dx'-cy',ay'-bx') \delta(ay'-bx'-1) \\
&= \int d\tau' dy' f(d\tau' y'-cy',ay'-b\tau'y') \vert y'\vert \delta(ay'-b\tau'y'-1) \\
&= \int d\tau' f\left(\frac{d\tau'-c}{a-b\tau'},1\right) \frac{1}{\vert -b\tau'+a\vert^2}.
\end{align}
In the first equality we integrated in the coordinate $y$. In the second equality we performed the change of coordinates and used that the Jacobian is $\text{det} (g)$ which is $1$ for SL$(2,\mathbb{R})$. In the third line we defined the projective coordinate $\tau'\equiv x'/y'$. The factor in the last line is also directly identified as the Jacobian in going from $\tau$ to $\tau'$ directly.

If now we generalize this argument to $\mathcal{N}=1$ and $\mathcal{N}=2$, we simply need to add the integrals over the fermions and similarly renormalize $\vartheta'_{i,\text{new}}=\vartheta_i'/y'$, which produces a factor of $y'$ in the \emph{denominator} for fermionic measures $d\vartheta_i$. For $\mathcal{N}$ fermionic coordinates, this leads in the end to the Berezinian
\begin{equation}\label{eq:generalberezinian}
\frac{\text{Ber} (g)}{y^{2-\mathcal{N}}\vert_{y'=1}},
\end{equation}
where $\text{Ber} (g)$ is the Berezinian of the (super)group element acting as a (super)M\"obius transformation on the coordinates. In particular for OSp$(2 \vert 2,\mathbb{R})$ ($\mathcal{N}=2$), the Berezinian is $1$, leading to $j$ purely imaginary for unitarity.\footnote{Considering O(2) instead of SO(2), the Berezinian is $\text{sign}\left(\text{det}\left[\begin{array}{cc} w & y \\ z & u \end{array} \right]\right)$, which when combined with the factor of footnote \ref{fnO2}, can be shown to be consistent with unitarity.}
To complete the proof, we further need the following identity:
\begin{equation}
b\tau+d+\beta_2\vartheta_1 + \delta_2 \vartheta_2 = \frac{1}{-b\tau'+a-\alpha_1\vartheta_1'-\beta_1\vartheta_2'},
\end{equation}
and that the angles $\psi$ one distills from the action $g \cdot x$ versus $g^{-1} \cdot x'$ are related by a minus sign. This can be proven by comparing $g^{-1}\bar{n} = \bar{n}'p$ with $g \bar{n}' = \bar{n} p^{-1}$ where the matrix $p^{-1}$ has $\psi \to -\psi$ compared to $p$. This minus sign shows that \eqref{eq:uni} is consistent if the charge $q\in\mathbb{R}$.

 \subsubsection{Infinitesimal Level: Lie Superalgebra}
Next, we work out the infinitesimal level of the Lie supergroup and transfer to the associated Lie superalgebra. 
From (\ref{eq:gact}) we can deduce the group action for the one-parameter subgroups:
   \begin{align}
   \left(e^{2\phi H} \circ f \right)(\tau,\vartheta_1, \vartheta_2) &=e^{-2\phi j}f(e^{2\phi}\tau,e^{\phi}\vartheta_1,e^{\phi}\vartheta_2), \nonumber \\
\left(e^{\gamma E_+} \circ f \right)(\tau,\vartheta_1, \vartheta_2) & =|1+\gamma \tau|^{2j}\left(1-\frac{i\gamma\vartheta_1\vartheta_2}{1+\gamma \tau}\right)^{2q} f\left( \frac{\tau}{1+\gamma \tau},\frac{\vartheta_1}{1+\gamma \tau},\frac{\vartheta_2}{1+\gamma \tau}\right),\nonumber \\
   \left(e^{\beta E_-} \circ f \right)(\tau,\vartheta_1, \vartheta_2)&=f(\tau+\beta,\vartheta_1,\vartheta_2),\nonumber \\
   \left(e^{2i\theta Z} \circ f \right)(\tau,\vartheta_1, \vartheta_2) & = (\cos\theta+i\sin\theta)^{2q}f(\tau,\vartheta_1\cos \theta+\vartheta_2\sin \theta,\vartheta_2\cos\theta-\vartheta_1\sin\theta),\nonumber \\
   \left(e^{\sqrt{2}\theta_x^{+} F_{x}^{+}} \circ f \right)(\tau,\vartheta_1, \vartheta_2)&=(1\hspace{-0.05cm}-\hspace{-0.05cm}i\theta_x^{+}\vartheta_2)^{2q}|1\hspace{-0.05cm}+\hspace{-0.05cm}\theta_x^{+}\vartheta_1|^{2j}f\hspace{-0.05cm}\left(\hspace{-0.05cm}\frac{\tau}{1\hspace{-0.05cm}+\hspace{-0.05cm}\theta_x^{+}\vartheta_1},\frac{\vartheta_1\hspace{-0.05cm}-\hspace{-0.05cm}\tau\theta_x^{+}}{1\hspace{-0.05cm}+\hspace{-0.05cm}\theta_x^{+}\vartheta_1},\frac{\vartheta_2}{1\hspace{-0.05cm}+\hspace{-0.05cm}\theta_x^{+}\vartheta_1}\hspace{-0.05cm}\right),\nonumber \\
   \left(e^{\sqrt{2}\theta_x^{-} F_{x}^{-}} \circ f \right)(\tau,\vartheta_1, \vartheta_2)&=f(\tau+\theta_x^-\vartheta_1,\vartheta_1+\theta_x^-,\vartheta_2),\nonumber \\
    \left(e^{\sqrt{2}\theta_y^{+} F_{y}^{+}} \circ f \right)(\tau,\vartheta_1, \vartheta_2)  &=\left( 1\hspace{-0.05cm}-\hspace{-0.05cm}i\theta_y^{+}\vartheta_1\right)^{2q}\hspace{-0.05cm}|1\hspace{-0.05cm}-\hspace{-0.05cm}\theta_y^{+}\vartheta_2|^{2j}f\hspace{-0.05cm}\left(\hspace{-0.05cm}\frac{\tau}{1\hspace{-0.05cm}-\hspace{-0.05cm}\theta_y^{+}\vartheta_2},\frac{\vartheta_1}{1\hspace{-0.05cm}-\hspace{-0.05cm}\theta_y^{+}\vartheta_2},\frac{\vartheta_2\hspace{-0.05cm}+\hspace{-0.05cm}\tau\theta_y^{+}}{1\hspace{-0.05cm}-\hspace{-0.05cm}\theta_y^{+}\vartheta_2}\hspace{-0.05cm}\right),\nonumber \\
     \left(e^{\sqrt{2}\theta_y^{-} F_{y}^{-}}\circ f\right)(\tau,\vartheta_1, \vartheta_2)&=f(\tau-\theta_y^-\vartheta_2,\vartheta_1,\vartheta_2-\theta_y^-).
   \end{align}
At the infinitesimal level, these correspond to the superspace differential operators:
\begin{align}
\label{eq:infinitesimal}
2H&=-2j+2\tau\partial_\tau+\vartheta_2\partial_{\vartheta_2}+\vartheta_1\partial_{\vartheta_1},\\
    2iZ&=2iq-\vartheta_1\partial_{\vartheta_2}+\vartheta_2\partial_{\vartheta_1},\\
    E_+&=2j\tau-2iq\vartheta_1\vartheta_2-\tau\vartheta_2\partial_{\vartheta_2}-\tau\vartheta_1\partial_{\vartheta_1}-\tau^2\partial_\tau,\\
    E_-&=\partial_\tau,\\
    \sqrt{2}F_x^+&=2j \vartheta_1-2iq\vartheta_2-\vartheta_1\vartheta_2\partial_{\vartheta_2}-\tau\partial_{\vartheta_1}-\tau\vartheta_1\partial_{\tau},\\
    \sqrt{2}F_x^-&=\partial_{\vartheta_1}+\vartheta_1\partial_\tau,\\
    \sqrt{2}F_y^+&=-2j\vartheta_2 - 2iq\vartheta_1-\vartheta_1\vartheta_2\partial_{\vartheta_1}+\tau\partial_{\vartheta_2}+\tau\vartheta_2\partial_\tau,\\
    \label{eq:infinitesimalL}
    \sqrt{2}F_y^-&=-\partial_{\vartheta_2}-\vartheta_2\partial_\tau.
\end{align}
These generators satisfy the $\mathfrak{osp}(2|2)$ algebra, with the exception of the fermionic generators that differ for a sign factor in the anticommutation relations, e.g.:
\begin{equation}
\label{eq:fermions}
    \{F_x^{\pm},F_x^{\pm}\}=\{F_y^{\pm},F_y^{\pm}\}=\mp E^{\pm}.
\end{equation}
The infinitesimal group action leads to a representation of the \textit{opposite} Lie superalgebra.\footnote{This mirrors the analysis of \cite{Fan:2021wsb} in the context of $\mathcal{N}=1$ JT supergravity, where the infinitesimal $\mathfrak{osp}(1\vert 2)$ generators  also satisfy the opposite Lie superalgebra.}

In this language, we can appreciate the direct sum decomposition in irreducible $\mathfrak{sl}(2,\mathbb{R})\oplus \mathfrak{u}(1)$ representations according to \eqref{eq:branch2}. Indeed, working with $H$ on a purely bosonic function $f(\tau)$, the Cartan $H$ generator reduces to the standard spin-$j$ $H$ generator of $\mathfrak{sl}(2,\mathbb{R})$, with corresponding $\mathfrak{u}(1)$-charge of $q$ under $Z$. Acting on the doubly fermionic function $\vartheta_1\vartheta_2 f(\tau)$, $H$ reduces to the Cartan generator of $\mathfrak{sl}(2,\mathbb{R})$ with spin $(j-1)$ and $\mathfrak{u}(1)$ charge $q$. On the other hand, acting on fermionic functions $\vartheta_1 f(\tau)$ and $\vartheta_2 f(\tau)$, $H$ reduces to the spin $(j-1/2)$ $\mathfrak{sl}(2,\mathbb{R})$ $H$ generator. The degeneracy is lifted when working with the linear combinations $ (\vartheta_1-i\vartheta_2)f(\tau)$ and $ (\vartheta_1+i\vartheta_2)f(\tau)$, which form irreducible representations of $\mathfrak{sl}(2,\mathbb{R})$. The former has $\mathfrak{u}(1)$ charge $q+1/2$, while the latter has $q-1/2$. The irreducible representations of $\mathfrak{sl}(2,\mathbb{R})$ are however not unitary, since that would require $j=-1/2+ik$.\\

The quadratic and the cubic Casimirs, $\mathcal{C}_2$ and $\mathcal{C}_3$, commute with all the generators, and in an irreducible representation, are proportional to the identity matrix. We can compute $\mathcal{C}_2$ and $\mathcal{C}_3$ explicitly in the principal series representation using the differential operators in (\ref{eq:infinitesimal})-\eqref{eq:infinitesimalL}. In this case, however, the fermionic generators satisfy opposite anti-commutation relations (\ref{eq:fermions}). As a consequence, equations (\ref{eq:Casimir1})-(\ref{eq:cubicCasimir}) modify into
\begin{align}
\label{eq:Casimir2}
     \mathcal{C}_2 &=H^2-Z^2+E^-E^+-(F^-\bar{F}^+-\bar{F}^-F^+), \\
    \label{eq:cubicCasimir2}
    \mathcal{C}_3 &=(H^2-Z^{2})Z+E^-E^+(Z-\tfrac{1}{2})+\tfrac{1}{2}F^-\bar{F}^+(H-3Z+1)\\
    &\hspace{1cm}+\tfrac{1}{2}\bar{F}^-F^+(H+3Z+1)-\tfrac{1}{2}E^-\bar{F}^+F^+-\tfrac{1}{2}\bar{F}^-F^-E^+,\nonumber
\end{align}
where the doubly fermionic terms $\sim FF$ have swapped signs compared to the earlier expressions.
In order to compute the terms $F^-\bar{F}^+$ and $\bar{F}^-F^+$ appearing in (\ref{eq:Casimir2})-(\ref{eq:cubicCasimir2}), it is convenient to introduce the complex fermionic variables $\vartheta$ and $\bar{\vartheta}$: 
\begin{equation}
    \vartheta\equiv \vartheta_1-i\vartheta_2,\qquad \bar{\vartheta}\equiv\vartheta_1+i\vartheta_2, \qquad\partial_{\vartheta}\equiv\frac{1}{2}(\partial_{\vartheta_1}+i\partial_{\vartheta_2}),\qquad
\partial_{\bar{\vartheta}}\equiv\frac{1}{2}(\partial_{\vartheta_1}-i\partial_{\vartheta_2}).
\end{equation}
The fermionic generators \eqref{eq:generatorstransf} can then be written as:
\begin{alignat}{2}
&F^+=j\vartheta+q\vartheta-\frac{1}{2}\vartheta\bar{\vartheta}\partial_{\bar{\vartheta}}-\tau\partial_{\bar{\vartheta}}-\frac{\tau}{2}\vartheta\partial_\tau, \qquad &&F^-=\partial_{\bar{\vartheta}}+\frac{1}{2}\vartheta\partial_\tau,\\
&\bar{F}^+=j\bar{\vartheta}-q\bar{\vartheta}+\frac{1}{2}\vartheta\bar{\vartheta}\partial_{\vartheta}-\tau\partial_{\vartheta}-\frac{\tau}{2}\bar{\vartheta}\partial_\tau, \qquad 
    &&\bar{F}^-=-\partial_{\vartheta}-\frac{1}{2}\bar{\vartheta}\partial_\tau.
\end{alignat}
Plugging the expressions (\ref{eq:infinitesimal})-\eqref{eq:infinitesimalL} into \eqref{eq:Casimir2}-(\ref{eq:cubicCasimir2}), we have (painstakingly) checked that the final expressions for the quadratic Casimir $\mathcal{C}_2$ and the cubic Casimir $\mathcal{C}_3$ are
\begin{align}
\label{eq:quadcas}
  \mathcal{C}_2 &=j^2-q^2 = - k^2 - q^2, \\
    \mathcal{C}_3 &=q(j^2-q^2) =  -q(k^2 + q^2).
\end{align}
Both $\mathcal{C}_2$ and $\mathcal{C}_3$ are proportional to the identity operator; and since they form a basis for the center of the universal enveloping algebra, this essentially proves that the constructed representation is indeed irreducible. Note that $\mathcal{C}_2$ is strictly negative, and the sign of $\mathcal{C}_3$ is 1:1 with the sign of the quantum number $q$ for these representations.

\subsubsection{Discrete representations: monomial realization}
As in the SL$(2,\mathbb{R})$ and OSp$(1\vert 2,\mathbb{R})$ cases, it is possible and illuminating to realize both the finite-dimensional and the discrete highest and lowest weight irreps in a monomial basis on the same superline $\mathbb{R}^{1 \vert 2}$, acted on by the differential generators \eqref{eq:infinitesimal}-\eqref{eq:infinitesimalL}. 

A lowest weight state is annihilated by both $F_x^-$ and $F_y^-$ (and hence automatically by $E^-$). The solution is just a constant: 
\begin{equation}
\psi_{\text{LW},j,q}(\tau,\vartheta_1,\vartheta_2) = 1, \qquad H=-j, \quad Z = q.
\end{equation}
Similarly, a highest weight state is annihilated by $F_x^+$ and $F_y^+$:
\begin{equation}
\psi_{\text{HW},j,q}(\tau,\vartheta_1,\vartheta_2) = \tau^{2j}-2iq\tau^{2j-1}\vartheta_1\vartheta_2 = \tau^{2j}e^{-2i q \frac{\vartheta_1\vartheta_2}{\tau}}, \qquad H=j, \quad Z = q.
\end{equation}
If $2j \in \mathbb{N}$, the representation contains both lowest and highest weight states, and is hence finite-dimensional. The basis states are
\begin{equation}
\{1,\vartheta_1,\vartheta_2,\vartheta_1\vartheta_2,\tau,\hdots ,\vartheta_1\vartheta_2 \tau^{2j-2},\vartheta_1 \tau^{2j-1},\vartheta_2 \tau^{2j-1},\tau^{2j-1},\tau^{2j}e^{-2i q \frac{\vartheta_1\vartheta_2}{\tau}}\}.
\end{equation}
The states proportional to $\sim \vartheta_1+i\vartheta_2$ and $\sim \vartheta_1-i\vartheta_2$ directly correspond to the second and third irrep in the branching rule decomposition \eqref{eq:branch2}, whereas the others are linear combinations of the  Grassmann algebra basis elements $\sim 1, \,\vartheta_1\vartheta_2$.

If $2j \in - \mathbb{N}$,\footnote{Just as in the simpler cases of SL$(2,\mathbb{R})$ and OSp$(1\vert 2,\mathbb{R})$, this restriction to half integers is not visible at the level of our current construction, which only probes the universal cover of OSp$(2\vert 2,\mathbb{R})$.} the representation is unbounded either from above or from below, and is the discrete lowest (resp. highest) weight irrep. The above basis simply continues unboundedly on either side. 
In particular, we observe the same branching rule \eqref{eq:branch2} at work here, as mentioned earlier.

\subsubsection{Principal series character}
\label{principal_series_char}
Of particular interest in (super)gravity amplitudes are the characters of the different representations. In order to obtain the character of the principal series representations, we first use the same trick as used in subsection \ref{Section:finite_dim_rep}, but formally applied to the principal series representation by analytic continuation of the $j$-label. We have already observed that this analytic continuation works for $\mathcal{N}=1$ in \cite{Fan:2021wsb}, so we anticipate a similar outcome here. Afterwards, we will explicitly derive the character by brute force and show that the results indeed match.

We consider the principal series representation character of SL$(2,\mathbb{R})$, parametrized in terms of $j$:
\begin{equation}
\chi_j^{\mathfrak{sl}(2,\mathbb{R}) }(\phi) = \frac{\cosh (2j+1)\phi}{\sinh \phi},
\end{equation}
and insert it in (\ref{eq:branch2}). We then obtain:
\begin{align}
\label{eq:charps}
&\chi_{j,b}^{\mathcal{N}=2}(\phi,\theta) \nonumber\\
&= \frac{\cosh (2j+1)\phi}{\sinh \phi}e^{2iq \theta} - \frac{\cosh (2j)\phi}{\sinh \phi}e^{2i(q-\frac{1}{2}) \theta} - \frac{\cosh (2j)\phi}{\sinh \phi}e^{2i(q+\frac{1}{2}) \theta} + \frac{\cosh (2j-1)\phi}{\sinh \phi}e^{2i q \theta} \nonumber \\
&= 2\cosh \left(2j\phi\right)e^{2iq\theta}\frac{(\cosh \phi - \cos \theta)}{\sinh \phi} =  \frac{2\cosh \left(2j\phi\right)e^{2iq\theta}}{\sqrt{\Delta(\phi,\theta)}},
\end{align}
where we recognize the Weyl denominator \eqref{eq:WeylDen} in the last equality. Finally setting $j=ik, \, k\in\mathbb{R}^+$, we obtain a candidate expression for the principal series character of OSp$(2\vert 2,\mathbb{R})$:
\begin{equation}
\label{eq:n2char}
\boxed{
\chi_{k,q}^{\mathcal{N}=2}(\phi,\theta) = 2\cos \left(2k\phi\right)e^{2iq\theta}\frac{\cosh \phi - \cos \theta}{\sinh \phi}}.
\end{equation}

We next reproduce and prove the result  (\ref{eq:n2char}) starting from the Borel-Weil realization of the $\mathfrak{osp}(2\vert 2)$ algebra in terms of a kernel $K(\tau,\vartheta_1,\vartheta_2|\tau',\vartheta_1',\vartheta_2')$, following Appendix E of \cite{Fan:2021wsb}, where the same computation was done for the group OSp$(1\vert 2,\mathbb{R})$. We work in a coordinate basis on the carrier space of square integrable functions on the superline $L^2(\mathbb{R}^{1|2})$, i.e. the real line ``thickened'' in the fermionic directions $\vartheta_1$ and $\vartheta_2$. The principal series representation \eqref{eq:gact} can be written equivalently as
\begin{equation}
f'(\tau,\vartheta_1,\vartheta_2)=\int d\tau' d\vartheta_1' d\vartheta_2' K(\tau,\vartheta_1,\vartheta_2|\tau',\vartheta_1',\vartheta_2') f(\tau',\vartheta_1',\vartheta_2'),
    \label{eq:BorelWeil}
\end{equation}
where
 \begin{align}
K(\tau,&\vartheta_1,\vartheta_2|\tau',\vartheta_1',\vartheta_2') =  e^{2iq \psi} \, 
\vert b\tau+d+\beta_2\vartheta_1 + \delta_2 \vartheta_2 \vert^{2j}\delta\left(\frac{c+a\tau+\alpha_2\vartheta_1+\gamma_2\vartheta_2}{d+b\tau+\beta_2\vartheta_1+\delta_2\vartheta_2}-\tau'\right) \nonumber \\ 
&\times\delta\left(\frac{-\gamma_1+w\vartheta_1+z\vartheta_2-\alpha_1\tau}{d+b\tau+\beta_2\vartheta_1+\delta_2\vartheta_2}-\vartheta_1'\right)\delta\left(\frac{-\delta_1+y\vartheta_1+u\vartheta_2-\beta_1\tau}{d+b\tau+\beta_2\vartheta_1+\delta_2\vartheta_2}-\vartheta_2'\right),
    \label{eq:Kernel_OSP2}
\end{align}
and $e^{2iq\psi}$ can be computed starting from (\ref{eq:cos_psi}), (\ref{eq:sin_psi}).

The character $\chi_{j,q}(g)$ in representation $(j,q)$ is then determined by summing up the contribution in the vector space $L^2(\mathbb{R}^{1|2})$ of functions that get mapped to themselves:
\begin{align}
\label{eq:kernelOSP}
    \chi_{j,q}^{\mathcal{N}=2}(\phi,\theta)&\equiv  \int d\tau d\vartheta_1 d\vartheta_2 \,K(\tau,\vartheta_1,\vartheta_2|\tau,\vartheta_1,\vartheta_2)\\
    &= \int d\tau d\vartheta_1 d\vartheta_2 e^{2iq \psi} \vert b\tau+d+\beta_2\vartheta_1 + \delta_2 \vartheta_2 \vert^{2j}\delta\left(\frac{c+a\tau+\alpha_2\vartheta_1+\gamma_2\vartheta_2}{d+b\tau+\beta_2\vartheta_1+\delta_2\vartheta_2}-\tau\right) \nonumber \\
    &{}\hspace{0.75cm}\times\delta\left(\frac{-\gamma_1+w\vartheta_1+z\vartheta_2-\alpha_1\tau}{d+b\tau+\beta_2\vartheta_1+\delta_2\vartheta_2}-\vartheta_1\right)\delta\left(\frac{-\delta_1+y\vartheta_1+u\vartheta_2-\beta_1\tau}{d+b\tau+\beta_2\vartheta_1+\delta_2\vartheta_2}-\vartheta_2\right). \nonumber
\end{align}
Since the character is a class function, i.e. only depends on the conjugacy class of the group element, we can further simplify the calculation by considering a representative group element. For the hyperbolic conjugacy class, we set:
\begin{equation}
g=\left[\begin{array}{c c | c c} 
	e^{\phi} & \epsilon & 0 & 0\\ 
    0 & e^{-\phi} & 0 & 0\\	
 \hline 
	0 & 0 & \cos\theta & -\sin\theta\\
    0 & 0 & \sin\theta & \cos\theta\\
\end{array}\right],
\label{eq:OSPhyp}
\end{equation}
where (\ref{eq:OSPhyp}) is a group element in the maximal bosonic subgroup $\text{SL}(2,\mathbb{R})\otimes \text{SO}(2)$, labeled by $(\phi,\theta)$ respectively. The parameter $\epsilon$ can be viewed as a small regulator in the computation of the ``solution at infinity''.

The bosonic and fermionic delta-functions can be worked out explicitly. For the bosonic one, we have:
\begin{equation}
\label{eq:bosdel}
\delta\left(\frac{e^{\phi}\tau}{e^{-\phi}+\epsilon\tau}-\tau\right)=\frac{\delta(\tau)}{e^{2\phi}-1}+\frac{\delta\left(\tau-\frac{e^{\phi}-e^{-\phi}}{\epsilon}\right)}{1-e^{-2\phi}}.
\end{equation}
While the fermionic delta functions are by definition just the arguments of those functions:
\begin{equation}
\delta\left(\frac{\vartheta_1 \cos\theta+\vartheta_2\sin\theta}{e^{-\phi}+\epsilon\tau}-\vartheta_1\right)=\frac{\vartheta_1 \cos\theta+\vartheta_2\sin\theta}{e^{-\phi}+\epsilon\tau}-\vartheta_1,
\end{equation}
\begin{equation}
\delta\left(\frac{\vartheta_2\cos\theta-\vartheta_1\sin\theta}{e^{-\phi}+\epsilon\tau}-\vartheta_2\right)=\frac{\vartheta_2\cos\theta-\vartheta_1\sin\theta}{e^{-\phi}+\epsilon\tau}-\vartheta_2.
\end{equation}
The integral in (\ref{eq:kernelOSP}) then precisely boils down to the principal series character (\ref{eq:n2char}).\footnote{Note that if one considers an elliptic conjugacy class element instead where $g \sim \text{diag}(\text{SO}(2), \text{SO}(2))$, the bosonic delta-function (\ref{eq:bosdel}) yields zero, just like for $\mathcal{N}=0,1$ in earlier work. This means, just as in those cases, that elliptic defects in gravity are defined by analytically continuing the characters from the hyperbolic scenario.} 

In gravity amplitudes,  these characters are to be inserted in the JT disk amplitude to transfer to the single-trumpet amplitude.
It is moreover manifestly true that these characters satisfy orthonormality relations:
\begin{equation}
\int_{0}^{+\infty} d\phi \int_0^{2\pi} d\theta\, \Delta(\phi,\theta) \,\chi_{k,q}^{\mathcal{N}=2}(\phi,\theta) \chi_{k',q'}^{*\mathcal{N}=2}(\phi,\theta) = (2\pi)^2\delta(k-k') \delta_{qq'},
\end{equation}
where the Weyl denominator $\Delta(\phi,\theta)$ appears here as the natural measure of the space of conjugacy class elements. This orthonormality relation is required when gluing super-geometries together using the gluing procedure from $\mathcal{N}=2$ super-Teichm\"uller space.

Stripping the Weyl denominator of \eqref{eq:n2char}, the character can be identified with a certain limit of the $\mathcal{N}=2$ super-Virasoro modular S-matrix for non-degenerate characters, see e.g. eq (3.16)-(3.18) in \cite{Ahn:2003tt}:
\begin{equation}
\chi_{k,q}^{\mathcal{N}=2}(\phi,\theta) \sim  2\cos \left(2k\phi\right)e^{2iq\theta} \quad \sim \quad \lim_{c \to \infty} S_{\frac{\phi}{2\pi},\frac{2\theta}{\pi b^2}}{}^{k,-q}.
\end{equation}
The $c \to \infty$ limit is what we denoted as the Schwarzian limit in \cite {Mertens:2017mtv}. Here it does not change the functional form of this expression, as is familiar from the simpler cases of $\mathcal{N}=0,1$ as well.

\subsection{Alternative SU($1,1 \vert 1$) perspective on $\mathcal{N}=2$ character}
In this appendix, we consider a different enlightening perspective on the computation of the $\mathcal{N}=2$ character and exploit the 2:1 homomorphism between the real supergroup OSp$(2|2,\mathbb{R})$ and the complex supergroup SU$(1,1|1)$ \cite{Fu:2016vas}:
\begin{equation}
    \text{OSp}(2|2)/\mathbb{Z}_2\simeq \text{SU}(1,1|1).
\end{equation}
We will eventually show that the principal series character computed starting from a group element $g\in$ SU$(1,1\vert 1)$ is the same as the one computed in Appendix \ref{principal_series_char} for (the component connected to the identity of) OSp$(2\vert 2,\mathbb{R})$.

SU$(1,1\vert 1)$ is a group of $3\times 3$ complex matrices with 5 bosonic and 4 fermionic variables
\begin{equation}
    g=\left[\begin{array}{c c | c } 
	a & b & \alpha \\ 
    c & d & \gamma \\	
 \hline 
	\beta & \delta & e\\
    
\end{array}\right],
\end{equation}
preserving the orthosymplectic form $\Omega = \text{diag}\left(\left[\begin{array}{cc}
0 & 1 \\
-1 & 0 \\ 
 \end{array} \right], 2\right)$ as $g\Omega g^{{\dagger}^3}=\Omega$.\footnote{The $\dagger$ operation consists of a super-transposition and complex conjugation of all the entries of the matrix.} 
 The maximal bosonic subgroup of SU$(1,1\vert 1)$ is
 \begin{equation}
  \text{SU}(1,1)\otimes \text{U}(1) \,\subset\, \text{SU}(1,1|1),    
 \end{equation}
with  SU$(1,1)$ being the group of $2\times 2$ complex matrices with unit determinant, 
 \begin{equation}
     h=\left(\begin{array}{cc}
a & b \\
\bar{b} & \bar{a} \\
 \end{array}\right), \qquad \vert a\vert^2 - \vert b \vert^2 = 1.
 \end{equation}
A well-known isomorphism relates SU$(1,1)$ and SL$(2,\mathbb{R})$ matrices \cite{Vilenkin}:
\begin{equation}
    g\in \text{SL}(2,\mathbb{R}) \,\, \longleftrightarrow \,\, h \in \text{SU}(1,1),\;\;h=t^{-1}gt,\;\;
    \label{eq:isomorphism}
\end{equation}
where $t=\frac{1}{\sqrt{2}}\left(\begin{array}{cc}
1 & i \\
i & 1 \\ 
 \end{array}\right)$. 
  
The linear fractional transformation that realizes SU$(1,1)$ maps a complex coordinate $z$ into:
 \begin{equation}
       z\mapsto\frac{az+b}{\bar{b}z+\bar{a}}.
       \label{eq:SU11}
 \end{equation}
Whereas the SL$(2,\mathbb{R})$ transformation maps the upper halfplane into itself (preserving the real axis), (\ref{eq:SU11}) maps the unit disk into itself, preserving the unit circle.  The isomorphism \eqref{eq:isomorphism} between  SL$(2,\mathbb{R})$ and SU$(1,1)$ is connected to the \textit{Cayley transform} in the form:
 \begin{equation}
 \left(\begin{array}{c} \tau \\ 1 \end{array}\right) \mapsto t^{-1}\cdot \left(\begin{array}{c} \tau \\ 1 \end{array}\right): \qquad 
          \tau\mapsto z \equiv i\, \frac{\tau-i}{\tau+i}.
     \label{eq:Cayley}
 \end{equation}
(\ref{eq:Cayley}) maps the upper halfplane (with coordinate $\tau$) into the unit circle with coordinate $z$.

In order to find the superconformal transformations for the complex bosonic and fermionic variables $\tau$ and $\vartheta$, we first act with $g\in \text{SU}(1,1\vert 1)$ on the complex vector $(z_+ \vert \vartheta)$. This results in the linear fractional transformations
  \begin{align}
      z_+'=\frac{a z_+ +\alpha\vartheta+b}{c z_+ +\gamma\vartheta+d}~, \qquad \vartheta'=\frac{\beta z_+ +e\vartheta+\delta}{c z_+ +\gamma\vartheta+d},
      \label{eq:transftautheta}
  \end{align}
and their complex conjugates (where we denote $z_-$ as the complex conjugate of $z_+$):
\begin{align}
           z_-'=\frac{\bar{a}z_- +\bar{\alpha}\bar{\vartheta}+\bar{b}}{\bar{c}z_-+\bar{\gamma}\bar{\vartheta}+\bar{d}}~,\qquad  \bar{\vartheta}'=\frac{\bar{\beta}z_- +\bar{e}\bar{\vartheta}+\bar{\delta}}{\bar{c}z_-+\bar{\gamma}\bar{\vartheta}+\bar{d}}.
           \label{eq:transftaumintheta}
  \end{align}
  These are the same transformations as those in Appendix A of \cite{Fu:2016vas}.

We can equivalently define a new complex bosonic variable $z$ through $z_+ \equiv z+\vartheta\bar{\vartheta}$. Its complex conjugate is immediately found to be $z_-=1/z-\vartheta\bar{\vartheta}$.\footnote{Using our convention that complex conjugation of Grassmann variables preserves the order.}
 The transformation for $z$ is a U$(1)$ phase transformation on the unit circle, $\displaystyle \bar{z}=\frac{1}{z}$.
Then the transformations for $z$, $\vartheta$, $\bar{\vartheta}$ are given by:
\begin{equation}
\begin{split}
&z'=\frac{b+{\alpha\vartheta}+a(z+\vartheta\bar{\vartheta})+\frac{(\bar{\delta}+\bar{e}\bar{\vartheta}+\bar{\beta}(1/z-\vartheta\bar{\vartheta}))(\delta+e\vartheta+\beta(z+\vartheta\bar{\vartheta}))}{\bar{d}+\bar{\gamma}\bar{\vartheta}+\bar{c}(1/z-\vartheta\bar{\vartheta})}}{d+\gamma\vartheta+c(z+\vartheta\bar{\vartheta})},\\
& \vartheta'=\frac{\delta+e\vartheta+\beta(z+\vartheta\bar{\vartheta})}{d+\gamma\vartheta+c(z+\vartheta\bar{\vartheta})},\qquad \bar{\vartheta}'=\frac{\bar{\beta}+z(\bar{\delta}+\bar{e}\bar{\vartheta}-\bar{\beta}\vartheta\bar{\vartheta})}{\bar{c}+z(\bar{d}+\bar{\gamma}\bar{\vartheta}-\bar{c}\vartheta\bar{\vartheta})}.
    \label{eq:superconformalSU11}
\end{split}
\end{equation}
The kernel expression in this case is given by:
\begin{align}
\label{eq:KernelSU111}
K(z,&\vartheta,\bar{\vartheta}|z',\vartheta',\bar{\vartheta}') =  e^{2iq \psi} \, 
\vert d+\gamma\vartheta+c(z+\vartheta\bar{\vartheta}) \vert^{2j}\nonumber\\&\times\delta\left(\frac{b+\alpha\vartheta+a(z+\vartheta\bar{\vartheta})+\frac{(\bar{\delta}+\bar{e}\bar{\vartheta}+\bar{\beta}(1/z-\vartheta\bar{\vartheta}))(\delta+e\vartheta+\beta(z+\vartheta\bar{\vartheta}))}{\bar{d}+\bar{\gamma}\bar{\vartheta}+\bar{c}(1/z-\vartheta\bar{\vartheta})}}{d+\gamma\vartheta+c(z+\vartheta\bar{\vartheta})}-z'\right) \nonumber \\ 
&\times\delta\left(\frac{\delta+e\vartheta+\beta(z+\vartheta\bar{\vartheta})}{d+\gamma\vartheta+c(z+\vartheta\bar{\vartheta})}-\vartheta'\right)\delta\left(\frac{\bar{\beta}+z(\bar{\delta}+\bar{e}\bar{\vartheta}-\bar{\beta}\vartheta\bar{\vartheta})}{\bar{c}+z(\bar{d}+\bar{\gamma}\bar{\vartheta}-\bar{c}\vartheta\bar{\vartheta})}-\bar{\vartheta}'\right).
\end{align} 
Using the isomorphism (\ref{eq:isomorphism}), a hyperbolic conjugacy class element in SU$(1,1)$ is of the form:
\begin{equation}
    m=\left[\begin{array}{c c } 
	e^{\phi} & 0 \\ 
    0 & e^{-\phi}\\	
\end{array}\right],\;m \in \text{SL}(2,\mathbb{R})\longmapsto \; h=\left[\begin{array}{c c } 
	\cosh \phi & i\sinh\phi \\ 
    -i\sinh\phi & \cosh \phi\\	
\end{array}\right],\;\;h\in \text{SU}(1,1).
\end{equation}
Therefore a generic group element $g\in \text{SU}(1,1|1)$ in the hyperbolic conjugacy class is of the form\footnote{Note that the off-diagonal $\epsilon$ factor that was present in (\ref{eq:OSPhyp}) does not appear in $m$ $\in$ SL$(2,\mathbb{R})$. In that case it was needed in order to have a quadratic equation with two solutions in the bosonic delta function.}
\begin{equation}
    g=\left[\begin{array}{c c | c } 
	\cosh \phi & i\sinh\phi & 0\\ 
    -i\sinh\phi & \cosh \phi & 0\\
    \hline
    0 & 0 & e^{i\theta} \\
\end{array}\right],
\label{SUhyp}
\end{equation}
where the bottom-right block is just a U$(1)$ phase factor. 

Using that the fermionic delta functions in \eqref{eq:KernelSU111} evaluate to their arguments and integrating over the fermionic variables $\vartheta,\bar{\vartheta}$, the character expression is given by:

\begin{align}
    \chi_{j,q}^{\mathcal{N}=2}(\phi,\theta)=&\int dz \;e^{2iq\theta}|\cosh\phi-iz\sinh\phi|^{2j}\left( \frac{e^{i\theta}}{\cosh\phi-iz\sinh\phi}-1\right) \nonumber\\&\qquad\times\left(\frac{ze^{-i\theta}}{z\cosh\phi+i\sinh\phi}-1\right)\delta\left(\frac{z\cosh\phi+i\sinh\phi}{\cosh\phi-iz\sinh\phi}-z\right).
    \label{eq:charSU11}
    \end{align}
The bosonic delta can be decomposed as:
\begin{equation}
  \delta \left(\frac{z\cosh\phi+i\sinh\phi}{\cosh\phi-iz\sinh\phi}-z\right)=\frac{\delta(z-i)}{1-e^{-2\phi}}+\frac{\delta(z+i)}{e^{2\phi}-1},
    \label{eq:delta_tau}
\end{equation} whose roots $z=\pm i$ correctly live on the unit circle. Also note that the Cayley transformation \eqref{eq:Cayley} maps the roots of the bosonic delta $z=\pm i$ in SU$(1,1)$ to the roots of the bosonic delta found for SL$(2,\mathbb{R})$, which in the $\epsilon=0$ case become $\tau=0$ and $\tau=\infty$.

Evaluating the terms in (\ref{eq:charSU11}) at $z=\pm i$ yields:
\begin{align}
   \chi_{j,q}^{\mathcal{N}=2}(\phi,\theta)=e^{2iq\theta}\left[e^{2j\phi}\frac{(e^{-\phi+i\theta}-1)(e^{-\phi-i\theta}-1)}{1-e^{-2\phi}}+e^{-2j\phi}\frac{(e^{\phi+i\theta}-1)(e^{\phi-i\theta}-1)}{e^{2\phi}-1} \right]. 
\end{align}
Here, we have exactly the same factors that appeared in the Weyl denominator (\ref{eq:WeylDen}), with the same bosonic and fermionic roots $\alpha_B=\pm2\phi$, $\alpha_F=\pm\phi\pm i\theta$ as the ones in (\ref{eq:roots}). Simplifying yields:
\begin{equation}
    \chi_{j,q}^{\mathcal{N}=2}(\phi,\theta)=2e^{2iq\theta} \cosh(2j\phi)\frac{\cosh\phi-\cos\theta}{\sinh\phi},
\end{equation}
which agrees with the expression of the principal series character in \eqref{eq:n2char}. 

\subsection{Other component of OSp$(2\vert 2,\mathbb{R})$}
\label{sec:othercomponent}
In spite of not describing gravity, it is illuminating to describe some properties of the second connected component of the supergroup. In the $\mathcal{N}=1$ case of OSp$(1\vert 2,\mathbb{R})$, both sectors play a role and are thought of as describing \textbf{R} and \textbf{NS} sectors. For the higher supersymmetric models, this is no longer true, but structurally it is interesting to observe the analogy.

If one writes the decomposition of a group element $g$ in the component of the supergroup that is \emph{not} connected to the identity, as a conjugated element in the maximal torus $T$, we can write:
\begin{equation}
g = c M t c^{-1}, \quad t \in T,
\end{equation}
where the fixed element $M = \text{diag}(1,1\vert -1,1)$ causes a flip between the two components of the supergroup. It can be regarded as a reflection operation on the 2-plane that is acted on by the bosonic O$(2)$ subgroup. 

We have the relations:
\begin{align}
M^{-1}F^\pm M = \mp \bar{F}^\pm, \qquad M^{-1}\bar{F}^\pm M = \mp F^\pm.
\end{align}
Hence the $(2 \vert 4)$-dimensional Jacobian matrix in the Weyl integration formula, acting on the super-vectorspace spanned by $E^+,E^-,F^+,\bar{F}^+,F^-,\bar{F}^-$ in this ordering of basis vectors, has the form:
\begin{equation}
 \text{Ad}(t^{-1}M{})-\mathbf{1} \equiv \text{diag}\left(e^{2\phi}-1, e^{-2\phi}-1, \left(\begin{array}{cc} -1 & - e^{\phi + i\theta}\\ -e^{\phi - i\theta}& -1 \end{array} \right), \left(\begin{array}{cc} -1 &  e^{-\phi + i\theta}\\ e^{-\phi - i\theta}& -1 \end{array} \right)\right).
\end{equation}
In particular, $F^\pm$ and $\bar{F}^\pm$ are no longer eigenvectors, but they transform in the above simple way. Hence the super-Jacobian is readily computed: 
\begin{equation}
\label{eq:wdother}
\Delta(t) \equiv \text{sdet}'( \text{Ad}(t^{-1}M{})-\mathbf{1})_{\mathfrak{g}/\mathfrak{t}} = \frac{(e^{2\phi}-1)(1-e^{-2\phi})}{(1-e^{2\phi})(1-e^{-2\phi})} =-1,
\end{equation}
the Weyl denominator is trivial in this sector of the supergroup.

One can redo the calculation of the principal series character in this sector. The calculation proceeds very similarly as in subsection \ref{principal_series_char}. The result is:
\begin{equation}
\chi_{k,q}^{\mathcal{N}=2}(\phi,\theta) = 2i\sin (2 k \phi) e^{2iq \theta}.
\end{equation}
Notice in particular that no non-trivial Weyl denominator is present here, in agreement with the explicit calculation \eqref{eq:wdother} above.
These characters also form an orthonormal set. Moreover, we also observe that upon insertion in a super-JT amplitude, these would not correspond to one-loop exact gravitational path integrals, but yield a similar all-loop perturbative expansion like $\mathcal{N}=1$ discussed in subsection \ref{s:amn1} in the main text.

\section{Supergravity and the BF formulation}
\label{app:tsg}
A BF gauge theory has a superspace action of the form:
\begin{equation}
S_\text{SJT} = \int_\Sigma \operatorname{STr}(\mathbf{B}\mathbf{F}).
\label{eq:BFaction}
\end{equation}
It is worthwhile to highlight precisely how the gauge transformations in the BF formulation \eqref{eq:BFaction} of supergravity correspond to the gravitational superdiffeomorphisms and local Lorentz transformations \cite{Gomis:1991cc}. This matching will be illuminating for what follows later on.

The superspace action \eqref{eq:BFaction} is invariant under gauge transformations, with $\mathbf{B}$ transforming (homogeneously) in the adjoint representation, and $\mathbf{A}$ as a gauge connection. Infinitesimally this reads directly in superspace: 
\begin{equation}
\delta \mathbf{A}_M = \partial_M \epsilon + [\mathbf{A}_M, \epsilon], \qquad \delta \mathbf{B} = [\mathbf{B}, \epsilon].
\end{equation}

For the applications to $\mathcal{N}=2
$ supergravity, the superspace manifold $\Sigma$ is $(2 \vert 4)$-dimensional and the gauge group is OSp$(2\vert 2,\mathbb{R})$.
For OSp$(2\vert 2,\mathbb{R})$, the dictionary between gauge theory and gravity in superspace in Lorentzian signature is
\begin{equation}
\label{eq:N2gc}
\mathbf{A}_M = \Omega_M H + \sum_{\pm}(e_M^{\;\;\;\pm} E^\pm + f_M^{\;\;\;\pm} F^\pm + \bar{f}_M^{\;\;\;\pm} \bar{F}^\pm ) + \sigma_M Z, \quad M = +,-, \alpha, \quad \alpha=1,2,3,4,
\end{equation}
where we used the generators in the form of \eqref{eq:superalgebra}. The components of $\mathbf{A}_M$ are again interpreted 
in terms of the (super) spin connection $\Omega$, super-zweibein $E^A \equiv (e^\pm \vert f^\pm,\bar{f}^\pm)$ (the index $A$ takes on $2 \vert 4$ possible values), and the gauge potential $\sigma$. Notice the distinction of this expansion compared to \eqref{eq:n2exp} and \eqref{C6:gaugefieldexpansion}. This is due to the fact that here we work with a Lorentzian target space instead.\footnote{It is straightforward to change the current discussion to the Euclidean target space of (\ref{C6:gaugefieldexpansion}), but for illustrative purposes we focus on Lorentzian signature here.} Both signatures are captured by the same OSp$(2\vert 2,\mathbb{R})$ supergroup, but with a different coset to describe the bulk superspace. This reflects the difference between hyperbolic superspace H$_{2 \vert 4}$ and AdS superspace AdS$_{2 \vert 4}$, as described in equation \eqref{eq:supads}.

Similarly expanding the gauge parameter as
\begin{equation}
\label{eq:gpexp}
\epsilon = l H +\sum_\pm (\epsilon^\pm E^\pm + u^\pm F^\pm + \bar{u}^\pm \bar{F}^\pm) + s Z,
\end{equation}
the superspace gauge transformations $\delta \mathbf{A}_M = \partial_M \epsilon + [\mathbf{A}_M, \epsilon]$, have the following component form:
\begin{align}
\delta_G e_M^{\;\;\;\pm} &= \partial_M \epsilon^\pm \pm \Omega_M \epsilon^\pm \mp e_M^{\;\;\;\pm} l + f_M^{\;\;\;\pm} \bar{u}^\pm + \bar{f}_M^{\;\;\;\pm} u^\pm, \\
\delta_G f_M^{\;\;\;\pm} &= \partial_M u^\pm \pm \frac{1}{2} \Omega_M u^\pm + \frac{1}{2} \sigma_M u^\pm \mp \frac{1}{2} f_M^{\;\;\;\pm} l - \frac{1}{2} f_M^{\;\;\;\pm} s - e_M^{\;\;\;\pm} u^\mp +f_M^{\;\;\;\mp} \epsilon^\pm, \\
\delta_G \bar{f}_M^{\;\;\;\pm} &= \partial_M \bar{u}^\pm \pm \frac{1}{2} \Omega_M \bar{u}^\pm - \frac{1}{2} \sigma_M \bar{u}^\pm \mp \frac{1}{2} \bar{f}_M^{\;\;\;\pm} l + \frac{1}{2} \bar{f}_M^{\;\;\;\pm} s + e_M^{\;\;\;\pm} \bar{u}^\mp + \bar{f}_M^{\;\;\;\pm} \epsilon^\pm, \\
\delta_G \Omega &= \partial_M l + 2 e_M^{\;\;\;+} \epsilon^- - 2 e_M^{\;\;\;-} \epsilon^+ - f_M^{\;\;\;+} \bar{u}^- -\bar{f}_M^{\;\;\;-} u^+ + f_M^{\;\;\;-} \bar{u}^+ + \bar{f}_M^{\;\;\;+} u^-, \\
\delta_G \sigma_M &= \partial_M s + f_M^{\;\;\;+} \bar{u}^- + \bar{f}^-_M u^+ + \bar{f}_M^{\;\;\;-} \bar{u}^+ + \bar{f}_M^{\;\;\;+} u^-.
\end{align}
The parameter $s$ in \eqref{eq:gpexp} parametrizes a (compact) U(1) gauge transformation under which the $f^\pm$ have charge $+1/2$ and the $\bar{f}^\pm$ have charge $-1/2$, the remaining fields uncharged. The parameter $l$ in \eqref{eq:gpexp} parametrizes the (bosonic) SO(1,1) local Lorentz transformation, simultaneously boosting $e^\pm$ as a vector, and $f^\pm, \bar{f}^\pm$ as spin-$1/2$ spinors.

The remaining $2 \vert 4$ parameters parametrize the super-diffeomorphisms on-shell:\footnote{This is exact in these topological field theories, since the flatness condition holds off-shell as well through the Lagrange multipliers.}
\begin{equation}
\epsilon^\pm = \xi^M e_M^{\;\;\;\pm}, \quad  u^\pm = \xi^M f_M^{\;\;\;\pm}, \quad \bar{u}^\pm = \xi^M \bar{f}_M^{\;\;\;\pm}, \quad l = \xi^M \Omega_M, \quad s= \xi^M \sigma_M,
\end{equation}
in terms of $ 2 \vert 4$ functions $\xi^M$. To see that super-diffeomorphisms can indeed cover this remaining $2\vert 4$ parameter family of gauge transformations, we rewrite the first three of the above relations in superspace as a $2 \vert 4$ equation $\epsilon^A = \xi^M E_M^{\;\;\; A} = E_{\;\;\;M}^A \xi^M$ to obtain the inverse:
\begin{equation}
\label{eq:solxi}
 E_N^{\;\;\;B}\kappa_{BA} \epsilon^A = g_{NM}\xi^M \equiv \xi_N,
\end{equation}
where we used the relation between the vielbein and metric:
\begin{equation}
\label{eq:smetr}
g_{MN} = E_M^{\;\;\;A}\;\kappa_{AB}\;E_{\;\;\;N}^B.
\end{equation}
So given any fixed $\epsilon^A$, if we choose $\xi_M$ as in \eqref{eq:solxi}, we can interpret the transformation as a super-diffeomorphism. 

We can summarize the following physical decomposition of the full gauge group of \eqref{eq:BFaction}:
\begin{equation}
\text{Gauge group = (super-diffeo)} \otimes \text{SO}(1,1)_{\text{Lorentz}} \otimes \text{U}(1)
\end{equation}

\section{Some representation theory of PSU$(1,1 \vert 2)$}
 \label{AppendixE}
The relevant superalgebra for the $\mathcal{N}=4$ JT supergravity model is $\mathfrak{psl}(2|2)$, with its noncompact real form $\mathfrak{psu}(1,1|2)$ \cite{Gotz:2005ka,Gotz:2006qp}. The $\mathfrak{psu}(1,1|2)$ algebra has $\mathfrak{sl}(2,\mathbb{R})\oplus \mathfrak{su}(2)$ as maximal bosonic subalgebra. It has 6 bosonic generators $H,E^\pm,Z,Z^\pm$ and 8 fermionic ones $F^\pm_\alpha, \bar{F}^\pm_\beta$ with $\alpha,\beta = 1,2$ spinor indices.
In the Chevalley basis, the Lie superalgebra has the form:
\begin{alignat}{2}
[H, E^\pm] &= \pm E^\pm, \qquad  &[Z,Z^\pm] &= \pm Z^\pm, \\
[H,F^\pm_\alpha] &= \pm \frac{1}{2}F^\pm_\alpha, \qquad &[H,\bar{F}^\pm_\alpha] &= \pm \frac{1}{2}\bar{F}^\pm_\alpha, \\
[Z,F^\pm_\alpha] &= \pm \frac{1}{2}F^\pm_\alpha, \qquad &[Z,\bar{F}^\pm_\alpha] &= \mp \frac{1}{2}\bar{F}^\pm_\alpha, \\
\{F^\pm_\alpha,\bar{F}^\pm_\beta\} &= \mp 2 \epsilon_{\alpha \beta} E^\pm, \qquad &\{F^\pm_\alpha,\bar{F}^\mp_\beta\} &= \pm 2 \epsilon_{\alpha \beta} Z^\pm, \\
\{F^+_\alpha,F^-_\beta\} &= 2 \epsilon_{\alpha\beta}(H-Z), \qquad &\{\bar{F}^+_\alpha,\bar{F}^-_\beta\} &= 2 \epsilon_{\alpha\beta}(H+Z), \\
[E^\pm,F^\mp_\alpha] &= \pm \bar{F}^\pm_\alpha, \qquad &[E^\pm,\bar{F}^\mp_\alpha] &= \mp F^\pm_\alpha, \\
[Z^\pm,F^\mp_\alpha] &= \pm \bar{F}^\mp_\alpha, \qquad &[Z^\pm,\bar{F}^\pm_\alpha] &= \mp F^\pm_\alpha, \\
[E^+,E^-] &= 2H, \qquad &[Z^+,Z^-] &= 2Z.
\end{alignat}

There are two Cartan generators $H$ and $Z$, with coordinates $\phi \in [0,\infty)$ and $\theta \in [0,\pi)$. The $4\vert 8$ roots are:
\begin{equation}
\alpha_B(t) = \pm 2\phi, \pm 2 i\theta, \quad \alpha_F(t) = \pm \phi \pm i \theta,
\end{equation}
where each fermionic root is counted twice. The Weyl denominator hence becomes:
\begin{align}
\Delta(\phi,\theta) &= \frac{(e^{2\phi}-1)(1-e^{-2\phi})\vert e^{2i\theta}-1\vert \vert e^{-2i\theta}-1\vert}{(e^{\phi+i\theta}-1)^2(e^{\phi-i\theta}-1)^2(e^{-\phi+i\theta}-1)^2(e^{-\phi-i\theta}-1)^2} \\
&= \frac{\sinh^2 \phi \sin^2 \theta}{(\cosh \phi - \cos \theta)^4}.\label{eq:N=4weyl}
\end{align}
There is a quadratic and higher Casimir. The quadratic Casimir is given by the explicit expression:
\begin{align}
\mathcal{C}_2 &= H^2 + \frac{1}{2}(E^-E^+ + E^+ E^-) - Z^2 - \frac{1}{2}(Z^-Z^++Z^+Z^-) \nonumber \\
&{}\hspace{0.7cm}- \frac{1}{4} \epsilon_{\alpha\beta}\left(F_\alpha^+F_\beta^- + F_\alpha^- F_\beta^+ + \bar{F}_\alpha^+\bar{F}_\beta^- + \bar{F}_\alpha^-\bar{F}_\beta^+\right).
\end{align}
On a highest weight state  (that is annihilated by all raising ${}^+$ generators) with eigenvalues $H=j_1+1$ and $Z=j_2$, the above expression reduces to:\footnote{Note the relative minus sign for the $H$-term. This is due to the fermionic contribution. The $j_1$ quantum number is introduced to match with SL$(2,\mathbb{R})$ notation, but it is \emph{not} the highest value of $H$ in the representation as denoted.}
\begin{equation}
\label{eq:N=4highestweightcasimir}
\mathcal{C}_2 = H^2 -H - Z^2 - Z = j_1(j_1+1) - j_2 (j_2+1).
\end{equation}
Finite-dimensional irreducible representations are characterized by this maximal value of both Cartan generators (on the highest state), denoted by the half-integers $j_1$ and $j_2$. Restricting to the maximal bosonic subgroup SL$(2,\mathbb{R})\otimes $ SU(2), with (tensor product) irreps denoted as $(j_1,j_2)$, the typical finite-dimensional irreps have the branching rule decomposition \cite{Gotz:2005ka}:
\begin{equation}
\label{eq:branch}
(j_1,j_2) \, \otimes \, \Big[2\, (0,0) \, \oplus \, 2\, (\frac{1}{2},\frac{1}{2}) \, \oplus \, (0,1) \, \oplus \, (1,0)\Big],
\end{equation}
with total dimension $16(2j_1+1)(2j_2+1)$ as readily checked.
If we use the finite-dimensional characters of $\mathfrak{sl}(2,\mathbb{R})$ and $\mathfrak{su}(2)$ 
\begin{equation}
    \chi_j^{\mathfrak{sl}(2,\mathbb{R})}(\phi)=\frac{\sinh(2j+1)\phi}{\sinh\phi}~,\qquad \chi_j^{\mathfrak{su}(2)}(\theta)=\frac{\sin(2j+1)\theta}{\sin\theta},
\end{equation} 
and the branching rule (\ref{eq:branch}), the $\mathcal{N}=4$ finite rep character is found to be
\begin{equation}
     \chi^{\mathcal{N}=4}_{j_1,j_2}(\phi,\theta)=4\frac{\sinh(2j_1+1)\phi}{\sinh\phi}\frac{\sin(2j_2+1)\theta}{\sin\theta}\left(\cos\theta-\cosh\phi\right)^2.
\end{equation}
Likewise, the discrete highest weight character can be computed using the same branching rules \eqref{eq:branch} as:
\begin{align}
\chi^{\mathcal{N}=4}_{j_1,j_2}(\phi,\theta)&= \frac{e^{(2j_1+1)\phi}}{2\sinh \phi} \frac{\sin (2j_2+1)\theta}{\sin \theta}\left[2 - 4(e^\phi + e^{-\phi}) \cos \theta + \frac{\sin 3 \theta}{\sin \theta} + e^{2\phi} +1+ e^{-2\phi}\right] \nonumber \\
&= 2 e^{(2j_1+1)\phi}\sin (2j_2+1)\theta \frac{(\cos\theta - \cosh \phi)^2}{ \sinh\phi \sin \theta}.\label{eq:N=4discretecharacter}
\end{align}
The principal series character, in turn, requires the formal analytic continuation of the branching rule decomposition (\ref{eq:branch}), using the principal series characters of SL$(2,\mathbb{R})$ instead. To induce a unitary representation, the analytically continued weight for general supersymmetry can be found from the Jacobian rule \eqref{eq:generalberezinian}, which for $\mathcal{N}=4$ yields: $j=1/2+ik, \,\, k\in \mathbb{R}^+$. 
Using the shifted value of $j_1=-1/2+ik, \,\, k\in \mathbb{R}^+$ instead,\footnote{If we would have denoted the highest value of $H$ in the highest weight irreps as $j_1$ (instead of $j_1+1$), we would have had $j_1=+1/2+ik$. This $+1/2$ shift is the resulting Weyl vector for this superalgebra with $\rho = \frac{1}{2} \sum_i \alpha_i \equiv \rho_B - \rho_F$ and $\rho_B=1$ and $\rho_F=2$. The fermionic contribution is hence effectively flipping the sign of the Weyl vector compared to the bosonic $\mathfrak{sl}(2,\mathbb{R})$ algebra, resulting in $-1/2 \to +1/2$.} we find: \begin{align}
\label{eq:charN4}
\chi_{k,j_2}^{\mathcal{N}=4}(\phi,\theta) = 4\cos(2k\phi) \sin (2j_2+1)\theta\frac{(\cos\theta - \cosh \phi)^2}{ \sinh\phi \sin \theta}.
\end{align} Character orthogonality is again manifest:
\begin{equation}
\int d\phi d\theta\, \Delta(\phi,\theta) \,\chi_{k,j}^{\mathcal{N}=4}(\phi,\theta) \chi_{k',j'}^{*\mathcal{N}=4}(\phi,\theta) = (2\pi)^2\delta(k-k') \delta_{jj'}.
\end{equation}

\subsection{Check via super-Virasoro modular S-matrix}
As before, the principal series character \eqref{eq:charN4} can be found in the Schwarzian (classical) limit of the $\mathcal{N}=4$ super-Virasoro modular S-matrix for non-degenerate characters. Several calculations along these lines have been made in the literature \cite{Heydeman:2020hhw,Iliesiu:2021are}. The relevant characters can be found in eqns (1) and (2) of \cite{Eguchi:1988af}, e.g. in the NS-sector:
\begin{align}
\label{eq:KMchar}
\text{ch}^{NS}_P(k,j;\nu,\tau) = q^{h-(j+1/2)^2/(k+1)+1/4}\frac{\theta_3(q;z)^2}{\eta(\tau)^3}\chi_{k-1}^j(\nu,\tau), \quad j = 0,\frac{1}{2},1, \hdots \frac{k}{2},
\end{align}
where we parametrize $h=P^2 +(j+1/2)^2/(k+1)-1/4$, and where $\chi_{k-1}^j(\nu,\tau)$ is the affine character of the $\widehat{\text{SU}}(2)_{k-1}$ affine algebra. The latter transforms under modular S-transformations as
\begin{equation}
\chi_{k-1}^j(\nu/\tau,-1/\tau) = \sum_{j'=0}^{k/2} S_j{}^{j'} \chi_{k-1}^{j'}(\nu,\tau), 
\quad j' = \text{half-integer},
\end{equation}
where 
\begin{equation}
S_j{}^{j'} = \sqrt{\frac{1}{k+1}}\sin (\frac{\pi(2j+1)(2j'+1)}{k+1}).
\end{equation}
Performing a modular S-transform on \eqref{eq:KMchar} leads to a linear combination of the same types of characters. Without loss of generality we focus on the simpler case where $\nu=0$:\footnote{We used
\begin{align}
\theta_3(0,-1/\tau) &= (-i\tau)^{1/2}\theta_3(0,\tau), \\
\eta(-1/\tau) &= (-i\tau)^{1/2} \eta(\tau).
\end{align}
}
\begin{align}
\text{ch}^{NS}_P(k,j;0,-1/\tau) = \int_{-\infty}^{+\infty} dP' \, \sum_{j'=0}^{k/2}\, \left( \frac{1}{\sqrt{2}}\cos (4\pi PP') S_j{}^{j'} \right)\, \text{ch}^{NS}_{P'}(k,j';0,\tau).
\end{align}
Introducing the central charge $c=6k$, we read off the total modular S-matrix:
\begin{equation}
S_{P,j}{}^{P',j'} = \frac{1}{\sqrt{2}}\cos (4\pi PP') \sqrt{\frac{1}{c/6+1}}\sin (\frac{\pi(2j+1)(2j'+1)}{c/6+1}).
\end{equation}
In the Schwarzian limit, we let $c\to \infty$ in a double-scaled fashion, where we set $P=bk$ and $2\pi b P'=\phi$ where $b\sim 1/\sqrt{c} \to 0$. The quantity $\pi(2j'+1)/(c/6)$ effectively becomes a continuous real number, between $0$ and $\pi$, denoted by $\theta$. Hence we obtain the $\mathcal{N}=4$ character \eqref{eq:charN4}, up to an irrelevant proportionality factor:
\begin{equation}
\lim_{c\to \infty} S_{P,j}{}^{P',j'} \, \sim \, \cos (2k\phi) \sin ((2j+1)\theta).
\end{equation}
As in all other cases, we also note that the Weyl denominator in \eqref{eq:charN4} is not produced when coming from the 2d (S)CFT perspective.

\subsection{Higher rank Casimir}
\label{app:hrcas}

The higher rank Casimir can be found by taking the contraction of the Lie superalgebra of $D(2,1;\alpha)$ superalgebra down to $\mathfrak{psu}(1,1|2) \oplus \mathfrak{u}(1)^3$ \cite{Aoyama:2015gna}, and then setting to zero the three U(1) generators. For the parent algebra $D(2,1;\alpha)$, an expression is known for the second Casimir operator as a quartic combination of operators \cite{VanDerJeugt:1985hq}. Denoting the bosonic subalgebra quadratic Casimirs as:
\begin{align}
S \equiv H^2 + \frac{1}{2}(E^+E^-+E^-E^+), \qquad
T \equiv Z^2 + \frac{1}{2}(Z^+Z^-+Z^-Z^+),
\end{align}
the quartic Casimir of PSU$(1,1|2)$ can be written as:
\begin{equation}
\mathcal{C}_4 = S^4 - T^4 + \frac{1}{2}\mathcal{C}_2(S^2+T^2) - 2(S^2-T^2) + \text{fermion bilinear}.
\end{equation}
For a highest weight irrep, starting with the explicit expression (4.11) of \cite{VanDerJeugt:1985hq}, one can show that $-\mathcal{C}_4$ evaluates to
\begin{equation}
(j_1+1)^2(j_1+2)^2 - j_2^2(j_2+1)^2 + 2 j_2(j_2+1)(2j_1+3)-2(j_1+1)[2j_2(j_2+1)+(j_1+2)(2j_1+1)].
\end{equation}
The quadratic and quartic Casimir can be taken as a basis for the center of the Lie superalgebra, and hence serve the purpose of fully specifying the representation.

\section{Perturbative analysis}
\label{app:Perturbationtheory}
A standard way to get physical insight for a non-Gaussian path integral is to expand it perturbatively. In the case of $\mathcal{N}=2$, we have a second Lagrange multiplier in the EOW brane action equation (\ref{eq:eowN2}), due to the fact that we have a second Casimir $\mathcal{C}_3$, cubic in the generators. We present a simple perturbative analysis where we will use purely bosonic dimensions for ease of notation. It is not hard to incorporate fermionic dimensions into the arguments below.

In terms of the generators $P_a$ of the Lie algebra, we have explicitly:
\begin{equation}
    \mathcal{C}_3 = P^aP^bP^c h_{abc}, \qquad h_{abc}\equiv \text{Tr}(P_{(a}P_bP_{c)}).
\end{equation}
Consequently, the worldline action takes the form
\begin{equation}
    S=\int d\tau \, \left(\Lambda^a(g^{-1}D_A g)_a+\Theta_1\Lambda_a^2+\Theta_2\Lambda_a\Lambda_b\Lambda_c h^{abc}\right).
    \label{eq:actionlambda}
\end{equation}
This is the action of a quantum field theory with fields $\Lambda_0,\Lambda_1$.\footnote{$\Lambda_2$ is set to zero, following the discussion in section \ref{Section4}.} In the familiar language of Feynman diagrams, we can depict each term appearing in (\ref{eq:actionlambda}). \\
\textbf{Source term:}
    \begin{equation}
\begin{tikzpicture}
    \draw[thick] (-0.1, -0.1) -- (0.1, 0.1);
    \draw[thick] (-0.1, 0.1) -- (0.1, -0.1);   
    \draw[thick] (0, 0) -- (1.8, 0); 
    \node[below] at (0.0, -0.1) {$J^a$};
    \node[below] at (1.8, -0.1) {$b$};
    \node[left] at (-0.6,0) {$J_a:$};
    \node[right] at (1.9,0) {$= (g^{-1}D_A g)_a \delta_{ab}$
    };
\end{tikzpicture}
\end{equation}
\textbf{Propagator}:
\begin{equation}
\begin{tikzpicture}
    \draw[thick] (0, 0) -- (1.8, 0);
    \node[below] at (0.0, -0.1) {$a$};
    \node[below] at (1.8, -0.1) {$b$};
    \node[left] at (-0.6,0) {$\langle\Lambda_a \Lambda_b\rangle:$};
    \node[right] at (1.9,0) {$\displaystyle= \frac{\delta_{ab}}{\Theta_1} \delta(x-y)$};
\end{tikzpicture}
\end{equation}
\textbf{Interaction vertex}: 
\begin{equation}
\begin{tikzpicture}
    \draw[thick] (-1, 1) -- (0,0);
    \draw[thick] (1, 1) -- (0,0);
    \draw[thick] (0, 0) -- (0,-1.1);
    \node[above] at (-1,1) {$a$};
    \node[above] at (1,1) {$b$};
    \node[below] at (0,-1.1) {$c$};
    \node[right] at (1.3,0) {$=h_{abc}\Theta_2$};
\end{tikzpicture}
\end{equation}

The path integral for the effective theory can be written as $e^{-S_{\text{eff}}[J]}=\int [\mathcal{D}\Lambda_a]e^{-S}$.
We consider several terms contributing in $S_{\text{eff}}[J]$ to the quadratic term $\sim J^2$ explicitly:
 \begin{align*}
    &\begin{tikzpicture}
    \draw[thick] (-0.1, -0.1) -- (0.1, 0.1);
    \draw[thick] (-0.1, 0.1) -- (0.1, -0.1);
    \draw[thick] (0, 0) -- (1.4, 0);
    \draw[thick, red] (1.4, 0) -- (2.6,0);
    \draw[thick] (2.6, 0) -- (4, 0);
    \node at (1.4,0)[circle,fill,inner sep=1.5pt]{};
    \node at (2.6,0)[circle,fill,inner sep=1.5pt]{};
    \draw[thick] (3.9, -0.1) -- (4.1, 0.1);
    \draw[thick] (3.9, 0.1) -- (4.1, -0.1);
    \node[left] at (-0.5,0) {$\mathcal{O}(\Theta_2^0): \,\, \sum\limits_{a,b}$};
    \node[below] at (0.0, -0.2) {$J^a$};
    \node[below] at (4, -0.2) {$J^b$};
    \node[above] at (1.95, 0.1) {$\textcolor{red}{\frac{1}{\Theta_1}}$};
    \node[right] at (4.2,0){$=\sum\limits_{a,b}\frac{1}{\Theta_1}\int J_a J_b\delta_{ab}$};
\end{tikzpicture}  \\
&~\mathcal{O}(\Theta_2^1): \text{does not exist.} \\
    &\begin{tikzpicture}
     \draw[thick] (-0.1, -0.1) -- (0.1, 0.1);
    \draw[thick] (-0.1, 0.1) -- (0.1, -0.1);
    \draw[thick] (0, 0) -- (1.4, 0);
    \draw[thick, red](2,0) circle (0.6);
    \node[below] at (1.3,-0.17){$x$};
    \node[below] at (2.7,-0.17){$y$};
    \node[above] at (2,0.6){$c$};
    \node[below] at (2,-0.6){$d$};
    \node[above] at (0.7,-0.1)[circle,fill,inner sep=1.5pt]{};
    \node[above] at (3.3,-0.1)[circle,fill,inner sep=1.5pt]{};
    \node[above] at (0.7,0.1){$\tau_1$};
    \node[above] at (3.3,0.1){$\tau_2$};
    \draw[thick] (2.6, 0) -- (4, 0);
    \draw[thick] (3.9, -0.1) -- (4.1, 0.1);
    \draw[thick] (3.9, 0.1) -- (4.1, -0.1);
    \node at (1.4,0)[circle,fill,inner sep=1.5pt]{};
    \node at (2.6,0)[circle,fill,inner sep=1.5pt]{};
    \node[left] at (-0.5,0) {$\mathcal{O}(\Theta_2^2): \,\, \sum\limits_{a,b}$ };
    \node[below] at (0,-0.2){$J^a$};
    \node[below] at (4,-0.2){$J^b$};
    \node[right] at (4.3,0){$=\frac{\Theta_2^2}{\Theta_1^2}\int d\tau_1 d\tau_2\int dx~ dy [\frac{1}{\Theta_1^2}\delta(\tau_1-x)\delta(\tau_2-y)] $};
    \node[right] at (4.8,-0.8){$\times \delta(x-y)^2\sum\limits_{a,b}h_{acd}h_{bcd}J_a(\tau_1)J_b(\tau_2)$};
\end{tikzpicture}
\end{align*}

Recalling that $\sum\limits_{a,b}J_a J_b\delta_{ab} = (g^{-1}D_A g)^a(g^{-1}D_A g)_a \to \dot{X}^Mg_{MN}\dot{X}^N$, we would a priori aim at writing a second order geometric Lagrangian, i.e. a functional of  $g_{MN}$ only.  The higher-order terms however, cannot be written purely in terms of the metric $g_{MN}$.

This is particularly transparent for the three-vertex itself, contributing to a $\sim J^3$-contribution at tree level as:
\begin{equation}
\begin{tikzpicture}
    \node at (0.5,0.5)[circle,fill,inner sep=1.5pt]{};
    \node at (0,-0.5)[circle,fill,inner sep=1.5pt]{};
    \node at (-0.5,0.5)[circle,fill,inner sep=1.5pt]{};
    \draw[thick] (-1, 1) -- (0,0);
    \draw[thick] (1, 1) -- (0,0);
    \draw[thick] (0, 0) -- (0,-1);
    \draw[thick] (1.1, 1.1) -- (0.9, 0.9);
    \draw[thick] (0.9, 1.1) -- (1.1, 0.9);
    \draw[thick] (-1.1, 1.1) -- (-0.9, 0.9);
    \draw[thick] (-0.9, 1.1) -- (-1.1, 0.9);
    \draw[thick] (0.1, -1.1) -- (-0.1, -0.9);
    \draw[thick] (-0.1, -1.1) -- (0.1, -0.9);
    \node[above] at (-1,1.1) {$J^a$};
    \node[above] at (1,1.1) {$J^b$};
    \node[below] at (0,-1.1) {$J^c$};
    \node[right] at (1.3,0) {$= \, \Theta_2\sum_{a,b,c}h_{abc}J_aJ_bJ_c$};
\end{tikzpicture}
\end{equation}
which since $J \sim e$, the vielbein, is not writable in terms of $g\sim e^2$ solely.

\bibliographystyle{ourbst}
\bibliography{NotesN2.bib}

\providecommand{\href}[2]{#2}\begingroup\raggedright\begin{thebibliography}{100}

\bibitem{Jackiw:1984je}
R.~Jackiw, ``{Lower Dimensional Gravity},''
  \href{http://dx.doi.org/10.1016/0550-3213(85)90448-1}{{\em Nucl. Phys. B}
  {\bfseries 252} (1985) 343--356}.

\bibitem{Teitelboim:1983ux}
C.~Teitelboim, ``{Gravitation and Hamiltonian Structure in Two Space-Time
  Dimensions},'' \href{http://dx.doi.org/10.1016/0370-2693(83)90012-6}{{\em
  Phys. Lett. B} {\bfseries 126} (1983) 41--45}.

\bibitem{Nayak:2018qej}
P.~Nayak, A.~Shukla, R.~M. Soni, S.~P. Trivedi, and V.~Vishal, ``{On the
  Dynamics of Near-Extremal Black Holes},''
  \href{http://dx.doi.org/10.1007/JHEP09(2018)048}{{\em JHEP} {\bfseries 09}
  (2018) 048}, \href{http://arxiv.org/abs/1802.09547}{{\ttfamily
  arXiv:1802.09547 [hep-th]}}.

\bibitem{Iliesiu:2020qvm}
L.~V. Iliesiu and G.~J. Turiaci, ``{The statistical mechanics of near-extremal
  black holes},'' \href{http://dx.doi.org/10.1007/JHEP05(2021)145}{{\em JHEP}
  {\bfseries 05} (2021) 145}, \href{http://arxiv.org/abs/2003.02860}{{\ttfamily
  arXiv:2003.02860 [hep-th]}}.

\bibitem{Castro:2021csm}
A.~Castro, V.~Godet, J.~Sim\'on, W.~Song, and B.~Yu, ``{Gravitational
  perturbations from NHEK to Kerr},''
  \href{http://dx.doi.org/10.1007/JHEP07(2021)218}{{\em JHEP} {\bfseries 07}
  (2021) 218}, \href{http://arxiv.org/abs/2102.08060}{{\ttfamily
  arXiv:2102.08060 [hep-th]}}.

\bibitem{Iliesiu:2022kny}
L.~V. Iliesiu, S.~Murthy, and G.~J. Turiaci, ``{Black hole microstate counting
  from the gravitational path integral},''
  \href{http://arxiv.org/abs/2209.13602}{{\ttfamily arXiv:2209.13602
  [hep-th]}}.

\bibitem{Castro:2022cuo}
A.~Castro, F.~Mariani, and C.~Toldo, ``{Near-extremal limits of de Sitter black
  holes},'' \href{http://dx.doi.org/10.1007/JHEP07(2023)131}{{\em JHEP}
  {\bfseries 07} (2023) 131}, \href{http://arxiv.org/abs/2212.14356}{{\ttfamily
  arXiv:2212.14356 [hep-th]}}.

\bibitem{Almheiri:2014cka}
A.~Almheiri and J.~Polchinski, ``{Models of AdS$_{2}$ backreaction and
  holography},'' \href{http://dx.doi.org/10.1007/JHEP11(2015)014}{{\em JHEP}
  {\bfseries 11} (2015) 014}, \href{http://arxiv.org/abs/1402.6334}{{\ttfamily
  arXiv:1402.6334 [hep-th]}}.

\bibitem{Jensen:2016pah}
K.~Jensen, ``{Chaos in AdS$_2$ Holography},''
  \href{http://dx.doi.org/10.1103/PhysRevLett.117.111601}{{\em Phys. Rev.
  Lett.} {\bfseries 117} no.~11, (2016) 111601},
  \href{http://arxiv.org/abs/1605.06098}{{\ttfamily arXiv:1605.06098
  [hep-th]}}.

\bibitem{Maldacena:2016upp}
J.~Maldacena, D.~Stanford, and Z.~Yang, ``{Conformal symmetry and its breaking
  in two dimensional Nearly Anti-de-Sitter space},''
  \href{http://dx.doi.org/10.1093/ptep/ptw124}{{\em PTEP} {\bfseries 2016}
  no.~12, (2016) 12C104}, \href{http://arxiv.org/abs/1606.01857}{{\ttfamily
  arXiv:1606.01857 [hep-th]}}.

\bibitem{Engelsoy:2016xyb}
J.~Engels\"oy, T.~G. Mertens, and H.~Verlinde, ``{An investigation of AdS$_{2}$
  backreaction and holography},''
  \href{http://dx.doi.org/10.1007/JHEP07(2016)139}{{\em JHEP} {\bfseries 07}
  (2016) 139}, \href{http://arxiv.org/abs/1606.03438}{{\ttfamily
  arXiv:1606.03438 [hep-th]}}.

\bibitem{Cotler:2016fpe}
J.~S. Cotler, G.~Gur-Ari, M.~Hanada, J.~Polchinski, P.~Saad, S.~H. Shenker,
  D.~Stanford, A.~Streicher, and M.~Tezuka, ``{Black Holes and Random
  Matrices},'' \href{http://dx.doi.org/10.1007/JHEP05(2017)118}{{\em JHEP}
  {\bfseries 05} (2017) 118}, \href{http://arxiv.org/abs/1611.04650}{{\ttfamily
  arXiv:1611.04650 [hep-th]}}. [Erratum: JHEP 09, 002 (2018)].

\bibitem{Stanford:2017thb}
D.~Stanford and E.~Witten, ``{Fermionic Localization of the Schwarzian
  Theory},'' \href{http://dx.doi.org/10.1007/JHEP10(2017)008}{{\em JHEP}
  {\bfseries 10} (2017) 008}, \href{http://arxiv.org/abs/1703.04612}{{\ttfamily
  arXiv:1703.04612 [hep-th]}}.

\bibitem{Kitaev:2018wpr}
A.~Kitaev and S.~J. Suh, ``{Statistical mechanics of a two-dimensional black
  hole},'' \href{http://dx.doi.org/10.1007/JHEP05(2019)198}{{\em JHEP}
  {\bfseries 05} (2019) 198}, \href{http://arxiv.org/abs/1808.07032}{{\ttfamily
  arXiv:1808.07032 [hep-th]}}.

\bibitem{Mertens:2017mtv}
T.~G. Mertens, G.~J. Turiaci, and H.~L. Verlinde, ``{Solving the Schwarzian via
  the Conformal Bootstrap},''
  \href{http://dx.doi.org/10.1007/JHEP08(2017)136}{{\em JHEP} {\bfseries 08}
  (2017) 136}, \href{http://arxiv.org/abs/1705.08408}{{\ttfamily
  arXiv:1705.08408 [hep-th]}}.

\bibitem{Mertens:2018fds}
T.~G. Mertens, ``{The Schwarzian theory \textemdash{} origins},''
  \href{http://dx.doi.org/10.1007/JHEP05(2018)036}{{\em JHEP} {\bfseries 05}
  (2018) 036}, \href{http://arxiv.org/abs/1801.09605}{{\ttfamily
  arXiv:1801.09605 [hep-th]}}.

\bibitem{Lam:2018pvp}
H.~T. Lam, T.~G. Mertens, G.~J. Turiaci, and H.~Verlinde, ``{Shockwave S-matrix
  from Schwarzian Quantum Mechanics},''
  \href{http://dx.doi.org/10.1007/JHEP11(2018)182}{{\em JHEP} {\bfseries 11}
  (2018) 182}, \href{http://arxiv.org/abs/1804.09834}{{\ttfamily
  arXiv:1804.09834 [hep-th]}}.

\bibitem{Harlow:2018tqv}
D.~Harlow and D.~Jafferis, ``{The Factorization Problem in Jackiw-Teitelboim
  Gravity},'' \href{http://dx.doi.org/10.1007/JHEP02(2020)177}{{\em JHEP}
  {\bfseries 02} (2020) 177}, \href{http://arxiv.org/abs/1804.01081}{{\ttfamily
  arXiv:1804.01081 [hep-th]}}.

\bibitem{Yang:2018gdb}
Z.~Yang, ``{The Quantum Gravity Dynamics of Near Extremal Black Holes},''
  \href{http://dx.doi.org/10.1007/JHEP05(2019)205}{{\em JHEP} {\bfseries 05}
  (2019) 205}, \href{http://arxiv.org/abs/1809.08647}{{\ttfamily
  arXiv:1809.08647 [hep-th]}}.

\bibitem{Blommaert:2018oro}
A.~Blommaert, T.~G. Mertens, and H.~Verschelde, ``{The Schwarzian Theory - A
  Wilson Line Perspective},''
  \href{http://dx.doi.org/10.1007/JHEP12(2018)022}{{\em JHEP} {\bfseries 12}
  (2018) 022}, \href{http://arxiv.org/abs/1806.07765}{{\ttfamily
  arXiv:1806.07765 [hep-th]}}.

\bibitem{Blommaert:2018iqz}
A.~Blommaert, T.~G. Mertens, and H.~Verschelde, ``{Fine Structure of
  Jackiw-Teitelboim Quantum Gravity},''
  \href{http://dx.doi.org/10.1007/JHEP09(2019)066}{{\em JHEP} {\bfseries 09}
  (2019) 066}, \href{http://arxiv.org/abs/1812.00918}{{\ttfamily
  arXiv:1812.00918 [hep-th]}}.

\bibitem{Iliesiu:2019xuh}
L.~V. Iliesiu, S.~S. Pufu, H.~Verlinde, and Y.~Wang, ``{An exact quantization
  of Jackiw-Teitelboim gravity},''
  \href{http://dx.doi.org/10.1007/JHEP11(2019)091}{{\em JHEP} {\bfseries 11}
  (2019) 091}, \href{http://arxiv.org/abs/1905.02726}{{\ttfamily
  arXiv:1905.02726 [hep-th]}}.

\bibitem{Saad:2019lba}
P.~Saad, S.~H. Shenker, and D.~Stanford, ``{JT gravity as a matrix integral},''
  \href{http://arxiv.org/abs/1903.11115}{{\ttfamily arXiv:1903.11115
  [hep-th]}}.

\bibitem{Saad:2019pqd}
P.~Saad, ``{Late Time Correlation Functions, Baby Universes, and ETH in JT
  Gravity},'' \href{http://arxiv.org/abs/1910.10311}{{\ttfamily
  arXiv:1910.10311 [hep-th]}}.

\bibitem{Blommaert:2019wfy}
A.~Blommaert, T.~G. Mertens, and H.~Verschelde, ``{Eigenbranes in
  Jackiw-Teitelboim gravity},''
  \href{http://dx.doi.org/10.1007/JHEP02(2021)168}{{\em JHEP} {\bfseries 02}
  (2021) 168}, \href{http://arxiv.org/abs/1911.11603}{{\ttfamily
  arXiv:1911.11603 [hep-th]}}.

\bibitem{Okuyama:2019xbv}
K.~Okuyama and K.~Sakai, ``{JT gravity, KdV equations and macroscopic loop
  operators},'' \href{http://dx.doi.org/10.1007/JHEP01(2020)156}{{\em JHEP}
  {\bfseries 01} (2020) 156}, \href{http://arxiv.org/abs/1911.01659}{{\ttfamily
  arXiv:1911.01659 [hep-th]}}.

\bibitem{Blommaert:2020seb}
A.~Blommaert, ``{Dissecting the ensemble in JT gravity},''
  \href{http://dx.doi.org/10.1007/JHEP09(2022)075}{{\em JHEP} {\bfseries 09}
  (2022) 075}, \href{http://arxiv.org/abs/2006.13971}{{\ttfamily
  arXiv:2006.13971 [hep-th]}}.

\bibitem{Saad:2021rcu}
P.~Saad, S.~H. Shenker, D.~Stanford, and S.~Yao, ``{Wormholes without
  averaging},'' \href{http://arxiv.org/abs/2103.16754}{{\ttfamily
  arXiv:2103.16754 [hep-th]}}.

\bibitem{Post:2022dfi}
B.~Post, J.~van~der Heijden, and E.~Verlinde, ``{A universe field theory for JT
  gravity},'' \href{http://dx.doi.org/10.1007/JHEP05(2022)118}{{\em JHEP}
  {\bfseries 05} (2022) 118}, \href{http://arxiv.org/abs/2201.08859}{{\ttfamily
  arXiv:2201.08859 [hep-th]}}.

\bibitem{Altland:2022xqx}
A.~Altland, B.~Post, J.~Sonner, J.~van~der Heijden, and E.~P. Verlinde,
  ``{Quantum chaos in 2D gravity},''
  \href{http://dx.doi.org/10.21468/SciPostPhys.15.2.064}{{\em SciPost Phys.}
  {\bfseries 15} no.~2, (2023) 064},
  \href{http://arxiv.org/abs/2204.07583}{{\ttfamily arXiv:2204.07583
  [hep-th]}}.

\bibitem{Jafferis:2022wez}
D.~L. Jafferis, D.~K. Kolchmeyer, B.~Mukhametzhanov, and J.~Sonner, ``{JT
  gravity with matter, generalized ETH, and Random Matrices},''
  \href{http://arxiv.org/abs/2209.02131}{{\ttfamily arXiv:2209.02131
  [hep-th]}}.

\bibitem{Blommaert:2021fob}
A.~Blommaert, L.~V. Iliesiu, and J.~Kruthoff, ``{Gravity factorized},''
  \href{http://dx.doi.org/10.1007/JHEP09(2022)080}{{\em JHEP} {\bfseries 09}
  (2022) 080}, \href{http://arxiv.org/abs/2111.07863}{{\ttfamily
  arXiv:2111.07863 [hep-th]}}.

\bibitem{Griguolo:2023aem}
L.~Griguolo, L.~Guerrini, R.~Panerai, J.~Papalini, and D.~Seminara,
  ``{Supersymmetric localization of (higher-spin) JT gravity: a bulk
  perspective},'' \href{http://arxiv.org/abs/2307.01274}{{\ttfamily
  arXiv:2307.01274 [hep-th]}}.

\bibitem{Mertens:2022irh}
T.~G. Mertens and G.~J. Turiaci, ``{Solvable models of quantum black holes: a
  review on Jackiw\textendash{}Teitelboim gravity},''
  \href{http://dx.doi.org/10.1007/s41114-023-00046-1}{{\em Living Rev. Rel.}
  {\bfseries 26} no.~1, (2023) 4},
  \href{http://arxiv.org/abs/2210.10846}{{\ttfamily arXiv:2210.10846
  [hep-th]}}.

\bibitem{Penington:2019kki}
G.~Penington, S.~H. Shenker, D.~Stanford, and Z.~Yang, ``{Replica wormholes and
  the black hole interior},''
  \href{http://dx.doi.org/10.1007/JHEP03(2022)205}{{\em JHEP} {\bfseries 03}
  (2022) 205}, \href{http://arxiv.org/abs/1911.11977}{{\ttfamily
  arXiv:1911.11977 [hep-th]}}.

\bibitem{Almheiri:2019qdq}
A.~Almheiri, T.~Hartman, J.~Maldacena, E.~Shaghoulian, and A.~Tajdini,
  ``{Replica Wormholes and the Entropy of Hawking Radiation},''
  \href{http://dx.doi.org/10.1007/JHEP05(2020)013}{{\em JHEP} {\bfseries 05}
  (2020) 013}, \href{http://arxiv.org/abs/1911.12333}{{\ttfamily
  arXiv:1911.12333 [hep-th]}}.

\bibitem{Kourkoulou:2017zaj}
I.~Kourkoulou and J.~Maldacena, ``{Pure states in the SYK model and
  nearly-$AdS_2$ gravity},'' \href{http://arxiv.org/abs/1707.02325}{{\ttfamily
  arXiv:1707.02325 [hep-th]}}.

\bibitem{Gao:2021uro}
P.~Gao, D.~L. Jafferis, and D.~K. Kolchmeyer, ``{An effective matrix model for
  dynamical end of the world branes in Jackiw-Teitelboim gravity},''
  \href{http://dx.doi.org/10.1007/JHEP01(2022)038}{{\em JHEP} {\bfseries 01}
  (2022) 038}, \href{http://arxiv.org/abs/2104.01184}{{\ttfamily
  arXiv:2104.01184 [hep-th]}}.

\bibitem{Moitra:2021uiv}
U.~Moitra, S.~K. Sake, and S.~P. Trivedi, ``{Jackiw-Teitelboim gravity in the
  second order formalism},''
  \href{http://dx.doi.org/10.1007/JHEP10(2021)204}{{\em JHEP} {\bfseries 10}
  (2021) 204}, \href{http://arxiv.org/abs/2101.00596}{{\ttfamily
  arXiv:2101.00596 [hep-th]}}.

\bibitem{Mertens:2020hbs}
T.~G. Mertens and G.~J. Turiaci, ``{Liouville quantum gravity -- holography, JT
  and matrices},'' \href{http://dx.doi.org/10.1007/JHEP01(2021)073}{{\em JHEP}
  {\bfseries 01} (2021) 073}, \href{http://arxiv.org/abs/2006.07072}{{\ttfamily
  arXiv:2006.07072 [hep-th]}}.

\bibitem{Fan:2021bwt}
Y.~Fan and T.~G. Mertens, ``{From quantum groups to Liouville and dilaton
  quantum gravity},'' \href{http://dx.doi.org/10.1007/JHEP05(2022)092}{{\em
  JHEP} {\bfseries 05} (2022) 092},
  \href{http://arxiv.org/abs/2109.07770}{{\ttfamily arXiv:2109.07770
  [hep-th]}}.

\bibitem{Goel:2020yxl}
A.~Goel, L.~V. Iliesiu, J.~Kruthoff, and Z.~Yang, ``{Classifying boundary
  conditions in JT gravity: from energy-branes to $\alpha$-branes},''
  \href{http://dx.doi.org/10.1007/JHEP04(2021)069}{{\em JHEP} {\bfseries 04}
  (2021) 069}, \href{http://arxiv.org/abs/2010.12592}{{\ttfamily
  arXiv:2010.12592 [hep-th]}}.

\bibitem{Suzuki:2021zbe}
K.~Suzuki and T.~Takayanagi, ``{JT gravity limit of Liouville CFT and matrix
  model},'' \href{http://dx.doi.org/10.1007/JHEP11(2021)137}{{\em JHEP}
  {\bfseries 11} (2021) 137}, \href{http://arxiv.org/abs/2108.12096}{{\ttfamily
  arXiv:2108.12096 [hep-th]}}.

\bibitem{Collier:2023cyw}
S.~Collier, L.~Eberhardt, B.~M\"uhlmann, and V.~A. Rodriguez, ``{The Virasoro
  Minimal String},'' \href{http://arxiv.org/abs/2309.10846}{{\ttfamily
  arXiv:2309.10846 [hep-th]}}.

\bibitem{Blommaert:2023wad}
A.~Blommaert, T.~G. Mertens, and S.~Yao, ``{The q-Schwarzian and Liouville
  gravity},'' \href{http://arxiv.org/abs/2312.00871}{{\ttfamily
  arXiv:2312.00871 [hep-th]}}.

\bibitem{Gliozzi:1976qd}
F.~Gliozzi, J.~Scherk, and D.~I. Olive, ``{Supersymmetry, Supergravity Theories
  and the Dual Spinor Model},''
  \href{http://dx.doi.org/10.1016/0550-3213(77)90206-1}{{\em Nucl. Phys. B}
  {\bfseries 122} (1977) 253--290}.

\bibitem{Chamseddine:1991fg}
A.~H. Chamseddine, ``{Superstrings in arbitrary dimensions},''
  \href{http://dx.doi.org/10.1016/0370-2693(91)91215-H}{{\em Phys. Lett. B}
  {\bfseries 258} (1991) 97--103}.

\bibitem{Astorino:2002bj}
M.~Astorino, S.~Cacciatori, D.~Klemm, and D.~Zanon, ``{AdS(2) supergravity and
  superconformal quantum mechanics},''
  \href{http://dx.doi.org/10.1016/S0003-4916(03)00008-3}{{\em Annals Phys.}
  {\bfseries 304} (2003) 128--144},
  \href{http://arxiv.org/abs/hep-th/0212096}{{\ttfamily arXiv:hep-th/0212096}}.

\bibitem{Forste:2017kwy}
S.~Forste and I.~Golla, ``{Nearly AdS$_2$ sugra and the super-Schwarzian},''
  \href{http://dx.doi.org/10.1016/j.physletb.2017.05.039}{{\em Phys. Lett. B}
  {\bfseries 771} (2017) 157--161},
  \href{http://arxiv.org/abs/1703.10969}{{\ttfamily arXiv:1703.10969
  [hep-th]}}.

\bibitem{Forste:2017apw}
S.~F\"orste, J.~Kames-King, and M.~Wiesner, ``{Towards the Holographic Dual of
  N = 2 SYK},'' \href{http://dx.doi.org/10.1007/JHEP03(2018)028}{{\em JHEP}
  {\bfseries 03} (2018) 028}, \href{http://arxiv.org/abs/1712.07398}{{\ttfamily
  arXiv:1712.07398 [hep-th]}}.

\bibitem{CamposDelgado:2022cwu}
R.~Campos~Delgado and S.~Forste, ``{Lyapunov exponents in N=2 supersymmetric
  Jackiw-Teitelboim gravity},''
  \href{http://dx.doi.org/10.1016/j.physletb.2022.137550}{{\em Phys. Lett. B}
  {\bfseries 835} (2022) 137550},
  \href{http://arxiv.org/abs/2209.15456}{{\ttfamily arXiv:2209.15456
  [hep-th]}}.

\bibitem{Fan:2021wsb}
Y.~Fan and T.~G. Mertens, ``{Supergroup structure of Jackiw-Teitelboim
  supergravity},'' \href{http://dx.doi.org/10.1007/JHEP08(2022)002}{{\em JHEP}
  {\bfseries 08} (2022) 002}, \href{http://arxiv.org/abs/2106.09353}{{\ttfamily
  arXiv:2106.09353 [hep-th]}}.

\bibitem{Stanford:2019vob}
D.~Stanford and E.~Witten, ``{JT gravity and the ensembles of random matrix
  theory},'' \href{http://dx.doi.org/10.4310/ATMP.2020.v24.n6.a4}{{\em Adv.
  Theor. Math. Phys.} {\bfseries 24} no.~6, (2020) 1475--1680},
  \href{http://arxiv.org/abs/1907.03363}{{\ttfamily arXiv:1907.03363
  [hep-th]}}.

\bibitem{Lin:2022zxd}
H.~W. Lin, J.~Maldacena, L.~Rozenberg, and J.~Shan, ``{Looking at
  supersymmetric black holes for a very long time},''
  \href{http://dx.doi.org/10.21468/SciPostPhys.14.5.128}{{\em SciPost Phys.}
  {\bfseries 14} no.~5, (2023) 128},
  \href{http://arxiv.org/abs/2207.00408}{{\ttfamily arXiv:2207.00408
  [hep-th]}}.

\bibitem{Turiaci:2023jfa}
G.~J. Turiaci and E.~Witten, ``{$\mathcal{N}=2$ JT Supergravity and Matrix
  Models},'' \href{http://arxiv.org/abs/2305.19438}{{\ttfamily arXiv:2305.19438
  [hep-th]}}.

\bibitem{Boruch:2023trc}
J.~Boruch, L.~V. Iliesiu, and C.~Yan, ``{Constructing all BPS black hole
  microstates from the gravitational path integral},''
  \href{http://arxiv.org/abs/2307.13051}{{\ttfamily arXiv:2307.13051
  [hep-th]}}.

\bibitem{Johnson:2019eik}
C.~V. Johnson, ``{Nonperturbative Jackiw-Teitelboim gravity},''
  \href{http://dx.doi.org/10.1103/PhysRevD.101.106023}{{\em Phys. Rev. D}
  {\bfseries 101} no.~10, (2020) 106023},
  \href{http://arxiv.org/abs/1912.03637}{{\ttfamily arXiv:1912.03637
  [hep-th]}}.

\bibitem{Johnson:2020heh}
C.~V. Johnson, ``{Jackiw-Teitelboim supergravity, minimal strings, and matrix
  models},'' \href{http://dx.doi.org/10.1103/PhysRevD.103.046012}{{\em Phys.
  Rev. D} {\bfseries 103} no.~4, (2021) 046012},
  \href{http://arxiv.org/abs/2005.01893}{{\ttfamily arXiv:2005.01893
  [hep-th]}}.

\bibitem{Johnson:2023ofr}
C.~V. Johnson, ``{Non-Perturbative ${\cal N}=2$ JT Supergravity},''
  \href{http://arxiv.org/abs/2306.10139}{{\ttfamily arXiv:2306.10139
  [hep-th]}}.

\bibitem{Fukuyama:1985gg}
T.~Fukuyama and K.~Kamimura, ``{Gauge Theory of Two-dimensional Gravity},''
  \href{http://dx.doi.org/10.1016/0370-2693(85)91322-X}{{\em Phys. Lett. B}
  {\bfseries 160} (1985) 259--262}.

\bibitem{Isler:1989hq}
K.~Isler and C.~A. Trugenberger, ``{A Gauge Theory of Two-dimensional Quantum
  Gravity},'' \href{http://dx.doi.org/10.1103/PhysRevLett.63.834}{{\em Phys.
  Rev. Lett.} {\bfseries 63} (1989) 834}.

\bibitem{Chamseddine:1989yz}
A.~H. Chamseddine and D.~Wyler, ``{Gauge Theory of Topological Gravity in
  (1+1)-Dimensions},''
  \href{http://dx.doi.org/10.1016/0370-2693(89)90528-5}{{\em Phys. Lett. B}
  {\bfseries 228} (1989) 75--78}.

\bibitem{Blommaert:2021etf}
A.~Blommaert and M.~Usatyuk, ``{Microstructure in matrix elements},''
  \href{http://dx.doi.org/10.1007/JHEP09(2022)070}{{\em JHEP} {\bfseries 09}
  (2022) 070}, \href{http://arxiv.org/abs/2108.02210}{{\ttfamily
  arXiv:2108.02210 [hep-th]}}.

\bibitem{Ammon:2013hba}
M.~Ammon, A.~Castro, and N.~Iqbal, ``{Wilson Lines and Entanglement Entropy in
  Higher Spin Gravity},'' \href{http://dx.doi.org/10.1007/JHEP10(2013)110}{{\em
  JHEP} {\bfseries 10} (2013) 110},
  \href{http://arxiv.org/abs/1306.4338}{{\ttfamily arXiv:1306.4338 [hep-th]}}.

\bibitem{Castro:2018srf}
A.~Castro, N.~Iqbal, and E.~Llabr\'es, ``{Wilson lines and Ishibashi states in
  AdS$_{3}$/CFT$_{2}$},'' \href{http://dx.doi.org/10.1007/JHEP09(2018)066}{{\em
  JHEP} {\bfseries 09} (2018) 066},
  \href{http://arxiv.org/abs/1805.05398}{{\ttfamily arXiv:1805.05398
  [hep-th]}}.

\bibitem{Mertens:2019tcm}
T.~G. Mertens and G.~J. Turiaci, ``{Defects in Jackiw-Teitelboim Quantum
  Gravity},'' \href{http://dx.doi.org/10.1007/JHEP08(2019)127}{{\em JHEP}
  {\bfseries 08} (2019) 127}, \href{http://arxiv.org/abs/1904.05228}{{\ttfamily
  arXiv:1904.05228 [hep-th]}}.

\bibitem{Vilenkin}
N.~Y. Vilenkin and A.~U. Klimyk, ``{Representation of Lie groups and Special
  Functions: Volume 1},'' {\em Kluwer Academic Publishers} (1991) .

\bibitem{PSHowe_1979}
P.~S. Howe, ``Super weyl transformations in two dimensions,''
  \href{https://dx.doi.org/10.1088/0305-4470/12/3/015}{{\em Journal of Physics
  A: Mathematical and General} {\bfseries 12} no.~3, (Mar, 1979) 393}.

\bibitem{Moore:1989yh}
G.~W. Moore and N.~Seiberg, ``{Taming the Conformal Zoo},''
  \href{http://dx.doi.org/10.1016/0370-2693(89)90897-6}{{\em Phys. Lett. B}
  {\bfseries 220} (1989) 422--430}.

\bibitem{Beasley:2009mb}
C.~Beasley, ``{Localization for Wilson Loops in Chern-Simons Theory},''
  \href{http://dx.doi.org/10.4310/ATMP.2013.v17.n1.a1}{{\em Adv. Theor. Math.
  Phys.} {\bfseries 17} no.~1, (2013) 1--240},
  \href{http://arxiv.org/abs/0911.2687}{{\ttfamily arXiv:0911.2687 [hep-th]}}.

\bibitem{Fan:2018wya}
Y.~Fan, ``{Localization and Non-Renormalization in Chern-Simons Theory},''
  \href{http://dx.doi.org/10.1007/JHEP01(2019)065}{{\em JHEP} {\bfseries 01}
  (2019) 065}, \href{http://arxiv.org/abs/1805.11076}{{\ttfamily
  arXiv:1805.11076 [hep-th]}}.

\bibitem{Penkov:1986ht}
I.~B. Penkov, ``{An introduction to geometric representation theory for complex
  simple Lie superalgebras},'' in {\em {XIII International Conference on
  Differential Geometric Methods in Theoretical Physics}}.
\newblock 1986.

\bibitem{Mikhaylov:2014aoa}
V.~Mikhaylov and E.~Witten, ``{Branes And Supergroups},''
  \href{http://dx.doi.org/10.1007/s00220-015-2449-y}{{\em Commun. Math. Phys.}
  {\bfseries 340} no.~2, (2015) 699--832},
  \href{http://arxiv.org/abs/1410.1175}{{\ttfamily arXiv:1410.1175 [hep-th]}}.

\bibitem{Gomis:1991cc}
J.~Gomis and J.~Roca, ``{Superfield description of N=2 topological
  supergravity},'' \href{http://dx.doi.org/10.1016/0370-2693(91)90803-X}{{\em
  Phys. Lett. B} {\bfseries 268} (1991) 197--202}.

\bibitem{Merbis:2023uax}
W.~Merbis, T.~Neogi, and A.~Ranjbar, ``{Asymptotic dynamics of three
  dimensional supergravity and higher spin gravity revisited},''
  \href{http://dx.doi.org/10.1007/JHEP06(2023)121}{{\em JHEP} {\bfseries 06}
  (2023) 121}, \href{http://arxiv.org/abs/2304.06761}{{\ttfamily
  arXiv:2304.06761 [hep-th]}}.

\bibitem{West:1990tg}
P.~C. West, ``{Introduction to supersymmetry and supergravity},'' {\em World
  Scientific} (1990) .

\bibitem{Okuyama:2023byh}
K.~Okuyama, ``{End of the world brane in double scaled SYK},''
  \href{http://dx.doi.org/10.1007/JHEP08(2023)053}{{\em JHEP} {\bfseries 08}
  (2023) 053}, \href{http://arxiv.org/abs/2305.12674}{{\ttfamily
  arXiv:2305.12674 [hep-th]}}.

\bibitem{Berkooz:2018jqr}
M.~Berkooz, M.~Isachenkov, V.~Narovlansky, and G.~Torrents, ``{Towards a full
  solution of the large N double-scaled SYK model},''
  \href{http://dx.doi.org/10.1007/JHEP03(2019)079}{{\em JHEP} {\bfseries 03}
  (2019) 079}, \href{http://arxiv.org/abs/1811.02584}{{\ttfamily
  arXiv:1811.02584 [hep-th]}}.

\bibitem{Berkooz:2022mfk}
M.~Berkooz, M.~Isachenkov, M.~Isachenkov, P.~Narayan, and V.~Narovlansky,
  ``{Quantum groups, non-commutative AdS$_{2}$, and chords in the double-scaled
  SYK model},'' \href{http://dx.doi.org/10.1007/JHEP08(2023)076}{{\em JHEP}
  {\bfseries 08} (2023) 076}, \href{http://arxiv.org/abs/2212.13668}{{\ttfamily
  arXiv:2212.13668 [hep-th]}}.

\bibitem{Blommaert:2023opb}
A.~Blommaert, T.~G. Mertens, and S.~Yao, ``{Dynamical actions and
  q-representation theory for double-scaled SYK},''
  \href{http://arxiv.org/abs/2306.00941}{{\ttfamily arXiv:2306.00941
  [hep-th]}}.

\bibitem{Lin:2022rbf}
H.~W. Lin, ``{The bulk Hilbert space of double scaled SYK},''
  \href{http://dx.doi.org/10.1007/JHEP11(2022)060}{{\em JHEP} {\bfseries 11}
  (2022) 060}, \href{http://arxiv.org/abs/2208.07032}{{\ttfamily
  arXiv:2208.07032 [hep-th]}}.

\bibitem{Goel:2023svz}
A.~Goel, V.~Narovlansky, and H.~Verlinde, ``{Semiclassical geometry in
  double-scaled SYK},'' \href{http://arxiv.org/abs/2301.05732}{{\ttfamily
  arXiv:2301.05732 [hep-th]}}.

\bibitem{Lin:2023trc}
H.~W. Lin and D.~Stanford, ``{A symmetry algebra in double-scaled SYK},''
  \href{http://arxiv.org/abs/2307.15725}{{\ttfamily arXiv:2307.15725
  [hep-th]}}.

\bibitem{Mertens:2022aou}
T.~G. Mertens, ``{Quantum exponentials for the modular double and applications
  in gravity models},'' \href{http://dx.doi.org/10.1007/JHEP09(2023)106}{{\em
  JHEP} {\bfseries 09} (2023) 106},
  \href{http://arxiv.org/abs/2212.07696}{{\ttfamily arXiv:2212.07696
  [hep-th]}}.

\bibitem{Ponsot:1999uf}
B.~Ponsot and J.~Teschner, ``{Liouville bootstrap via harmonic analysis on a
  noncompact quantum group},''
  \href{http://arxiv.org/abs/hep-th/9911110}{{\ttfamily arXiv:hep-th/9911110}}.

\bibitem{Teschner:2001rv}
J.~Teschner, ``{Liouville theory revisited},''
  \href{http://dx.doi.org/10.1088/0264-9381/18/23/201}{{\em Class. Quant.
  Grav.} {\bfseries 18} (2001) R153--R222},
  \href{http://arxiv.org/abs/hep-th/0104158}{{\ttfamily arXiv:hep-th/0104158}}.

\bibitem{Teschner:2013tqy}
J.~Teschner and G.~S. Vartanov, ``{Supersymmetric gauge theories, quantization
  of $\mathcal{M}_{\mathrm{flat}}$, and conformal field theory},''
  \href{http://dx.doi.org/10.4310/ATMP.2015.v19.n1.a1}{{\em Adv. Theor. Math.
  Phys.} {\bfseries 19} (2015) 1--135},
  \href{http://arxiv.org/abs/1302.3778}{{\ttfamily arXiv:1302.3778 [hep-th]}}.

\bibitem{Balasubramanian:2020jhl}
V.~Balasubramanian, A.~Kar, S.~F. Ross, and T.~Ugajin, ``{Spin structures and
  baby universes},'' \href{http://dx.doi.org/10.1007/JHEP09(2020)192}{{\em
  JHEP} {\bfseries 09} (2020) 192},
  \href{http://arxiv.org/abs/2007.04333}{{\ttfamily arXiv:2007.04333
  [hep-th]}}.

\bibitem{Shuji}
S.~{Matsumoto}, S.~{Uehara}, and Y.~{Yasui}, ``{A superparticle on the super
  Riemann surface},'' \href{http://dx.doi.org/10.1063/1.528882}{{\em Journal of
  Mathematical Physics} {\bfseries 31} no.~2, (Feb., 1990) 476--501}.

\bibitem{Frappat:1996pb}
L.~Frappat, P.~Sorba, and A.~Sciarrino, ``{Dictionary on Lie superalgebras},''
  \href{http://arxiv.org/abs/hep-th/9607161}{{\ttfamily arXiv:hep-th/9607161}}.

\bibitem{Scheunert:1976wj}
M.~Scheunert, W.~Nahm, and V.~Rittenberg, ``{Irreducible Representations of the
  OSP(2,1) and SPL(2,1) Graded Lie Algebras},''
  \href{http://dx.doi.org/10.1063/1.523149}{{\em J. Math. Phys.} {\bfseries 18}
  (1977) 155}.

\bibitem{Gotz:2005jz}
G.~Gotz, T.~Quella, and V.~Schomerus, ``{Representation theory of sl(2|1)},''
  \href{http://dx.doi.org/10.1016/j.jalgebra.2007.03.012}{{\em J. Algebra}
  {\bfseries 312} (2007) 829--848},
  \href{http://arxiv.org/abs/hep-th/0504234}{{\ttfamily arXiv:hep-th/0504234}}.

\bibitem{GelfandNaimark}
I.~M. Gel'fand and M.~A. Naimark, ``Unitary representations of the classical
  groups,'' {\em Acad. Sci. USSR} (1950) .

\bibitem{knapptrapa}
A.~Knapp and P.~Trapa, {\em Representation Theory of Lie Groups},
  \href{http://dx.doi.org/10.1090/pcms/008/02}{ch.~Representations of
  Semisimple Lie Groups, pp.~5--87}.
\newblock IAS/Park City Mathematics Series, 01, 2000.

\bibitem{Jeffrey:2000}
A.~Jeffrey, ``{Representation Theory of Lie Groups},'' {\em IAS/Park City
  Mathematics Series} {\bfseries 8} (2000) .

\bibitem{Ahn:2003tt}
C.~Ahn, M.~Stanishkov, and M.~Yamamoto, ``{One point functions of N = 2
  superLiouville theory with boundary},''
  \href{http://dx.doi.org/10.1016/j.nuclphysb.2004.02.007}{{\em Nucl. Phys. B}
  {\bfseries 683} (2004) 177--195},
  \href{http://arxiv.org/abs/hep-th/0311169}{{\ttfamily arXiv:hep-th/0311169}}.

\bibitem{Fu:2016vas}
W.~Fu, D.~Gaiotto, J.~Maldacena, and S.~Sachdev, ``{Supersymmetric
  Sachdev-Ye-Kitaev models},''
  \href{http://dx.doi.org/10.1103/PhysRevD.95.026009}{{\em Phys. Rev. D}
  {\bfseries 95} no.~2, (2017) 026009},
  \href{http://arxiv.org/abs/1610.08917}{{\ttfamily arXiv:1610.08917
  [hep-th]}}. [Addendum: Phys.Rev.D 95, 069904 (2017)].

\bibitem{Gotz:2005ka}
G.~Gotz, T.~Quella, and V.~Schomerus, ``{Tensor products of psl(2$\vert$2)
  representations},'' \href{http://arxiv.org/abs/hep-th/0506072}{{\ttfamily
  arXiv:hep-th/0506072}}.

\bibitem{Gotz:2006qp}
G.~Gotz, T.~Quella, and V.~Schomerus, ``{The WZNW model on PSU(1,1$\vert$2)},''
  \href{http://dx.doi.org/10.1088/1126-6708/2007/03/003}{{\em JHEP} {\bfseries
  03} (2007) 003}, \href{http://arxiv.org/abs/hep-th/0610070}{{\ttfamily
  arXiv:hep-th/0610070}}.

\bibitem{Heydeman:2020hhw}
M.~Heydeman, L.~V. Iliesiu, G.~J. Turiaci, and W.~Zhao, ``{The statistical
  mechanics of near-BPS black holes},''
  \href{http://dx.doi.org/10.1088/1751-8121/ac3be9}{{\em J. Phys. A} {\bfseries
  55} no.~1, (2022) 014004}, \href{http://arxiv.org/abs/2011.01953}{{\ttfamily
  arXiv:2011.01953 [hep-th]}}.

\bibitem{Iliesiu:2021are}
L.~V. Iliesiu, M.~Kologlu, and G.~J. Turiaci, ``{Supersymmetric indices
  factorize},'' \href{http://dx.doi.org/10.1007/JHEP05(2023)032}{{\em JHEP}
  {\bfseries 05} (2023) 032}, \href{http://arxiv.org/abs/2107.09062}{{\ttfamily
  arXiv:2107.09062 [hep-th]}}.

\bibitem{Eguchi:1988af}
T.~Eguchi and A.~Taormina, ``{On the Unitary Representations of $N=2$ and $N=4$
  Superconformal Algebras},''
  \href{http://dx.doi.org/10.1016/0370-2693(88)90360-7}{{\em Phys. Lett. B}
  {\bfseries 210} (1988) 125--132}.

\bibitem{Aoyama:2015gna}
S.~Aoyama and Y.~Honda, ``{Spin-chain with PSU$(2\vert 2)\otimes$U(1)$^3$ and
  Non-linear $\Sigma$-model with D(2,1;$\gamma$)},''
  \href{http://dx.doi.org/10.1016/j.physletb.2015.03.006}{{\em Phys. Lett. B}
  {\bfseries 743} (2015) 531},
  \href{http://arxiv.org/abs/1502.03684}{{\ttfamily arXiv:1502.03684
  [hep-th]}}.

\bibitem{VanDerJeugt:1985hq}
J.~Van Der~Jeugt, ``{Irreducible representations of the exceptional Lie
  superalgebra $D(2,1 ;\alpha)$},''
  \href{http://dx.doi.org/10.1063/1.526547}{{\em J. Math. Phys.} {\bfseries 26}
  (1985) 913--924}.

\end{thebibliography}\endgroup
\end{document}